\documentclass[11pt,a4paper,twoside,openright]{book}

\usepackage{amsmath, amssymb, bm}
\usepackage{geometry}
\usepackage{xcolor}

\usepackage{graphicx}
\usepackage{subcaption}
\usepackage{booktabs}
\usepackage[hidelinks]{hyperref}

\usepackage{comment}
\usepackage[authoryear,round]{natbib}
\usepackage{algorithm}
\usepackage{algpseudocode}

\usepackage{listings}
\lstdefinestyle{pythonstyle}{
	language=Python,
	basicstyle=\ttfamily\small,
	numbers=left,
	numberstyle=\tiny,
	numbersep=8pt,
	showstringspaces=false,
	frame=single,
	tabsize=4,
	breaklines=true,
	keywordstyle=\color{blue!70!black}\bfseries,
	commentstyle=\color{green!40!black}\itshape,
	stringstyle=\color{red!60!black},
	columns=fullflexible,
	keepspaces=true
}
\lstnewenvironment{pycode}[1][]%
{\lstset{style=pythonstyle,#1}}{}

\newcommand{\Gop}{\mathcal{G}}   
\newcommand{\Dop}{\mathcal{D}}   
\newcommand{\Lop}{\mathcal{L}}   
\newcommand{\Aop}{\mathcal{A}}
\newcommand{\Bop}{\mathcal{B}}

\title{
	Motif: A Modular Finite-Volume Framework for Transient Incompressible Flow
	\\[1.5cm]
	\includegraphics[width=0.25\textwidth, angle=15]{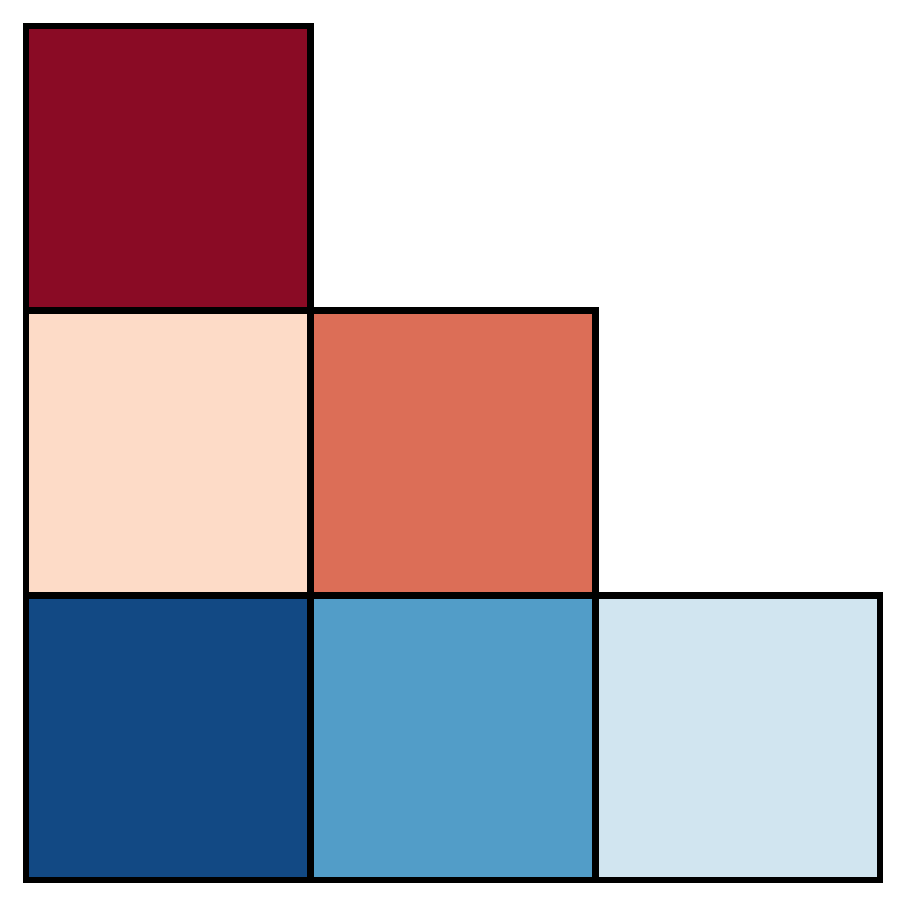}
}

\author{Utku Şentürk\\
	Ege University, Faculty of Engineering, \\
	Department of Mechanical
	Engineering, \\
	35040 Bornova, Izmir, Türkiye\\
	\texttt{utku.senturk@ege.edu.tr}}

\date{September 9, 2026}

\begin{document}
	
\pagestyle{plain}
	
\frontmatter	

\maketitle

\section*{Abstract}
	This document presents the theory, implementation, and verification of
	Motif, a two-dimensional incompressible flow solver developed as a
	teaching and research framework. The governing equations are discretized
	using the finite-volume method on staggered Cartesian grids. Convection
	is treated explicitly using the second-order Adams--Bashforth method,
	diffusion implicitly using the Crank--Nicolson method, and
	pressure--velocity coupling using projection methods. Both the Standard
	projection and the approximate second-order projection of Perot (1993)
	are considered. Particular attention is given to the discrete operators,
	boundary conditions, temporal treatment of pressure, conservation
	properties, and numerical dissipation and dispersion. The implementation is verified using a sequence of periodic, wall-bounded,
	and inlet--outlet flow problems on uniform and smoothly stretched grids. Spatial and temporal convergence,
	discrete mass conservation, kinetic-energy behavior, enstrophy, and
	energy spectra are examined. The results demonstrate the expected
	accuracy of the spatial discretization and distinguish the temporal
	behavior of the two projection approaches. The document is intended both
	to describe Motif in sufficient detail for independent implementation
	and to examine the numerical properties relevant to its continued
	development toward direct numerical simulation.


\cleardoublepage
\section*{Preface}
\addcontentsline{toc}{section}{Preface}

This document is intended to make the theory and implementation of a fluid solver named Motif publicly available. Motif grew out of the graduate-level courses CFD-I and CFD-II that I have taught since 2018. Incompressible flow, finite-volume discretization, grid layouts, time integration, and pressure--velocity coupling have been central topics in these courses. Hands-on sessions were carried out with students to help them write their own codes for the lid-driven cavity and Taylor--Green vortex problems. These codes are generally sufficient for teaching purposes. However, as more features are added, they become messy and difficult to organize.

Motif began as a small project in the summer of  2026 to restructure these teaching codes while experimenting with the recent AI tools that possess substantial coding capabilities. The grid, finite-volume operators, boundary
conditions, linear solvers, time-integration methods, and
pressure--velocity coupling algorithms were separated into components. In this way, each of the pieces was
tested independently. The initial
objective was therefore not to develop a new numerical method, but to
reorganize into a more modular framework. 

A second motivation came from personal experience as both a developer and a practitioner of CFD codes. General-purpose CFD software is designed to solve a very wide range of problems. But it is not developed with the numerical requirements of DNS as its primary objective. DNS places particular importance on properties such as discrete conservation, low numerical dissipation and dispersion, controlled numerical errors. The importance of these requirements became increasingly apparent as Motif developed. So they influenced many of the numerical choices examined in this document.

Projection method is one of the approaches for pressure--velocity coupling explained in my courses. It forms the basis of the present implementation. A common finite-volume structure like Motif makes it possible to implement other popular methods such as SIMPLE and PISO. The framework can also be extended to additional
physical problems. For example, natural-convection heat transfer or flow in non-inertial reference frames can be considered by incorporating the
required transport equations and source terms. By representing bodies using the immersed boundary techniques, external flow aerodynamics can be studied. In this sense, Motif is
intended to remain both an extensible tool and a teaching tool.

The progress in AI capabilities since they became accessible around 2022 has been breathtaking. This development also influenced the way of the presentation of the current work. Rather than simply deploying another CFD code for others to use, I chose to document the numerical formulation and implementation in sufficient detail that interested readers could construct their own solvers with the guidance of modern AI chatbots. This approach seems appropriate to the period in which this document is being written.

\vspace{1.5em}

\begin{flushright}
	İzmir, September 2026\\
	Utku Şentürk
\end{flushright}

\clearpage
\tableofcontents
\clearpage

\mainmatter

\chapter{Introduction}
\label{sec:introduction}

Motif is a finite-volume framework for the numerical solution of transient
incompressible flows on orthogonal Cartesian grids. It was developed with
three main objectives:
\begin{enumerate}
	\item to provide a teaching-friendly implementation,
	
	\item to provide a numerical framework suitable for direct numerical
	simulation (DNS),
	
	\item to provide a modular structure in which alternative solution
	techniques or additional physics can be implemented without changing
	the grid, variable arrangement, or basic discretization principles.
\end{enumerate}
The numerical formulation is therefore kept sufficiently simple that its
individual components can be examined, tested, and modified independently.

\section{Elements of incompressible CFD}
\label{sec:elements_incompressible_cfd}

A beginner in CFD, already familiar with basic numerical methods, might
expect the following recipe to work: Write down the governing
differential equations, choose a discretization method such as the finite
difference, finite volume, or finite element method, discretize the
equations on a sufficiently fine grid, and march the resulting algebraic
system forward in time. Why should solving the Navier--Stokes equations
require anything more? But it does.

The Navier--Stokes equations contain a remarkably rich range of mathematical
and physical behavior. One of the major distinctions in
CFD is that compressible and incompressible flows have developed into two
different worlds of numerical solution. In compressible flow,
density is a thermodynamic variable and pressure disturbances propagate at
a finite acoustic wave speed. The compressible equations possess a
wave-propagation character, and pressure can be obtained as part of the
thermodynamic state.

The situation is fundamentally different for incompressible flow. Density
is constant and the velocity field must be divergence-free. Pressure is no longer supplied by an independent thermodynamic relation.
Instead, it acts as the quantity that enforces this divergence constraint.
A change in the velocity field influences the pressure throughout the flow domain. The pressure field
must adjust so that the resulting velocity field remains divergence-free.
This is one reason why pressure--velocity coupling occupies such a central
place in incompressible CFD.

Once this difficulty is recognized, several other pieces of the numerical
method become closely connected. Where should pressure and velocity be
stored on the grid? How should the discrete divergence and pressure
gradient be constructed? How should pressure be calculated when there is
no pressure evolution equation? How should boundary conditions enter the
pressure problem? How should convection and diffusion be advanced in time?
These are not secondary implementation details. Together, they define the character of an incompressible flow solver. Different choices
lead to different families of incompressible flow algorithms. Motif provides one particular combination of these elements.

\section{Overview of Motif}
\label{sec:motif_overview}

Motif is a solver for transient, incompressible fluid flow. The current
implementation is limited to two-dimensional rectangular domains. The
computational grid is Cartesian and may be uniform or nonuniform. A
staggered arrangement is used for the pressure and velocity variables.

The present implementation focuses on projection methods for
pressure--velocity coupling. Both convection and diffusion are discretized using conservative finite-volume operators. Convective transport is evaluated from momentum fluxes through control-volume faces, while diffusive fluxes use face-normal velocity gradients obtained from a piecewise-linear velocity variation. 

Time advancement uses the second-order Adams--Bashforth method
for convection and the Crank--Nicolson method for diffusion. This semi-implicit treatment has a long history in incompressible DNS-based flow solvers
\citep{kim1985application,KimMoinMoser1987,perot1993analysis, VerziccoOrlandi1996,taira2007immersed,Mittal2008,brunton2013reduced,  Kajishima2017}. For teaching purposes, a fully explicit Euler method is implemented as well. Both the standard projection method \citep{Chorin1968} and a second-order approximate
projection based on the formulation of \citet{perot1993analysis} are
implemented. The same finite-volume structure is intended to support future implementations of alternative pressure--velocity coupling methods, such as
SIMPLE \citep{PatankarSpalding1972} and PISO \citep{Issa1986}.

The spatial operators, boundary conditions, linear solvers,
time-integration methods, and pressure--velocity coupling algorithms are
implemented as separate components. This modular structure permits different
solution algorithms to use the same grid and finite-volume operators.

\section{DNS suitability of Motif}

In broad terms, DNS refers to a numerical
solution of the viscous flow equations without the use of turbulence
modeling. This definition alone, however, does not establish the numerical
suitability of a method for DNS. The dynamically relevant spatial and
temporal scales must be resolved directly. Numerical errors arising from
the spatial discretization, pressure--velocity coupling, and time
integration must therefore remain sufficiently small that they do not
significantly alter the physics being resolved. The numerical
properties of the complete discretization must be examined.

General-purpose finite-volume solvers commonly favor robustness and
geometrical flexibility. Collocated pressure--velocity arrangements,
momentum interpolation, pressure-correction iterations, upwind-biased
convective schemes, and implicit time integration are examples
of methods commonly used for this purpose. These approaches are appropriate
for many CFD applications. However, they do not in general guarantee the discrete conservation and the low numerical dissipation and dispersion desirable for DNS.

Motif uses the classical staggered arrangement introduced by
\citet{HarlowWelch1965}, in which pressure is stored at pressure-cell centers
and the normal velocity components are stored at control-volume faces.
This arrangement avoids the spurious pressure modes associated with a
straightforward collocated discretization and provides a natural basis for
discrete conservation of mass, momentum, and kinetic energy
\citep{Morinishi1998,Perot2002}. These conservation properties also depend
on the particular discrete operators and are therefore examined explicitly
later in this work.

The pressure--velocity arrangement is particularly relevant. On collocated
grids, special momentum interpolation is commonly required to prevent
pressure--velocity decoupling. Rhie--Chow interpolation is widely used for
this purpose \citep{RhieChow1983}. Conventional collocated
pressure--velocity coupling can introduce errors in discrete kinetic-energy
conservation through the pressure and velocity-interpolation terms
\citep{HamIaccarino2004,FeltenLund2006}. These errors are not an inherent
consequence of collocation. Modified collocated formulations have been
developed that recover exact kinetic-energy conservation in the inviscid
limit \citep{Shashank2010}. The staggered arrangement used in Motif avoids
the need for such momentum interpolation because the normal velocity
components are already located at the faces where the finite-volume mass
fluxes are required.

With this in mind, DNS suitability is outlined separately. It is then assessed using a more challenging test problem than those used for verification. Discrete
mass and momentum conservation, pressure-work neutrality, kinetic-energy
and enstrophy behavior, spatial and temporal refinement, and the
distribution of kinetic energy over resolved wavenumbers are examined.

The remainder of this work develops the formulation from the governing
equations to its discrete implementation. The finite-volume operators,
boundary conditions, time-integration methods, and projection algorithms
are described first. Verification problems are then used to establish
spatial and temporal accuracy. Finally, conservation and spectral tests are
used to assess the suitability of the numerical method for DNS.

\chapter{Finite-Volume Discretization}

The finite-volume method expresses the governing conservation equations
through balances over discrete control volumes. Formal derivations and
detailed implementations of the method can be found in the classical works of
\citet{PatankarSpalding1972} and \citet{Ferziger2020}. The spatial
discretization adopted in Motif is designed to provide second-order accuracy
on uniform and smoothly stretched Cartesian grids. This is achieved through
1) second-order midpoint approximation of the control-volume face integrals,
2) linear interpolation of variables, 3) second-order approximations of
the face gradients required by the diffusive and pressure terms and 4) 
second-order boundary treatments implemented through ghost cells.
A staggered variable arrangement is used on a rectangular Cartesian domain,
as detailed in Section~\ref{sec:grid}.

\section{Governing Equations solved by Motif}

Motif solves the transient, two-dimensional incompressible Navier--Stokes equations on a Cartesian domain. The equations are written in conservative form,

\begin{align}
	\nabla \cdot \mathbf{V} &= 0,
	\label{eq:continuity}
	\\
	\frac{\partial \mathbf{V}}{\partial t}
	+
	\nabla \cdot
	\left(
	\mathbf{V}\mathbf{V}
	\right)
	&=
	-\frac{1}{\rho}\nabla p
	+
	\nabla \cdot
	\left(
	\nu \nabla \mathbf{V}
	\right),
	\label{eq:momentum}
\end{align}

where $\mathbf{V}=(u,v)^T$ denotes the velocity vector, $p$ is the pressure, $\rho$ is the constant density, and $\nu$ is the constant kinematic viscosity.

Expanding Eq.~\eqref{eq:momentum} into its Cartesian components gives

\begin{align}
	\frac{\partial u}{\partial t}
	+
	\nabla\cdot(\mathbf{V}u)
	&=
	-\frac{1}{\rho}
	\nabla p\cdot\mathbf{i}
	+
	\nabla\cdot(\nu\nabla u),
	\label{eq:xmomentum}
	\\
	\frac{\partial v}{\partial t}
	+
	\nabla\cdot(\mathbf{V}v)
	&=
	-\frac{1}{\rho}
	\nabla p\cdot\mathbf{j}
	+
	\nabla\cdot(\nu\nabla v).
	\label{eq:ymomentum}
\end{align}

These equations form the basis of the finite-volume discretization described in the following sections.

\section{Computational grid} \label{sec:grid}

The computational domain is the two-dimensional rectangular region

\begin{equation}
	\Omega
	=
	[0,L_x]\times[0,L_y].
	\label{eq:computational_domain}
\end{equation}

\begin{figure}[htbp]
	\centering
	\includegraphics[width=0.95\textwidth]{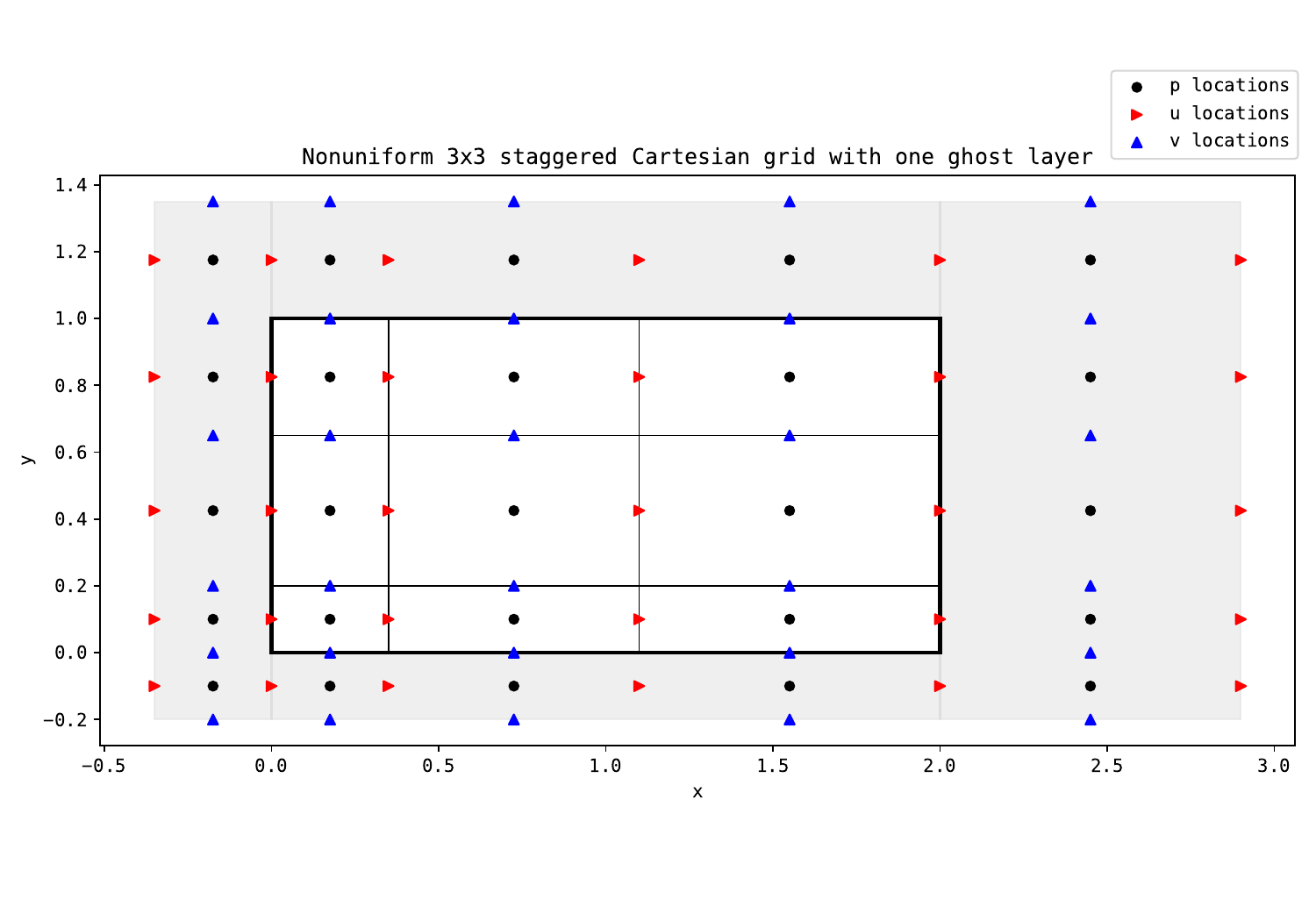}
	\caption{Staggered Cartesian grid used by Motif. Pressure is stored at
		cell centers, while the velocity components are stored on the
		corresponding pressure-cell faces. The shaded exterior region denotes
		the ghost layer used for boundary-condition implementation.}
	\label{fig:grid_layout}
\end{figure}

The domain is broken into $N_x$ pressure control volumes in
the $x$-direction and $N_y$ pressure control volumes in the $y$-direction.
The mesh is Cartesian and orthogonal. Both uniform and nonuniform meshes
are supported (Figure \ref{fig:grid_layout}).

One layer of ghost control volumes is added outside the physical domain for the implementation of boundary conditions. Accordingly,
the pressure field is stored on an $(N_x+2)\times(N_y+2)$ array. The
pressure-cell indices therefore span

\begin{equation}
	i=0,\ldots,N_x+1,
	\qquad
	j=0,\ldots,N_y+1.
	\label{eq:pressure_indices}
\end{equation}

The interior pressure control volumes occupy

\begin{equation}
	i=1,\ldots,N_x,
	\qquad
	j=1,\ldots,N_y,
	\label{eq:physical_indices}
\end{equation}

whereas $i=0$, $i=N_x+1$, $j=0$, and $j=N_y+1$ denote the ghost-cell
layer.

The staggered velocity components are stored on separate arrays. Since
the $u$-velocity is located on the pressure-cell faces normal to the
$x$-direction, there are $N_x+1$ physical $u$-locations in the
$x$-direction. Here, the term physical location refers to the interior and boundary locations combined. Including one ghost layer on each side, the resulting
array has dimensions

\begin{equation}
	(N_x+3)\times(N_y+2).
\end{equation}

Similarly, the $v$-velocity is stored on the pressure-cell faces normal
to the $y$-direction, yielding $N_y+1$ physical locations in the
$y$-direction. Including one ghost layer on each side, the corresponding
array dimensions become

\begin{equation}
	(N_x+2)\times(N_y+3).
\end{equation}

See the overview of this layout  in Table \ref{tab:array_dimensions}.

\begin{table}[!htbp]
	\centering
	\caption{Storage layout. One ghost layer is included on each side of every array.}
	\label{tab:array_dimensions}
	\begin{tabular}{lcc}
		\hline
		Variable & Physical locations & Stored array dimensions \\
		\hline
		Pressure $p$
		&
		$N_x \times N_y$
		&
		$(N_x+2)\times(N_y+2)$
		\\
		
		$x$-velocity $u$
		&
		$(N_x+1)\times N_y$
		&
		$(N_x+3)\times(N_y+2)$
		\\
		
		$y$-velocity $v$
		&
		$N_x\times(N_y+1)$
		&
		$(N_x+2)\times(N_y+3)$
		\\
		\hline
	\end{tabular}
\end{table}

Throughout this work, integer indices denote pressure-cell locations.
Variables stored on staggered velocity locations are identified using
half-integer indices relative to the neighboring pressure cells. Thus,
$u_{i+\frac12,j}$ is located on the east face of pressure cell $(i,j)$,
whereas $v_{i,j+\frac12}$ is located on its north face.

The pressure-cell faces are located at

\begin{equation}
	x_{\frac12},x_{\frac32},\ldots,x_{N_x+\frac12},
	\qquad
	y_{\frac12},y_{\frac32},\ldots,y_{N_y+\frac12},
	\label{eq:pressure_faces}
\end{equation}

where

\begin{equation}
	x_{\frac12}=0,
	\qquad
	x_{N_x+\frac12}=L_x,
\end{equation}

and

\begin{equation}
	y_{\frac12}=0,
	\qquad
	y_{N_y+\frac12}=L_y.
\end{equation}

The pressure-cell centers are located at the midpoints of the
adjacent faces,

\begin{equation}
	x_i
	=
	\frac{x_{i-\frac12}+x_{i+\frac12}}{2},
	\qquad
	i=1,\ldots,N_x,
	\label{eq:pressure_center_x}
\end{equation}

\begin{equation}
	y_j
	=
	\frac{y_{j-\frac12}+y_{j+\frac12}}{2},
	\qquad
	j=1,\ldots,N_y,
	\label{eq:pressure_center_y}
\end{equation}

with corresponding cell dimensions

\begin{equation}
	\Delta x_i
	=
	x_{i+\frac12}-x_{i-\frac12},
	\label{eq:pressure_cell_width}
\end{equation}

\begin{equation}
	\Delta y_j
	=
	y_{j+\frac12}-y_{j-\frac12},
	\label{eq:pressure_cell_height}
\end{equation}

and pressure control-volume area

\begin{equation}
	V^p_{i,j}
	=
	\Delta x_i\Delta y_j.
	\label{eq:pressure_cell_area}
\end{equation}

The term control-volume area in this work refers to the cell volume divided per unit depth as the grid is two-dimensional.

The active $u$-momentum control volumes are centered on the internal
vertical faces of the pressure cells. Their widths in the $x$-direction
are

\begin{equation}
	\Delta x^u_{i+\frac12}
	=
	x_{i+1}-x_i,
	\qquad
	i=1,\ldots,N_x-1.
	\label{eq:u_cv_width_internal_faces}
\end{equation}

The corresponding active $u$-momentum control-volume area is

\begin{equation}
	V^u_{i+\frac12,j}
	=
	\Delta x^u_{i+\frac12}\Delta y_j,
	\qquad
	i=1,\ldots,N_x-1.
	\label{eq:u_cv_area}
\end{equation}

Similarly, the active $v$-momentum control volumes are centered on the
internal horizontal faces of the pressure cells. Their heights are

\begin{equation}
	\Delta y^v_{j+\frac12}
	=
	y_{j+1}-y_j,
	\qquad
	j=1,\ldots,N_y-1.
	\label{eq:v_cv_height_internal_faces}
\end{equation}

The corresponding active $v$-momentum control-volume area is

\begin{equation}
	V^v_{i,j+\frac12}
	=
	\Delta x_i\Delta y^v_{j+\frac12},
	\qquad
	j=1,\ldots,N_y-1.
	\label{eq:v_cv_area}
\end{equation}

\section{Grid stretching} \label{sec:stretch}

In a nonuniform grid, the cell spacing varies with position and may change
arbitrarily between neighboring cells. This is particularly relevant to
the gradient approximations used in Motif. Consider, for example, the
pressure gradient at the east face $(i+1/2,j)$ of pressure control volume
$(i,j)$. Motif evaluates this gradient from the neighboring pressure values
as
\begin{equation}
	\left(
	\frac{\partial p}{\partial x}
	\right)_{i+1/2,j}
	\approx
	\frac{
		p_{i+1,j}-p_{i,j}
	}{
		x_{i+1,j}-x_{i,j}
	}.
\end{equation}
A Taylor-series expansion about the face location $x_{i+1/2,j}$ with the truncation error defined as the exact gradient minus its discrete approximation, gives
\begin{equation}
	\epsilon
	=
	\frac{
		(x_{i+1/2,j}-x_{i,j})^2
		-
		(x_{i+1,j}-x_{i+1/2,j})^2
	}{
		2(x_{i+1,j}-x_{i,j})
	}
	\left(
	\frac{\partial^2 p}{\partial x^2}
	\right)_{i+1/2,j}
	+
	O(\Delta x^2).
\end{equation}
If the face $(i+1/2,j)$ lies midway between the two pressure locations,
the two squared distances in the numerator are equal and the leading term
vanishes. The pressure-gradient approximation is then second-order
accurate. On a general nonuniform grid, however, the face need not coincide
with this midpoint, and the leading term does not vanish \citep{Ferziger2020}.

The purpose of smooth grid stretching is to make this midpoint mismatch decrease sufficiently rapidly under grid refinement. It can be shown that, for grids generated from a fixed smooth coordinate mapping, the difference between neighboring cell widths is of second order in the grid spacing. The corresponding face-to-midpoint displacement is therefore also of second order, allowing second-order convergence of the centered gradient approximation to be retained. This is particularly important in the present verification studies. Because first-order errors caused by grid nonuniformity could contaminate the measured spatial convergence of the numerical formulation. This is certainly not wanted.

\begin{figure}[!htbp]
	\centering
	
	\begin{subfigure}[t]{0.48\textwidth}
		\centering
		\includegraphics[width=\linewidth]{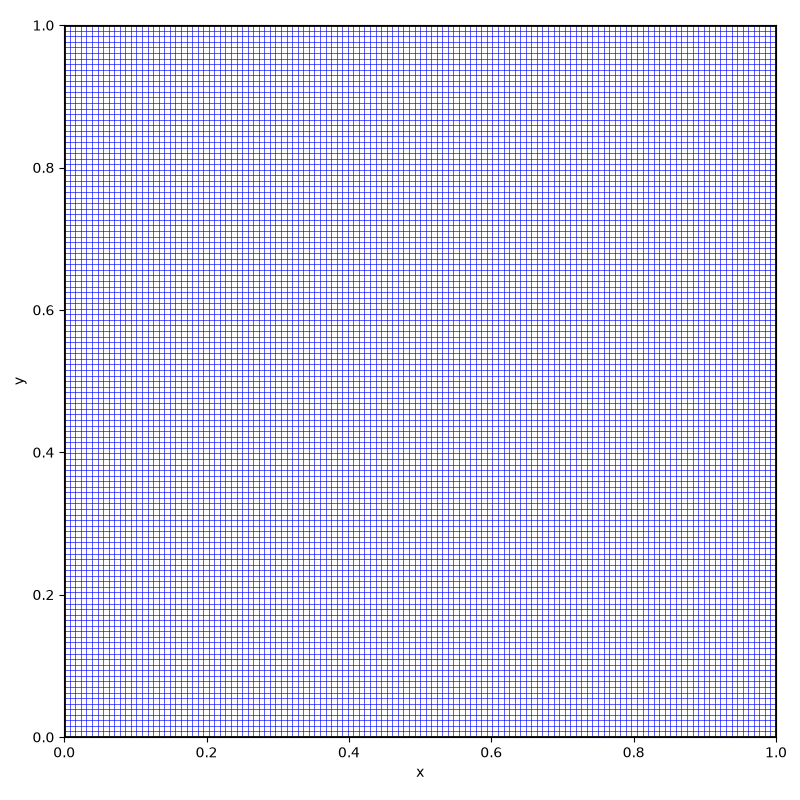}
		\caption{Uniform grid.}
		\label{fig:cavity_grid_uniform}
	\end{subfigure}
	\hfill
	\begin{subfigure}[t]{0.48\textwidth}
		\centering
		\includegraphics[width=\linewidth]{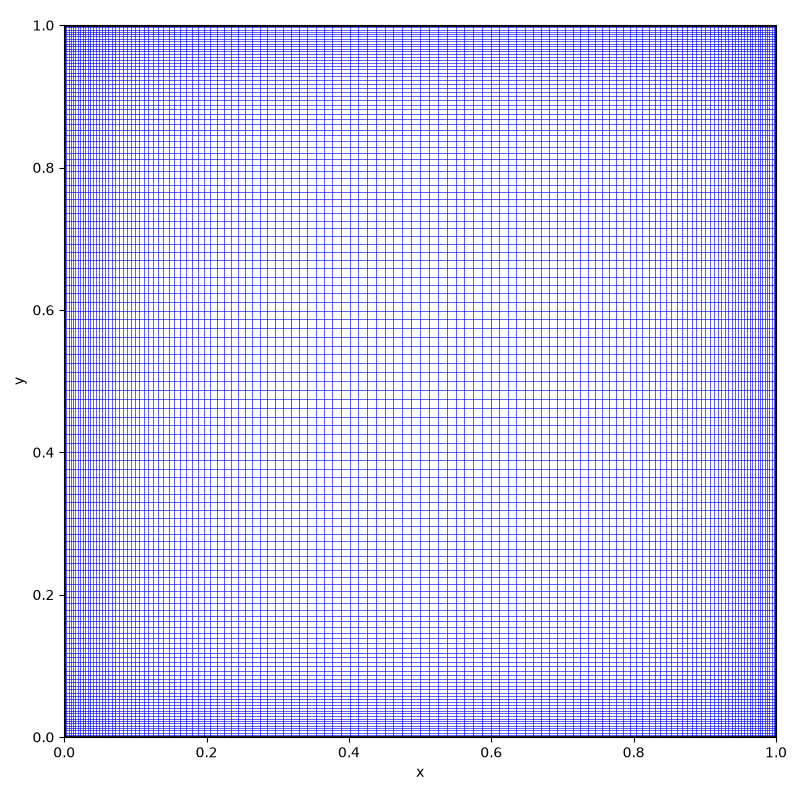}
		\caption{Stretched grid, $s_x=s_y=0.6$.}
		\label{fig:cavity_grid_nonuniform}
	\end{subfigure}
	
	\caption{Examples of $128\times128$ grids:
		(a) uniform and (b) stretched.}
	\label{fig:cavity_grids}
\end{figure}

For the smoothly stretched grids considered in the current code, the physical face coordinates are generated from uniformly distributed computational
coordinates $\xi,\eta\in[0,1]$ according to
\begin{align}
	x(\xi)
	&=
	L_x
	\left[
	\xi-\frac{s_x}{2\pi}\sin(2\pi\xi)
	\right],
	\\
	y(\eta)
	&=
	L_y
	\left[
	\eta-\frac{s_y}{2\pi}\sin(2\pi\eta)
	\right],
\end{align}
where $s_x$ and $s_y$ control the stretching in the two coordinate
directions \citep{Pandit2007}. Setting either stretching parameter to
zero recovers uniform spacing in the corresponding direction (Figure \ref{fig:cavity_grids}).

\section{Interpolating variables} \label{ssec:interp}

The staggered layout used in Motif requires spatial interpolations for the convective term. Distance-weighted linear interpolation is used on both uniform and nonuniform grids. For a generic variable $\phi$, if the target coordinate $x_t$ lies between the source coordinates $x_L$ and $x_R$, then the interpolated value is

\begin{equation}
	\phi_t
	=
	w_L\phi_L+w_R\phi_R,
	\label{eq:linear_interpolation}
\end{equation}

with

\begin{equation}
	w_L
	=
	\frac{x_R-x_t}{x_R-x_L},
	\qquad
	w_R
	=
	\frac{x_t-x_L}{x_R-x_L}.
	\label{eq:linear_interpolation_weights}
\end{equation}

These weights reproduce constant and linear fields exactly. On a uniform
mesh with the target located midway between the source points, they reduce
to

\begin{equation}
	w_L=w_R=\frac{1}{2}.
	\label{eq:uniform_interpolation_weights}
\end{equation}

Convective flux computations require interpolation in both coordinate directions. Motif performs these by applying one-dimensional linear interpolation successively in the $x$- and $y$-directions. For example, a value at the target point $(x_t,y_t)$ surrounded by the four source values $\phi_{LB}$, $\phi_{RB}$, $\phi_{LT}$, and $\phi_{RT}$ is evaluated as \citep{press2001numerical}:

\begin{align}
	\phi_t
	={}&
	w_L^x w_B^y \phi_{LB}
	+
	w_R^x w_B^y \phi_{RB}
	\nonumber\\
	&+
	w_L^x w_T^y \phi_{LT}
	+
	w_R^x w_T^y \phi_{RT}.
	\label{eq:bilinear_interpolation}
\end{align}

These interpolations are spatially second-order accurate.

\section{Spatial discretizations}

\subsection{Continuity equation}
\label{ssec:continuity}

The incompressibility constraint is integrated over the pressure
control volume $V^p_{i,j}$,

\begin{equation}
	\int_{V^p_{i,j}}
	\nabla\cdot\mathbf{V}\,dV
	=
	0.
\end{equation}

Application of the divergence theorem gives

\begin{equation}
	\int_{A^p_{i,j}}
	\mathbf{V}\cdot\mathbf{n}\,dA
	=
	0,
\end{equation}

where $A^p_{i,j}$ denotes the boundary of the pressure control
volume. Since the normal velocity components are stored directly at
the pressure-control-volume faces, no interpolation is required.
For the two-dimensional Cartesian grid, the discrete flux balance is

\begin{equation}
	u_{i+\frac12,j}\Delta y_j
	-
	u_{i-\frac12,j}\Delta y_j
	+
	v_{i,j+\frac12}\Delta x_i
	-
	v_{i,j-\frac12}\Delta x_i
	=
	0.
	\label{eq:continuity_flux_balance}
\end{equation}

Equivalently,

\begin{equation}
	\left(
	u_{i+\frac12,j}
	-
	u_{i-\frac12,j}
	\right)\Delta y_j
	+
	\left(
	v_{i,j+\frac12}
	-
	v_{i,j-\frac12}
	\right)\Delta x_i
	=
	0.
\end{equation}

Division by the pressure-control-volume area
$V^p_{i,j}=\Delta x_i\Delta y_j$ gives

\begin{equation}
	\frac{
		u_{i+\frac12,j}-u_{i-\frac12,j}
	}{
		\Delta x_i
	}
	+
	\frac{
		v_{i,j+\frac12}-v_{i,j-\frac12}
	}{
		\Delta y_j
	}
	=
	0.
	\label{eq:continuity_discrete}
\end{equation}

Thus, discrete mass conservation is imposed as a balance of
the volume fluxes through the four faces of each pressure control
volume. 

\subsection{Convective term}
\label{ssec:conv}

The convective terms are discretized in conservative form by evaluating
the net momentum flux through each staggered momentum control volume.
For the $u$-momentum equation,

\begin{equation}
	\int_{V^u_{i+\frac12,j}}
	\nabla\cdot(\mathbf{V}u)\,dV
	=
	\int_{A^u_{i+\frac12,j}}
	(\mathbf{V}u)\cdot\mathbf{n}\,dA,
	\label{eq:u_convective_integral}
\end{equation}

whereas for the $v$-momentum equation,

\begin{equation}
	\int_{V^v_{i,j+\frac12}}
	\nabla\cdot(\mathbf{V}v)\,dV
	=
	\int_{A^v_{i,j+\frac12}}
	(\mathbf{V}v)\cdot\mathbf{n}\,dA.
	\label{eq:v_convective_integral}
\end{equation}

For a $u$-momentum control volume centered at
$(x_{i+\frac12},y_j)$, the resulting integrated convective flux
imbalance is

\begin{align}
	\int_{A^u_{i+\frac12,j}}
	(\mathbf{V}u)\cdot\mathbf{n}\,dA
	={}&
	u_{i+1,j}^{\,2}\Delta y_j
	-
	u_{i,j}^{\,2}\Delta y_j
	\nonumber\\
	&+
	u_{i+\frac12,j+\frac12}
	v_{i+\frac12,j+\frac12}
	\Delta x^u_{i+\frac12}
	\nonumber\\
	&-
	u_{i+\frac12,j-\frac12}
	v_{i+\frac12,j-\frac12}
	\Delta x^u_{i+\frac12}.
	\label{eq:u_convective_flux}
\end{align}

Similarly, for a $v$-momentum control volume centered at
$(x_i,y_{j+\frac12})$,

\begin{align}
	\int_{A^v_{i,j+\frac12}}
	(\mathbf{V}v)\cdot\mathbf{n}\,dA
	={}&
	u_{i+\frac12,j+\frac12}
	v_{i+\frac12,j+\frac12}
	\Delta y^v_{j+\frac12}
	\nonumber\\
	&-
	u_{i-\frac12,j+\frac12}
	v_{i-\frac12,j+\frac12}
	\Delta y^v_{j+\frac12}
	\nonumber\\
	&+
	v_{i,j+1}^{\,2}\Delta x_i
	-
	v_{i,j}^{\,2}\Delta x_i.
	\label{eq:v_convective_flux}
\end{align}

Whenever a velocity is not stored at the required face location, it is obtained using the second-order, distance-weighted interpolation described in Section~\ref{ssec:interp}.

Equations~\eqref{eq:u_convective_flux} and
\eqref{eq:v_convective_flux} are integrated momentum-flux imbalances.
They are not divided by the corresponding momentum-control-volume
areas at this stage. The area normalization is introduced later when
the semi-discrete momentum equations are assembled.

\subsection{Diffusive term}
\label{ssec:diff}

The diffusive term is discretized by evaluating the net diffusive flux
through each staggered momentum control volume. For the $u$-momentum
equation,

\begin{equation}
	\int_{V^u_{i+\frac12,j}}
	\nabla\cdot\left(\nu\nabla u\right)\,dV
	=
	\int_{A^u_{i+\frac12,j}}
	\nu\nabla u\cdot\mathbf{n}\,dA,
	\label{eq:u_diffusive_integral}
\end{equation}

whereas for the $v$-momentum equation,

\begin{equation}
	\int_{V^v_{i,j+\frac12}}
	\nabla\cdot\left(\nu\nabla v\right)\,dV
	=
	\int_{A^v_{i,j+\frac12}}
	\nu\nabla v\cdot\mathbf{n}\,dA.
	\label{eq:v_diffusive_integral}
\end{equation}

Applying the finite-volume approximation, the integrated diffusive
flux imbalance over a $u$-momentum control volume centered at
$(x_{i+\frac12},y_j)$ becomes

\begin{align}
	\int_{A^u_{i+\frac12,j}}
	\nu\nabla u\cdot\mathbf{n}\,dA
	={}&
	\nu
	\frac{u_{i+\frac32,j}-u_{i+\frac12,j}}
	{x_{i+\frac32}-x_{i+\frac12}}
	\Delta y_j
	\nonumber\\
	&-
	\nu
	\frac{u_{i+\frac12,j}-u_{i-\frac12,j}}
	{x_{i+\frac12}-x_{i-\frac12}}
	\Delta y_j
	\nonumber\\
	&+
	\nu
	\frac{u_{i+\frac12,j+1}-u_{i+\frac12,j}}
	{y_{j+1}-y_j}
	\Delta x^u_{i+\frac12}
	\nonumber\\
	&-
	\nu
	\frac{u_{i+\frac12,j}-u_{i+\frac12,j-1}}
	{y_j-y_{j-1}}
	\Delta x^u_{i+\frac12}.
	\label{eq:u_diffusive_flux}
\end{align}

Similarly, for a $v$-momentum control volume centered at
$(x_i,y_{j+\frac12})$,

\begin{align}
	\int_{A^v_{i,j+\frac12}}
	\nu\nabla v\cdot\mathbf{n}\,dA
	={}&
	\nu
	\frac{v_{i+1,j+\frac12}-v_{i,j+\frac12}}
	{x_{i+1}-x_i}
	\Delta y^v_{j+\frac12}
	\nonumber\\
	&-
	\nu
	\frac{v_{i,j+\frac12}-v_{i-1,j+\frac12}}
	{x_i-x_{i-1}}
	\Delta y^v_{j+\frac12}
	\nonumber\\
	&+
	\nu
	\frac{v_{i,j+\frac32}-v_{i,j+\frac12}}
	{y_{j+\frac32}-y_{j+\frac12}}
	\Delta x_i
	\nonumber\\
	&-
	\nu
	\frac{v_{i,j+\frac12}-v_{i,j-\frac12}}
	{y_{j+\frac12}-y_{j-\frac12}}
	\Delta x_i.
	\label{eq:v_diffusive_flux}
\end{align}

The normal velocity gradients are evaluated directly at the momentum
control-volume faces using the actual geometric distances of the
nonuniform Cartesian mesh. No interpolations are needed.

Equations~\eqref{eq:u_diffusive_flux} and
\eqref{eq:v_diffusive_flux} are integrated diffusive flux imbalances.
They are not divided by the corresponding momentum-control-volume
areas at this stage. Division by these areas gives the diffusive terms
appearing in the semi-discrete momentum equations.

\subsection{Pressure term}
\label{ssec:pressure}

The pressure term is discretized by evaluating the corresponding
component of the net pressure force over each staggered momentum
control volume. For the $u$-momentum equation,

\begin{equation}
	\int_{V^u_{i+\frac12,j}}
	\nabla p\cdot\mathbf{i}\,dV
	=
	\int_{A^u_{i+\frac12,j}}
	p\mathbf{i}\cdot\mathbf{n}\,dA,
	\label{eq:u_pressure_integral}
\end{equation}

whereas for the $v$-momentum equation,

\begin{equation}
	\int_{V^v_{i,j+\frac12}}
	\nabla p\cdot\mathbf{j}\,dV
	=
	\int_{A^v_{i,j+\frac12}}
	p\mathbf{j}\cdot\mathbf{n}\,dA.
	\label{eq:v_pressure_integral}
\end{equation}

For a $u$-momentum control volume centered at
$(x_{i+\frac12},y_j)$, only the two faces normal to the $x$
direction contribute. The integrated pressure contribution is

\begin{equation}
	\int_{A^u_{i+\frac12,j}}
	p\mathbf{i}\cdot\mathbf{n}\,dA
	=
	\left(
	p_{i+1,j}-p_{i,j}
	\right)\Delta y_j.
	\label{eq:u_pressure_flux}
\end{equation}

Division by the $u$-momentum control-volume area

\begin{equation}
	V^u_{i+\frac12,j}
	=
	\left(x_{i+1}-x_i\right)\Delta y_j
\end{equation}

gives

\begin{equation}
	\frac{1}{V^u_{i+\frac12,j}}
	\int_{A^u_{i+\frac12,j}}
	p\mathbf{i}\cdot\mathbf{n}\,dA
	=
	\frac{p_{i+1,j}-p_{i,j}}
	{x_{i+1}-x_i}.
	\label{eq:u_pressure_discrete}
\end{equation}

Similarly, for a $v$-momentum control volume centered at
$(x_i,y_{j+\frac12})$, only the two faces normal to the $y$
direction contribute, giving

\begin{equation}
	\int_{A^v_{i,j+\frac12}}
	p\mathbf{j}\cdot\mathbf{n}\,dA
	=
	\left(
	p_{i,j+1}-p_{i,j}
	\right)\Delta x_i.
	\label{eq:v_pressure_flux}
\end{equation}

Division by the $v$-momentum control-volume area

\begin{equation}
	V^v_{i,j+\frac12}
	=
	\Delta x_i
	\left(y_{j+1}-y_j\right)
\end{equation}

gives

\begin{equation}
	\frac{1}{V^v_{i,j+\frac12}}
	\int_{A^v_{i,j+\frac12}}
	p\mathbf{j}\cdot\mathbf{n}\,dA
	=
	\frac{p_{i,j+1}-p_{i,j}}
	{y_{j+1}-y_j}.
	\label{eq:v_pressure_discrete}
\end{equation}

Thus, at internal staggered velocity locations, each component of the
pressure gradient is obtained directly from the pressures in the two
adjacent pressure control volumes. No pressure interpolation is
required.

\chapter{Discrete operators} \label{sec:discrete_operators}

An operator denotes a discrete mapping between fields. Operator notation is convenient for describing the
projection method and the associated linear systems. These operators will be presented in this section.

All expressions below refer to the nonuniform $3\times3$ staggered
Cartesian grid shown in Figure~\ref{fig:grid_layout}. 

\section{Divergence operator $\Dop$}
\label{ssec:divergence_op}

The $\Dop$ operator is a rule that takes the staggered velocity field
and returns one scalar value for every pressure control volume. It is
the volume-normalized form of the discretized continuity equation
derived in Section~\ref{ssec:continuity}.

For pressure control volume $V^p_{i,j}$, the integrated outward
volumetric flux imbalance is

\begin{equation}
	(\Dop_f\mathbf{V})_{i,j}
	=
	\left(
	u_{i+\frac12,j}
	-
	u_{i-\frac12,j}
	\right)\Delta y_j
	+
	\left(
	v_{i,j+\frac12}
	-
	v_{i,j-\frac12}
	\right)\Delta x_i,
	\label{eq:divergence_flux_operator}
\end{equation}

where $\Dop_f$ denotes the integrated flux form of the divergence.

The divergence operator $\Dop$ is obtained by dividing the integrated
flux imbalance by the pressure-control-volume area $V^p_{i,j} = \Delta x_i\Delta y_j$,

\begin{equation}
	(\Dop\mathbf{V})_{i,j}
	=
	\frac{
		(\Dop_f\mathbf{V})_{i,j}
	}{
		V^p_{i,j}
	},
	\label{eq:divergence_operator}
\end{equation}

consequently,

\begin{equation}
	(\Dop\mathbf{V})_{i,j}
	=
	\frac{
		u_{i+\frac12,j}
		-
		u_{i-\frac12,j}
	}{
		\Delta x_i
	}
	+
	\frac{
		v_{i,j+\frac12}
		-
		v_{i,j-\frac12}
	}{
		\Delta y_j
	}.
	\label{eq:divergence_operator_expanded}
\end{equation}

Equation~\eqref{eq:divergence_operator_expanded} is evaluated for

\begin{equation}
	i=1,\ldots,N_x,
	\qquad
	j=1,\ldots,N_y,
\end{equation}

and therefore produces one scalar divergence value at every physical
pressure control volume. Conceptually, the output of $\Dop$ may be
viewed as an $N_x\times N_y$ cell-centered array. For the
$3\times3$ grid considered here, the output therefore contains nine
cell-centered divergence values.

For nonperiodic boundaries, pressure control volumes adjacent to a
domain boundary use the velocity stored directly on the corresponding
physical boundary face. Ghost velocities do not enter the divergence
calculation directly. For periodic boundaries, the same
$N_x\times N_y$ set of pressure control volumes is retained, with the
velocity fluxes at the periodic seams supplied by the periodic
staggered field. Thus, periodicity changes the treatment of the
boundary-face velocities but not the number or location of the
divergence values.

In the Motif implementation, this operation is evaluated directly from
the staggered $u$- and $v$-velocity arrays whenever the divergence is
required. No discrete divergence matrix is assembled.

\section{Gradient operator $\Gop$} \label{ssec:gop}

The $\Gop$ operator is a rule that takes the cell-centered pressure
field and returns the two components of its gradient at the staggered
velocity locations. The $x$-component is evaluated at the $u$-locations,
whereas the $y$-component is evaluated at the $v$-locations. Both include the boundary faces.

At an internal $u$-location between the neighboring pressure control
volumes $(i,j)$ and $(i+1,j)$,

\begin{equation}
	(\Gop p)^u_{i+\frac12,j}
	=
	\frac{
		p_{i+1,j}-p_{i,j}
	}{
		x_{i+1}-x_i
	},
	\qquad
	i=1,\ldots,N_x-1,
	\qquad
	j=1,\ldots,N_y.
	\label{eq:gradient_u}
\end{equation}

Thus, the pressure gradient at an internal $u$-location is obtained
directly from the pressures in the two adjacent pressure control
volumes. The denominator is the actual distance between the two
pressure centers, and no pressure interpolation is required.

Similarly, at an internal $v$-location between pressure control
volumes $(i,j)$ and $(i,j+1)$,

\begin{equation}
	(\Gop p)^v_{i,j+\frac12}
	=
	\frac{
		p_{i,j+1}-p_{i,j}
	}{
		y_{j+1}-y_j
	},
	\qquad
	i=1,\ldots,N_x,
	\qquad
	j=1,\ldots,N_y-1.
	\label{eq:gradient_v}
\end{equation}

Again, the denominator is the actual distance between the adjacent
pressure centers.

The gradient operator is also evaluated at the physical
domain-boundary velocity locations. For a nonperiodic boundary, the
corresponding pressure ghost value imposed by the pressure boundary
condition is used together with the adjacent physical pressure value.
For a periodic boundary, the pressure values in the physical cells on
the two sides of the periodic seam are used, together with the
pressure-center distance across the seam.

Consequently, for a grid containing $N_x\times N_y$ physical pressure
control volumes, the complete physical $x$-component of the gradient
may be viewed as an $(N_x+1)\times N_y$ array,

\begin{equation}
	(\Gop p)^u
	=
	\left[
	(\Gop p)^u_{i+\frac12,j}
	\right],
	\qquad
	i=0,\ldots,N_x,
	\qquad
	j=1,\ldots,N_y.
\end{equation}

whereas the complete physical $y$-component may be viewed as an
$N_x\times(N_y+1)$ array,

\begin{equation}
	(\Gop p)^v
	=
	\left[
	(\Gop p)^v_{i,j+\frac12}
	\right],
	\qquad
	i=1,\ldots,N_x,
	\qquad
	j=0,\ldots,N_y.
\end{equation}

For the $3\times3$ pressure grid considered here, $\Gop p$
contains $4\times3$ physical $x$-gradient values at the $u$-locations
and $3\times4$ physical $y$-gradient values at the $v$-locations.
These physical array sizes are the same for nonperiodic and periodic
boundaries. 

In the Motif implementation, the two gradient components are returned
together as a staggered velocity-type field. The operation is evaluated
directly from pressure values and the corresponding geometric
pressure-center distances. No discrete gradient matrix is assembled
for this calculation.

\section{Pressure Laplacian operator $\Lop$}

The pressure Laplacian operator $\Lop$ is a rule that takes the
cell-centered pressure field and returns one scalar value for every
physical pressure control volume. In Motif, it is not implemented as
an independent discrete stencil. Instead, the pressure gradient is
first evaluated at the staggered velocity locations and the divergence
of this gradient field is then evaluated at the pressure cells,

\begin{equation}
	\Lop p
	=
	\Dop(\Gop p).
	\label{eq:pressure_laplacian}
\end{equation}

Consider pressure control volume $V^p_{i,j}$. Application of $\Gop$
provides the pressure-gradient components on its four faces. For an
interior pressure control volume, these are

\begin{equation}
	(\Gop p)^u_{i+\frac12,j}
	=
	\frac{p_{i+1,j}-p_{i,j}}
	{x_{i+1}-x_i},
\end{equation}

\begin{equation}
	(\Gop p)^u_{i-\frac12,j}
	=
	\frac{p_{i,j}-p_{i-1,j}}
	{x_i-x_{i-1}},
\end{equation}

\begin{equation}
	(\Gop p)^v_{i,j+\frac12}
	=
	\frac{p_{i,j+1}-p_{i,j}}
	{y_{j+1}-y_j},
\end{equation}

and

\begin{equation}
	(\Gop p)^v_{i,j-\frac12}
	=
	\frac{p_{i,j}-p_{i,j-1}}
	{y_j-y_{j-1}}.
\end{equation}

The divergence operator is then applied to these face-centered
gradient values. The resulting pressure Laplacian is

\begin{equation}
	(\Lop p)_{i,j}
	=
	\frac{
		\left[
		(\Gop p)^u_{i+\frac12,j}
		-
		(\Gop p)^u_{i-\frac12,j}
		\right]\Delta y_j
		+
		\left[
		(\Gop p)^v_{i,j+\frac12}
		-
		(\Gop p)^v_{i,j-\frac12}
		\right]\Delta x_i
	}{
		V^p_{i,j}
	}.
	\label{eq:pressure_laplacian_fv}
\end{equation}

Since $V^p_{i,j} = \Delta x_i\Delta y_j,$ this may also be written as

\begin{align}
	(\Lop p)_{i,j}
	={}&
	\frac{1}{\Delta x_i}
	\left(
	\frac{p_{i+1,j}-p_{i,j}}
	{x_{i+1}-x_i}
	-
	\frac{p_{i,j}-p_{i-1,j}}
	{x_i-x_{i-1}}
	\right)
	\nonumber\\
	&+
	\frac{1}{\Delta y_j}
	\left(
	\frac{p_{i,j+1}-p_{i,j}}
	{y_{j+1}-y_j}
	-
	\frac{p_{i,j}-p_{i,j-1}}
	{y_j-y_{j-1}}
	\right).
	\label{eq:pressure_laplacian_expanded}
\end{align}

Thus, the five-point pressure stencil that is familiar from finite-difference type discretizations emerges from applying
the same discrete gradient and divergence operators presented above. For the orthogonal Cartesian grids
considered in Motif, $\Lop=\Dop\Gop$ is algebraically equivalent to
the corresponding five-point pressure Laplacian on both uniform and
nonuniform grids. Both approaches produce the same numerical results and conservation
properties. The $\Dop\Gop$ form is used, because it follows naturally
from the projection formulation and reuses the existing discrete
operators.

Equation~\eqref{eq:pressure_laplacian_fv} is evaluated for

\begin{equation}
	i=1,\ldots,N_x,
	\qquad
	j=1,\ldots,N_y,
\end{equation}

so that one pressure-Laplacian value is obtained at every physical
pressure control volume. For the $3\times3$
grid considered here, $\Lop p$ contains nine scalar values.

For pressure control volumes adjacent to a nonperiodic boundary, one
or more of the face-centered gradient values in
Equation~\eqref{eq:pressure_laplacian_fv} are boundary-face values.
These are obtained by $\Gop$ using the pressure ghost values imposed
by the corresponding pressure boundary conditions. For a periodic
boundary, the corresponding gradient is evaluated across the periodic
seam. Thus, periodicity changes the evaluation of the boundary-face
gradients but not the $N_x\times N_y$ locations or number of
pressure-Laplacian values.

In the Motif implementation, the operation is evaluated directly in
two stages. The pressure field is first passed to the gradient
operator, which produces the two staggered pressure-gradient fields.
These are then passed to the divergence operator, which returns the
$N_x\times N_y$ pressure-Laplacian values. No discrete Laplacian
matrix is assembled by this operator.

\section{Velocity Laplacian operators $\Lop_u$ and $\Lop_v$}
\label{ssec:velocity_laplacian}

The operators $\Lop_u$ and $\Lop_v$ arise from the finite-volume
discretization of the diffusive terms given in
Section~\ref{ssec:diff}. The operator $\Lop_u$ is a rule that takes
the staggered $u$-velocity field and returns one scalar Laplacian
value at the $u$-velocity locations.
Similarly, $\Lop_v$ takes the staggered $v$-velocity field and returns
one scalar Laplacian value at the $v$-velocity locations.

For a $u$-momentum control volume centered at
$(x_{i+\frac12},y_j)$, Eq.~\eqref{eq:u_diffusive_flux} gives the
integrated diffusive flux imbalance. Dividing that equation by $\nu$
and by the momentum control-volume area

\begin{equation}
	V^u_{i+\frac12,j}
	=
	\Delta x^u_{i+\frac12}\Delta y_j
\end{equation}

defines

\begin{align}
	(\Lop_u u)_{i+\frac12,j}
	={}&
	\frac{1}{\Delta x^u_{i+\frac12}}
	\left(
	\frac{u_{i+\frac32,j}-u_{i+\frac12,j}}
	{x_{i+\frac32}-x_{i+\frac12}}
	-
	\frac{u_{i+\frac12,j}-u_{i-\frac12,j}}
	{x_{i+\frac12}-x_{i-\frac12}}
	\right)
	\nonumber\\
	&+
	\frac{1}{\Delta y_j}
	\left(
	\frac{u_{i+\frac12,j+1}-u_{i+\frac12,j}}
	{y_{j+1}-y_j}
	-
	\frac{u_{i+\frac12,j}-u_{i+\frac12,j-1}}
	{y_j-y_{j-1}}
	\right).
	\label{eq:velocity_laplacian_u}
\end{align}

The result is located at the same staggered $u$-location as the
velocity on which the $u$-momentum equation is advanced.

Similarly, for a $v$-momentum control volume centered at
$(x_i,y_{j+\frac12})$, division of
Eq.~\eqref{eq:v_diffusive_flux} by $\nu$ and by

\begin{equation}
	V^v_{i,j+\frac12}
	=
	\Delta x_i\Delta y^v_{j+\frac12}
\end{equation}

gives

\begin{align}
	(\Lop_v v)_{i,j+\frac12}
	={}&
	\frac{1}{\Delta x_i}
	\left( 
	\frac{v_{i+1,j+\frac12}-v_{i,j+\frac12}}
	{x_{i+1}-x_i}
	-
	\frac{v_{i,j+\frac12}-v_{i-1,j+\frac12}}
	{x_i-x_{i-1}}
	\right) 
	\nonumber\\
	&+
	\frac{1}{\Delta y^v_{j+\frac12}}
	\left( 
	\frac{v_{i,j+\frac32}-v_{i,j+\frac12}}
	{y_{j+\frac32}-y_{j+\frac12}}
	-
	\frac{v_{i,j+\frac12}-v_{i,j-\frac12}}
	{y_{j+\frac12}-y_{j-\frac12}}
	\right).
	\label{eq:velocity_laplacian_v}
\end{align}

The result is located at the corresponding staggered $v$-location.

For nonperiodic boundaries, the $u$-momentum equation is advanced at

\begin{equation}
	i=1,\ldots,N_x-1,
	\qquad
	j=1,\ldots,N_y,
\end{equation}

so that $\Lop_u u$ may be viewed as an
$(N_x-1)\times N_y$ array of Laplacian values at the internal
$u$-faces. Similarly, the $v$-momentum equation is advanced at

\begin{equation}
	i=1,\ldots,N_x,
	\qquad
	j=1,\ldots,N_y-1,
\end{equation}

and $\Lop_v v$ may be viewed as an
$N_x\times(N_y-1)$ array at the internal $v$-faces. The physical
velocity values on the corresponding domain boundaries are prescribed
by the boundary conditions and are not independent momentum unknowns.

When a direction is periodic, there is no physical boundary velocity
to exclude in that direction. The independent periodic $u$-locations
then number $N_x\times N_y$ when $x$ is periodic, whereas the
independent periodic $v$-locations number $N_x\times N_y$ when $y$ is
periodic. The periodic seam values required by the finite-volume
stencil are obtained from the corresponding values on the opposite
side of the domain.

Thus, for the $3\times3$ pressure grid considered here with
nonperiodic boundaries, $\Lop_u u$ contains $2\times3$ scalar values
and $\Lop_v v$ contains $3\times2$ scalar values. If the domain is
doubly periodic, each instead contains $3\times3$ independent
Laplacian values.

\section{Convective operators $\mathcal{C}^u$ and $\mathcal{C}^v$}
\label{ssec:convective_operator}

Unlike the velocity Laplacian
operators, each convective operator depends on both components of the
staggered velocity field because the transported momentum and the
normal advecting velocity are both required at the faces of a momentum
control volume.

For a $u$-momentum control volume centered at
$(x_{i+\frac12},y_j)$, Eq.~\eqref{eq:u_convective_flux} gives the
integrated convective momentum-flux imbalance. Division by the
momentum control-volume area

\begin{equation}
	V^u_{i+\frac12,j}
	=
	\Delta x^u_{i+\frac12}\Delta y_j
\end{equation}

defines

\begin{equation}
	\mathcal{C}^u(\mathbf{V})_{i+\frac12,j}
	=
	\frac{
		u_{i+1,j}^{\,2}-u_{i,j}^{\,2}
	}{
		\Delta x^u_{i+\frac12}
	}
	+
	\frac{
		u_{i+\frac12,j+\frac12}v_{i+\frac12,j+\frac12}
		-
		u_{i+\frac12,j-\frac12}v_{i+\frac12,j-\frac12}
	}{
		\Delta y_j
	}.
	\label{eq:convective_operator_u}
\end{equation}

Thus, $\mathcal{C}^u$ is a rule that takes the complete staggered
velocity field $\mathbf{V}=(u,v)^T$ and returns one scalar
area-normalized convective contribution at the
$u$-velocity locations.

Similarly, division of Eq.~\eqref{eq:v_convective_flux} by

\begin{equation}
	V^v_{i,j+\frac12}
	=
	\Delta x_i\Delta y^v_{j+\frac12}
\end{equation}

defines

\begin{equation}
	\mathcal{C}^v(\mathbf{V})_{i,j+\frac12}
	=
	\frac{
		u_{i+\frac12,j+\frac12}v_{i+\frac12,j+\frac12}
		-
		u_{i-\frac12,j+\frac12}v_{i-\frac12,j+\frac12}
	}{
		\Delta x_i
	}
	+
	\frac{
		v_{i,j+1}^{\,2}-v_{i,j}^{\,2}
	}{
		\Delta y^v_{j+\frac12}
	}.
	\label{eq:convective_operator_v}
\end{equation}

Accordingly, $\mathcal{C}^v$ takes the same staggered velocity field
and returns one scalar area-normalized convective contribution at the
$v$-velocity locations.

For nonperiodic boundaries, $\mathcal{C}^u$ is evaluated at

\begin{equation}
	i=1,\ldots,N_x-1,
	\qquad
	j=1,\ldots,N_y,
\end{equation}

and may therefore be viewed as an
$(N_x-1)\times N_y$ array of convective contributions at the internal
$u$-faces. Similarly, $\mathcal{C}^v$ is evaluated at

\begin{equation}
	i=1,\ldots,N_x,
	\qquad
	j=1,\ldots,N_y-1,
\end{equation}

and may be viewed as an $N_x\times(N_y-1)$ array at the internal
$v$-faces. The physical velocity values on the corresponding domain
boundaries are prescribed by the boundary conditions and are not
independently advanced by the momentum equations.

If the $x$ direction is periodic, the independent $u$-location at the
periodic seam is included, so that $\mathcal{C}^u$ contains
$N_x\times N_y$ independent values. If the $y$ direction is periodic,
the independent $v$-location at the periodic seam is included, so that
$\mathcal{C}^v$ contains $N_x\times N_y$ independent values. The
opposite representation of the same periodic seam is not counted as an
additional independent momentum location.

Thus, for the $3\times3$ pressure grid considered here with
nonperiodic boundaries, $\mathcal{C}^u$ contains $2\times3$ scalar
values and $\mathcal{C}^v$ contains $3\times2$ scalar values. For a
doubly periodic domain, each contains $3\times3$ independent values.

In the Motif implementation, the convection routine first evaluates
the integrated momentum-flux imbalances corresponding to
Eqs.~\eqref{eq:u_convective_flux} and
\eqref{eq:v_convective_flux}. These values are stored at the selected
staggered momentum locations. Division by the local momentum
control-volume areas is performed subsequently during assembly of the
momentum right-hand side.

A summary of all operators with their input and output locations are presented in Table \ref{tab:operator_summary}.

\begin{table}[!htbp]
	\centering
	\caption{Summary of the discrete operators for the case of $3\times3$
		pressure grid.}
	\label{tab:operator_summary}
	\begin{tabular}{lllll}
		\toprule
		Operator
		& \shortstack{Nonperiodic input\\locations}
		& \shortstack{Nonperiodic output\\locations}
		& \shortstack{Periodic input\\locations}
		& \shortstack{Periodic output\\locations} \\
		\midrule
		
		$\Dop$
		& $4\times3$ $u$; $3\times4$ $v$
		& $3\times3$ pressure
		& $3\times3$ $u$; $3\times3$ $v$
		& $3\times3$ pressure \\
		
		$\Gop$
		& $3\times3$ pressure
		& $4\times3$ $u$; $3\times4$ $v$
		& $3\times3$ pressure
		& $3\times3$ $u$; $3\times3$ $v$ \\
		
		$\Lop$
		& $3\times3$ pressure
		& $3\times3$ pressure
		& $3\times3$ pressure
		& $3\times3$ pressure \\
		
		$\Lop_u$
		& $2\times3$ $u$
		& $2\times3$ $u$
		& $3\times3$ $u$
		& $3\times3$ $u$ \\
		
		$\Lop_v$
		& $3\times2$ $v$
		& $3\times2$ $v$
		& $3\times3$ $v$
		& $3\times3$ $v$ \\
		
		$\mathcal{C}_u$
		& $4\times3$ $u$; $3\times4$ $v$
		& $2\times3$ $u$
		& $3\times3$ $u$; $3\times3$ $v$
		& $3\times3$ $u$ \\
		
		$\mathcal{C}_v$
		& $4\times3$ $u$; $3\times4$ $v$
		& $3\times2$ $v$
		& $3\times3$ $u$; $3\times3$ $v$
		& $3\times3$ $v$ \\
		
		$\Bop_2$
		& $2\times3$ $u$; $3\times2$ $v$
		& $2\times3$ $u$; $3\times2$ $v$
		& $3\times3$ $u$; $3\times3$ $v$
		& $3\times3$ $u$; $3\times3$ $v$ \\
		
		\bottomrule
	\end{tabular}
\end{table}

\chapter{Time Integration}

\section{Semi-discrete momentum equations}

In a semi-discrete formulation, spatial derivatives are discretized while time derivative remains continuous.
The semi-discrete momentum equations expressed in terms of the operators introduced above become

\begin{equation}
	\frac{\partial u}{\partial t}
	=
	-\mathcal{C}^u
	+
	\nu\Lop_u u
	-
	\frac{1}{\rho}
	(\Gop p)^u,
	\label{eq:semi_discrete_u}
\end{equation}

and

\begin{equation}
	\frac{\partial v}{\partial t}
	=
	-\mathcal{C}^v
	+
	\nu\Lop_v v
	-
	\frac{1}{\rho}
	(\Gop p)^v,
	\label{eq:semi_discrete_v}
\end{equation}

where $\mathcal{C}$ denotes the area-normalized convective operator
obtained from the integrated finite-volume flux balances presented in
Section~\ref{ssec:convective_operator}. The operators $\Lop_u$ and $\Lop_v$ are the
discrete Laplacian operators acting on the staggered $u$- and
$v$-velocity fields, respectively, as defined in
Section~\ref{ssec:velocity_laplacian}. Finally, pressure gradient operator is explained in Section~\ref{ssec:gop}.

These equations constitute a system of ordinary differential equations
in time for the staggered velocity field. Motif provides two temporal
integration schemes for their solution: the first-order Explicit Euler
method and a second-order Adams--Bashforth/Crank--Nicolson method. The latter is the default scheme.

\section{Explicit Euler scheme}

Motif provides a first-order Explicit Euler scheme primarily for
verification and algorithm development. Applying the Explicit Euler
method to Eqs.~(\ref{eq:semi_discrete_u}) and
(\ref{eq:semi_discrete_v}) gives

\begin{equation}
	u^{n+1}
	=
	u^{n}
	+
	\Delta t
	\left(
	-
	\mathcal{C}^{u,n}
	+
	\nu\Lop_u u^{n}
	-
	\frac{1}{\rho}
	(\Gop p)^{u,n}
	\right),
	\label{eq:euler_update_u}
\end{equation}

\begin{equation}
	v^{n+1}
	=
	v^{n}
	+
	\Delta t
	\left(
	-
	\mathcal{C}^{v,n}
	+
	\nu\Lop_v v^{n}
	-
	\frac{1}{\rho}
	(\Gop p)^{v,n}
	\right),
	\label{eq:euler_update_v}
\end{equation}

Here, the superscript $n$ denotes the current time level and $\Delta t$
is the time-step size. 

In the predictor step of the projection method, the pressure gradient is omitted from Equations \ref{eq:euler_update_u} and \ref{eq:euler_update_v}, yielding a pressure-free intermediate velocity field:
\begin{equation}
	u^{*}
	=
	u^{n}
	+
	\Delta t
	\left(
	-
	\mathcal{C}^{u,n}
	+
	\nu\Lop_u u^{n}
	\right),
	\label{eq:euler_update_u_no_p}
\end{equation}

\begin{equation}
	v^{*}
	=
	v^{n}
	+
	\Delta t
	\left(
	-
	\mathcal{C}^{v,n}
	+
	\nu\Lop_v v^{n}
	\right),
	\label{eq:euler_update_v_no_p}
\end{equation}
The superscript $*$ denotes the unknown intermediate velocity field. The pressure contribution is subsequently recovered through the projection procedure described in Section~\ref{sec:projection}.

\section{Adams--Bashforth/Crank--Nicolson scheme}

The default time-integration scheme implemented in Motif combines the
second-order Adams--Bashforth method for the convective term with the
second-order Crank--Nicolson method for the diffusive term \citep{kim1985application,perot1993analysis, taira2007immersed, brunton2013reduced,  Kajishima2017}. The convective and diffusive terms impose fundamentally different
stability constraints when treated explicitly. The explicit treatment of
convection is primarily limited by the Courant number,

\begin{equation}
	\mathrm{Co}
	=
	\max
	\left(
	\frac{|u|\Delta t}{\Delta x},
	\frac{|v|\Delta t}{\Delta y}
	\right),
	\label{eq:courant_number}
\end{equation}

whereas the explicit treatment of diffusion is subject to the more
restrictive parabolic stability condition,

\begin{equation}
	\nu\Delta t
	\left(
	\frac{1}{\Delta x^2}
	+
	\frac{1}{\Delta y^2}
	\right)
	\le \beta,
	\label{eq:diffusion_stability}
\end{equation}

where $\beta$ is a stability constant that depends on the temporal
discretization \citep{Ferziger2020}. Since the diffusive stability limit scales with the
square of the smallest grid spacing, it becomes restrictive more rapidly
than the convective stability limit as the mesh is refined.
Consequently, Motif advances the convective term explicitly using the
second-order Adams--Bashforth method while treating the diffusive term
implicitly using the second-order Crank--Nicolson method. 

The convective term is advanced explicitly because it is nonlinear in the
velocity field: $\nabla \cdot
\left(
\mathbf{V}\mathbf{V}\right) $. An implicit treatment would require the solution of a
nonlinear system at every time step through Newton iterations or
Picard-type linearizations, resulting in a higher
computational cost than the explicit Adams--Bashforth method.

Thus, the hybrid
approach retains second-order temporal accuracy while avoiding the
severe time-step restriction associated with explicit diffusion.

Applying these discretizations to the semi-discrete momentum equations gives

\begin{equation}
	u^{n+1}
	=
	u^{n}
	+
	\Delta t
	\left[
	-
	\left(
	\frac{3}{2}\mathcal{C}^{u,n}
	-
	\frac{1}{2}\mathcal{C}^{u,n-1}
	\right)
	+
	\frac{1}{2}
	\left(
	\nu\Lop_u u^{n+1}
	+
	\nu\Lop_u u^{n}
	\right)
	-
	\frac{1}{\rho}
	(\Gop p)^{u,n+1/2}
	\right],
	\label{eq:ab2_cn_u}
\end{equation}

and

\begin{equation}
	v^{n+1}
	=
	v^{n}
	+
	\Delta t
	\left[
	-
	\left(
	\frac{3}{2}\mathcal{C}^{v,n}
	-
	\frac{1}{2}\mathcal{C}^{v,n-1}
	\right)
	+
	\frac{1}{2}
	\left(
	\nu\Lop_v v^{n+1}
	+
	\nu\Lop_v v^{n}
	\right)
	-
	\frac{1}{\rho}
	(\Gop p)^{v,n+1/2}
	\right],
	\label{eq:ab2_cn_v}
\end{equation}

Because the diffusive operator depends on this unknown velocity, the
momentum equations are implicit.

As in the Explicit Euler scheme, the predictor step implemented in Motif omits the pressure gradient, producing a pressure-free intermediate velocity field:
\begin{equation}
	u^{*}
	=
	u^{n}
	-
	\Delta t
	\left(
	\frac{3}{2}\mathcal{C}^{u,n}
	-
	\frac{1}{2}\mathcal{C}^{u,n-1}
	\right)
	+
	\frac{\nu\Delta t}{2}
	\left(
	\Lop_u u^{*}
	+
	\Lop_u u^{n}
	\right),
	\label{eq:ab2_cn_u_laplacian}
\end{equation}

\begin{equation}
	v^{*}
	=
	v^{n}
	-
	\Delta t
	\left(
	\frac{3}{2}\mathcal{C}^{v,n}
	-
	\frac{1}{2}\mathcal{C}^{v,n-1}
	\right)
	+
	\frac{\nu\Delta t}{2}
	\left(
	\Lop_v v^{*}
	+
	\Lop_v v^{n}
	\right),
	\label{eq:ab2_cn_v_laplacian}
\end{equation}

The temporal location of the pressure term requires some attention.
The time integration of the momentum equation is performed over the
interval $[t^n,t^{n+1}]$. The Adams--Bashforth approximation represents
the time-integrated convective contribution with second-order accuracy,
while the Crank--Nicolson approximation does the same for the diffusive
contribution. Equivalently, these contributions are centered at the
midpoint of the time interval, $t^{n+1/2}$ \citep{ArmfieldStreet2003}.

For the complete momentum balance to retain the same second-order
temporal consistency, the pressure-gradient contribution must likewise
represent the pressure gradient at the midpoint of the time interval.
Accordingly, the pressure terms in Eqs.~(\ref{eq:ab2_cn_u}) and
(\ref{eq:ab2_cn_v}) are written as
\begin{equation}
	(\Gop p)^{u,n+1/2},
	\qquad
	(\Gop p)^{v,n+1/2}.
\end{equation}

This establishes the required temporal location of the pressure gradient
in the time-integrated momentum equations. It does not yet establish
that the pressure obtained from a particular projection approximation is
second-order accurate at this time level. The temporal accuracy of the
computed pressure is examined separately in the verification chapter.

The resulting equations are rearranged into the Helmholtz systems derived in the following subsection.

\section{Helmholtz equations}

Because the diffusive term is treated implicitly, the intermediate
velocity appears in the Crank--Nicolson contribution at the new time
level. Collecting the unknown diffusive contributions on the left-hand side
yields a Helmholtz equation for each velocity component,

\begin{equation}
	\Aop_u u^{*}=f^{u},
	\qquad
	\Aop_u
	=
	I-\frac{\nu\Delta t}{2}\Lop_u,
	\label{eq:helmholtz_u}
\end{equation}

and

\begin{equation}
	\Aop_v v^{*}=f^{v},
	\qquad
	\Aop_v
	=
	I-\frac{\nu\Delta t}{2}\Lop_v,
	\label{eq:helmholtz_v}
\end{equation}

where

\begin{equation}
	f^{u}
	=
	u^{n}
	-
	\Delta t
	\left(
	\frac{3}{2}\mathcal{C}^{u,n}
	-
	\frac{1}{2}\mathcal{C}^{u,n-1}
	\right)
	+
	\frac{\nu\Delta t}{2}
	\Lop_u u^{n},
	\label{eq:helmholtz_rhs_u}
\end{equation}

and

\begin{equation}
	f^{v}
	=
	v^{n}
	-
	\Delta t
	\left(
	\frac{3}{2}\mathcal{C}^{v,n}
	-
	\frac{1}{2}\mathcal{C}^{v,n-1}
	\right)
	+
	\frac{\nu\Delta t}{2}
	\Lop_v v^{n}.
	\label{eq:helmholtz_rhs_v}
\end{equation}

Here, $\Aop_u$ and $\Aop_v$ denote the Helmholtz operators for the
two staggered velocity components. Their corresponding linear-system
matrices are obtained after incorporating the velocity boundary
conditions through the ghost-layer relations, as described in
Chapter~\ref{sec:solvers}.

\chapter{Projection method}
\label{sec:projection}

The pressure field obtained during the projection step is denoted by
$p^{n+1/2}$. This notation reflects the temporal location of the
pressure-gradient contribution in the time-centered momentum equations
derived in the preceding chapter. It does not imply that the pressure
produced by a particular projection method is necessarily second-order
accurate at this time level.

\section{Standard projection}

Since the pressure gradient is omitted from the Helmholtz equations,
intermediate velocity is generally not divergence free:

\begin{equation}
	\Dop \mathbf{V}^{*}
	\neq
	0.
	\label{eq:intermediate_divergence}
\end{equation}

The purpose of the projection step is to construct a corrected velocity
field,
$\mathbf{V}^{n+1}$,
that satisfies the discrete incompressibility constraint,

\begin{equation}
	\Dop \mathbf{V}^{n+1}
	=
	0.
	\label{eq:projected_divergence}
\end{equation}

This is accomplished in two steps. In the first step, we derive and solve a pressure equation to impose a correction on the intermediate velocity. To derive the equation, we subtract the gradient of a scalar pressure field from the intermediate velocity,

\begin{equation}
	\mathbf{V}^{n+1}
	=
	\mathbf{V}^{*}
	-
	\frac{\Delta t}{\rho}
	\Gop p^{\,n+1/2}.
	\label{eq:velocity_projection}
\end{equation}

Applying the discrete
divergence operator to Equation~\ref{eq:velocity_projection} gives,

\begin{equation}
	\Dop\mathbf{V}^{n+1}
	=
	\Dop\mathbf{V}^{*}
	-
	\frac{\Delta t}{\rho}
	\Dop\left(
	\Gop p^{\,n+1/2}
	\right).
	\label{eq:projected_divergence_expanded}
\end{equation}

The corrected velocity must satisfy the discrete incompressibility
constraint in Equation \ref{eq:projected_divergence}, therefore,

\begin{equation}
	\Dop\left(
	\Gop p^{\,n+1/2}
	\right)
	=
	\frac{\rho}{\Delta t}
	\Dop\mathbf{V}^{*}.
	\label{eq:pressure_poisson_div_grad}
\end{equation}

Using the discrete pressure Laplacian introduced in
Equation~\ref{eq:pressure_laplacian},

\begin{equation}
	\Lop p
	=
	\Dop\left(
	\Gop p
	\right),
\end{equation}

the pressure Poisson equation is obtained:

\begin{equation}
	\Lop p^{\,n+1/2}
	=
	\frac{\rho}{\Delta t}
	\Dop\mathbf{V}^{*}.
	\label{eq:pressure_poisson}
\end{equation}

Equation~\ref{eq:pressure_poisson} is written using the
area-normalized divergence and Laplacian operators. In the linear-system
assembly, Motif uses the equivalent integrated finite-volume form,

\begin{equation}
	\Dop_f\left(
	\Gop p^{\,n+1/2}
	\right)
	=
	\frac{\rho}{\Delta t}
	\Dop_f\mathbf{V}^{*},
	\label{eq:pressure_poisson_flux_form}
\end{equation}

where $\Dop_f$ denotes the net outward flux imbalance over a pressure
control volume. The normalized and integrated forms are equivalent
because both sides of the equation are multiplied by the same local pressure
control-volume area.

The second step is the correction procedure. After the pressure Poisson equation has been solved, the pressure
gradient is evaluated at the staggered velocity locations and used to
correct the intermediate velocity according to
Equation~\ref{eq:velocity_projection}. In component form, the correction
at the physical $u$ locations is

\begin{equation}
	u^{n+1}_{i+\frac12,j}
	=
	u^{*}_{i+\frac12,j}
	-
	\frac{\Delta t}{\rho}
	\frac{
		p^{\,n+1/2}_{i+1,j}
		-
		p^{\,n+1/2}_{i,j}
	}{
		x_{i+1}-x_i
	},
	\label{eq:u_velocity_correction}
\end{equation}

whereas the correction at the physical $v$ locations is

\begin{equation}
	v^{n+1}_{i,j+\frac12}
	=
	v^{*}_{i,j+\frac12}
	-
	\frac{\Delta t}{\rho}
	\frac{
		p^{\,n+1/2}_{i,j+1}
		-
		p^{\,n+1/2}_{i,j}
	}{
		y_{j+1}-y_j
	}.
	\label{eq:v_velocity_correction}
\end{equation}

The correction is applied to all physical staggered velocity locations,
including the boundary-normal velocity locations. At wall 
boundaries, the normal pressure gradient is zero, so there is no actual
correction to the boundary-normal velocity. Including these locations
in the correction is therefore an implementation convenience. At
boundaries where pressure is prescribed, the boundary-normal velocity
may instead receive a nonzero projection correction. For doubly periodic domains, the duplicated boundary-face values and
the associated ghost layers are restored after the correction. 

\section{Perot projection}

In addition to the standard projection method, Motif implements the
second-order generalized projection method proposed by
\cite{perot1993analysis}. Rather than using the exact inverse of the
time-discrete momentum operator in the projection, Perot constructs a
second-order approximation to this inverse.

For the present Adams--Bashforth/Crank--Nicolson discretization, the
dimensionless Helmholtz operators for the staggered momentum equations
are

\begin{equation}
	\Aop_u
	=
	I
	-
	\frac{\nu\Delta t}{2}\Lop_u,
	\qquad
	\Aop_v
	=
	I
	-
	\frac{\nu\Delta t}{2}\Lop_v.
	\label{eq:perot_component_helmholtz}
\end{equation}

The corresponding second-order approximate inverse operators used in the Perot
projection are

\begin{equation}
	\Bop_2^u
	=
	\Delta t
	\left(
	2I-\Aop_u
	\right),
	\qquad
	\Bop_2^v
	=
	\Delta t
	\left(
	2I-\Aop_v
	\right).
	\label{eq:perot_component_b2}
\end{equation}

Substitution of Equation~\ref{eq:perot_component_helmholtz} gives

\begin{equation}
	\Bop_2^u
	=
	\Delta t
	\left(
	I
	+
	\frac{\nu\Delta t}{2}\Lop_u
	\right),
	\qquad
	\Bop_2^v
	=
	\Delta t
	\left(
	I
	+
	\frac{\nu\Delta t}{2}\Lop_v
	\right).
	\label{eq:perot_component_b2_expanded}
\end{equation}

When applied to the pressure-gradient field in the Perot projection,
these expressions become

\begin{equation}
	\Bop_2^u(\Gop p)^u
	=
	\Delta t
	\left[
	(\Gop p)^u
	+
	\frac{\nu\Delta t}{2}
	\Lop_u(\Gop p)^u
	\right],
\end{equation}

and

\begin{equation}
	\Bop_2^v(\Gop p)^v
	=
	\Delta t
	\left[
	(\Gop p)^v
	+
	\frac{\nu\Delta t}{2}
	\Lop_v(\Gop p)^v
	\right].
\end{equation}

The velocity Laplacians acting on the pressure gradients read,

\begin{align}
	\left[\Lop_u(\Gop p)^u\right]_{i+\frac12,j}
	={}&
	\frac{1}{\Delta x^u_{i+\frac12}}
	\left(
	\frac{
		(\Gop p)^u_{i+\frac32,j}
		-
		(\Gop p)^u_{i+\frac12,j}
	}{
		x_{i+\frac32}-x_{i+\frac12}
	}
	-
	\frac{
		(\Gop p)^u_{i+\frac12,j}
		-
		(\Gop p)^u_{i-\frac12,j}
	}{
		x_{i+\frac12}-x_{i-\frac12}
	}
	\right)
	\nonumber\\
	&+
	\frac{1}{\Delta y_j}
	\left(
	\frac{
		(\Gop p)^u_{i+\frac12,j+1}
		-
		(\Gop p)^u_{i+\frac12,j}
	}{
		y_{j+1}-y_j
	}
	-
	\frac{
		(\Gop p)^u_{i+\frac12,j}
		-
		(\Gop p)^u_{i+\frac12,j-1}
	}{
		y_j-y_{j-1}
	}
	\right),
\end{align}
and
\begin{align}
	\left[\Lop_v(\Gop p)^v\right]_{i,j+\frac12}
	={}&
	\frac{1}{\Delta x_i}
	\left(
	\frac{
		(\Gop p)^v_{i+1,j+\frac12}
		-
		(\Gop p)^v_{i,j+\frac12}
	}{
		x_{i+1}-x_i
	}
	-
	\frac{
		(\Gop p)^v_{i,j+\frac12}
		-
		(\Gop p)^v_{i-1,j+\frac12}
	}{
		x_i-x_{i-1}
	}
	\right)
	\nonumber\\
	&+
	\frac{1}{\Delta y^v_{j+\frac12}}
	\left(
	\frac{
		(\Gop p)^v_{i,j+\frac32}
		-
		(\Gop p)^v_{i,j+\frac12}
	}{
		y_{j+\frac32}-y_{j+\frac12}
	}
	-
	\frac{
		(\Gop p)^v_{i,j+\frac12}
		-
		(\Gop p)^v_{i,j-\frac12}
	}{
		y_{j+\frac12}-y_{j-\frac12}
	}
	\right).
\end{align}

Thus, $\Bop_2^u$ and $\Bop_2^v$ modify the corresponding staggered
pressure-gradient fields with the same velocity Laplacian operators
that appear diffusion term discretizations.

For compact notation, the two component operators may be collected
into the block-diagonal operator

\begin{equation}
	\Bop_2
	=
	\begin{bmatrix}
		\Bop_2^u & 0 \\
		0 & \Bop_2^v
	\end{bmatrix}.
	\label{eq:perot_global_b2}
\end{equation}

The pressure equation is then obtained by replacing the standard
velocity-correction operator with $\Bop_2$,

\begin{equation}
	\Dop
	\left(
	\Bop_2
	\Gop p^{\,n+1/2}
	\right)
	=
	\rho
	\Dop
	\mathbf{V}^{*},
	\label{eq:perot_pressure_equation}
\end{equation}

or, equivalently, in integrated finite-volume form,

\begin{equation}
	\Dop_f
	\left(
	\Bop_2
	\Gop p^{\,n+1/2}
	\right)
	=
	\rho
	\Dop_f
	\mathbf{V}^{*}.
	\label{eq:perot_pressure_flux}
\end{equation}

After the pressure field has been obtained, the corrected velocity is
computed as

\begin{equation}
	\mathbf{V}^{n+1}
	=
	\mathbf{V}^{*}
	-
	\frac{1}{\rho}
	\Bop_2
	\Gop p^{\,n+1/2}.
	\label{eq:perot_velocity_correction}
\end{equation}

Following Perot, we may define the first-order approximation to the inverse
momentum operator as,
\begin{equation}
	\Bop_1 = \Delta t\,I .
\end{equation}
For inviscid flow, $\nu=0$, the second-order approximation $\Bop_2$
reduces to $\Bop_1$ which is the standard projection.

\subsubsection{More on the operator $\Bop_2$}

The operator $\Bop_2$ is a rule that takes a staggered input field and
returns another field at the same staggered locations. In the Perot
projection, the input is the pressure-gradient field $\Gop p$. Its
$x$-component is acted on by $\Bop_2^u$ at the $u$-momentum locations,
whereas its $y$-component is acted on by $\Bop_2^v$ at the
$v$-momentum locations.

In Motif, $\Bop_2^u$ and $\Bop_2^v$ are represented as
sparse matrices. They are constructed according to
Equation~\ref{eq:perot_component_b2}. Thus, $\Bop_2$ is not evaluated
by separately applying the velocity Laplacian during every projection;
the two matrices are assembled once and subsequently applied to the
staggered input field.

For nonperiodic boundaries, $\Bop_2^u$ acts on the independent
$u$-momentum locations

\begin{equation}
	i=1,\ldots,N_x-1,
	\qquad
	j=1,\ldots,N_y,
\end{equation}

whereas $\Bop_2^v$ acts on

\begin{equation}
	i=1,\ldots,N_x,
	\qquad
	j=1,\ldots,N_y-1.
\end{equation}

The corresponding matrices therefore have dimensions
$(N_x-1)N_y\times(N_x-1)N_y$ and
$N_x(N_y-1)\times N_x(N_y-1)$, respectively. For the $3\times3$
pressure grid considered earlier, both matrices are therefore
$6\times6$. Conceptually, their outputs may be viewed as $2\times3$
and $3\times2$ arrays of independent staggered correction values.

For a doubly periodic domain, each velocity component has
$N_xN_y$ independent momentum locations. Consequently,
$\Bop_2^u$ and $\Bop_2^v$ are both
$N_xN_y\times N_xN_y$ matrices; for the $3\times3$ example, each is
$9\times9$. Periodic duplicate faces are restored after the matrix
application and are not additional independent unknowns.

\chapter{Boundary conditions}

Boundary conditions in Motif are imposed through the physical boundary values and the ghost cells surrounding the computational domain. The interior finite-volume
discretization is therefore applied uniformly throughout the domain,
including cells adjacent to the boundaries, without introducing
boundary-specific discretization stencils. Ghost-cell values are
constructed so that the desired velocity or pressure condition is
satisfied at the corresponding physical boundary.

The staggered arrangement leads to different array dimensions for the
pressure and velocity components. One ghost layer is allocated on every
side of each field, irrespective of the prescribed boundary-condition
type. Although some boundary conditions do not mathematically require
every ghost value, the uniform storage layout considerably simplifies
the implementation of the finite-volume operators and provides a common
framework for wall, inlet, outlet, and periodic boundaries.

Velocity boundary conditions are imposed twice during each time step.
Before the momentum predictor, ghost velocities are updated to provide
consistent boundary values for the convective, diffusive, and Helmholtz
operators. After the projection step, the velocity ghost cells are
updated again to enforce the final boundary conditions using the
corrected velocity field. Pressure boundary conditions are applied after
each pressure solve, immediately before evaluating the pressure gradient
used in the velocity correction.

\section{Velocity boundary conditions}

Because of the staggered arrangement, the normal and tangential velocity
components are not stored at the same locations relative to a physical
boundary. Consequently, they are treated differently.

Consider the west boundary. The normal velocity component is stored
directly on the boundary face at
$u_{\frac12,j}$, whereas the tangential velocity component is stored at
$v_{1,j+\frac12}$, one half-cell inside the computational domain.
Therefore, prescribed values of the normal velocity are assigned
directly, while prescribed tangential velocities are imposed through the
ghost layer.

For a stationary, no-slip west wall,

\begin{equation}
	u_{\frac12,j}=0,
	\label{eq:west_wall_u}
\end{equation}

is imposed directly on the boundary face. The corresponding tangential
ghost velocity is obtained by reflection,

\begin{equation}
	v_{0,j+\frac12}
	=
	-
	v_{1,j+\frac12},
	\label{eq:west_wall_v}
\end{equation}

which yields

\begin{equation}
	v_{\mathrm{wall}}
	=
	\frac{
		v_{0,j+\frac12}
		+
		v_{1,j+\frac12}
	}{2}
	=
	0,
\end{equation}

thereby satisfying the no-slip condition with second-order spatial
accuracy.

For a moving wall with prescribed tangential velocity
$V_{\mathrm{wall}}$,

\begin{equation}
	v_{0,j+\frac12}
	=
	2V_{\mathrm{wall}}
	-
	v_{1,j+\frac12},
	\label{eq:moving_wall_v}
\end{equation}

while the wall-normal velocity remains

\begin{equation}
	u_{\frac12,j}=0.
\end{equation}

For a prescribed velocity inlet,

\begin{equation}
	u_{\frac12,j}
	=
	U_{\mathrm{in}},
	\label{eq:inlet_u}
\end{equation}

is assigned directly, whereas the tangential component is imposed by

\begin{equation}
	v_{0,j+\frac12}
	=
	2V_{\mathrm{in}}
	-
	v_{1,j+\frac12}.
	\label{eq:inlet_v}
\end{equation}

Thus, only the tangential velocity requires a ghost-cell value. 

At a prescribed-pressure outlet, neither velocity component is known a
priori. Before the momentum predictor, both satisfy a zero normal
gradient,

\begin{equation}
	u_{N_x+\frac12,j}
	=
	u_{N_x-\frac12,j},
	\qquad
	v_{N_x+1,j+\frac12}
	=
	v_{N_x,j+\frac12},
	\label{eq:outlet_predictor}
\end{equation}

providing the ghost values required by the momentum operators.

After the pressure projection, however, the outlet-normal velocity is
updated through the pressure correction,

\begin{equation}
	u_{N_x+\frac12,j}^{\,n+1}
	=
	u_{N_x+\frac12,j}^{\,*}
	-
	\frac{\Delta t}{\rho}
	\left(
	\Gop p^{\,n+1/2}
	\right)_{N_x+\frac12,j},
	\label{eq:outlet_projection}
\end{equation}

and this corrected value is retained. The outlet tangential velocity
continues to satisfy the zero-normal-gradient condition,

\begin{equation}
	v_{N_x+1,j+\frac12}^{\,n+1}
	=
	v_{N_x,j+\frac12}^{\,n+1}.
	\label{eq:outlet_final}
\end{equation}

The remaining boundaries are treated analogously. At horizontal
boundaries, the roles of the $u$- and $v$-velocity components are
interchanged.

For periodic boundaries, ghost values are copied from the corresponding
interior values on the opposite side of the computational domain. This
construction allows the same finite-volume stencils to be applied across
periodic interfaces without modification.

\section{Pressure boundary conditions}

Pressure is stored at cell centers. Consequently, all pressure boundary
conditions are imposed through the surrounding ghost cells.

For a zero normal pressure gradient, the ghost value is copied from the
adjacent interior cell. For example, at the west boundary,

\begin{equation}
	p_{0,j}
	=
	p_{1,j},
	\label{eq:pressure_neumann_west}
\end{equation}

which gives

\begin{equation}
	\left.
	\frac{\partial p}{\partial x}
	\right|_{\mathrm{west}}
	=
	0.
\end{equation}

This is the case where the boundary velocity is prescribed. Another interpretation is that the projection does not modify
the prescribed boundary-normal velocity because the normal gradient of pressure is zero. The remaining boundaries are treated analogously.

For a prescribed pressure boundary with a pressure $p_b$ the ghost value is assigned with a linear interpolation. For example, at the west boundary,

\begin{equation}
	p_{0,j}
	=
	2p_b
	-
	p_{1,j},
	\label{eq:pressure_dirichlet}
\end{equation}

so that,

\begin{equation}
	p_b
	=
	\frac{
		p_{0,j}
		+
		p_{1,j}
	}{2}.
\end{equation}

This construction is spatially second-order accurate.

For periodic boundaries, the pressure ghost cells are obtained directly
from the corresponding interior cells on the opposite side of the
computational domain,

\begin{equation}
	p_{0,j}
	=
	p_{N_x,j},
	\qquad
	p_{N_x+1,j}
	=
	p_{1,j},
	\label{eq:pressure_periodic}
\end{equation}

with analogous relations in the $y$-direction.

When only Neumann or periodic pressure boundary
conditions are prescribed, the resulting linear system of Equation  \ref{eq:pressure_poisson_flux_form} determines the pressure up to an arbitrary additive constant. In such cases, Motif removes the null space
by fixing the pressure ($p_{\mathrm{ref}}$) at one reference cell. Unless otherwise stated, cell $(1,1)$ is used as the reference cell. When a pressure value is
prescribed on a boundary, the corresponding Dirichlet condition provides
the pressure reference directly.

The pressure ghost cells are refreshed immediately after each pressure
solve. The updated ghost values are then used to evaluate the staggered
pressure gradient during the subsequent velocity-correction step. In this way, the ghost values remain consistent
with the pressure field of the current projection step.

\chapter{Solution Procedure}

\section{Linear system solvers}\label{sec:solvers}

The finite-volume discretization and implicit time integration described
above generate sparse linear systems for the Helmholtz, Poisson, and
generalized Perot pressure equations. For a fixed grid, constant fluid
properties, and a constant time step, the corresponding coefficient
matrices remain unchanged throughout the simulation. Motif therefore
separates the construction and factorization of these matrices from the
repeated solution of the linear systems during time integration.

The pressure-free momentum predictor requires the two independent
Helmholtz systems introduced previously. After incorporation of the
ghost-layer boundary relations, their assembled forms are

\begin{equation}
	\Aop_u u^{*}=f^u+f^u_{\mathrm{BC}},
	\qquad
	\Aop_v v^{*}=f^v+f^v_{\mathrm{BC}},
\end{equation}

for the staggered $u$- and $v$-velocity components. Here,
$f^u_{\mathrm{BC}}$ and $f^v_{\mathrm{BC}}$ contain the known constant
contributions from the velocity boundary conditions. The coefficients
multiplying unknown velocities in the ghost-layer boundary relations are
incorporated into the corresponding Helmholtz matrix coefficients. The
two matrices are assembled independently because the velocity components
occupy different staggered momentum control volumes.

The pressure projection introduces a third linear system. For the
Standard projection, the coefficient matrix is assembled from the
discrete pressure operator

\begin{equation}
	\Dop_f\Gop .
\end{equation}

For the Perot projection, the second-order approximate inverse operators
$\Bop_2^u$ and $\Bop_2^v$ are first constructed from the assembled
Helmholtz matrices. The corresponding generalized pressure matrix is
then assembled from

\begin{equation}
	\Dop_f\Bop_2\Gop .
\end{equation}

In both cases, the pressure boundary conditions are incorporated into
the assembled linear system. Coefficients multiplying unknown pressures
in the boundary relations are incorporated into the pressure matrix,
while known prescribed pressure values contribute to the right-hand
side. Homogeneous Neumann conditions at solid walls introduce no
additional right-hand-side contribution. Periodic boundaries are
incorporated directly through the corresponding periodic discrete
operators. When the pressure problem has a constant null space, one
pressure value is prescribed as a reference.

The coefficient matrices are assembled using the sparse-matrix tools
provided by \texttt{scipy.sparse}. During construction they are stored
in compressed sparse row (CSR) format. Before factorization, each matrix
is converted to compressed sparse column (CSC) format and passed to the
SuperLU sparse direct solver through the
\texttt{scipy.sparse.linalg.splu} interface.

The resulting sparse LU factorizations are retained by the corresponding
solver objects. Thus, for an AB2--CN calculation, the $u$-Helmholtz
matrix, the $v$-Helmholtz matrix, and the pressure matrix associated
with the selected projection method are assembled and factorized during
initialization. These factorizations are then reused throughout time
integration.

At each time step, the explicit terms and the current velocity field,
together with any constant contributions from the velocity boundary
conditions, produce new right-hand-side vectors for the Helmholtz
systems. The stored LU factorizations are used to solve these systems
for the provisional velocity components. The resulting provisional
velocity then produces a new right-hand side for the pressure system,
which is solved using its stored factorization. No coefficient matrix is
reassembled or refactorized during this sequence.

This separation between matrix construction and repeated solution is
particularly advantageous for time integration. The relatively
expensive sparse matrix assembly and LU factorization are performed only
during initialization, while each time step requires only the solution
of the already factorized systems for new right-hand-side vectors.

\section{The algorithm}

The components described in the preceding chapters are combined
in Motif according to the following procedure:

\begin{enumerate}
	
	\item \textit{Initialization.}
	
	Before time integration begins, the grid, boundary conditions, linear
	solvers, and time-integration state are initialized.
	
	\begin{enumerate}
		
		\item The physical domain is discretized by specifying the pressure-cell
		face coordinates in the $x$- and $y$-directions. From these domain-edge
		coordinates, Motif constructs the pressure control volumes, staggered
		momentum control volumes, variable locations, and ghost layers.
		
		\item The initial staggered velocity field $\mathbf{V}^{0}$ is constructed
		at the corresponding $u$- and $v$-velocity locations.
		
		\item The velocity boundary-condition relations appropriate to the problem
		are specified.
		
		\item The pressure boundary-condition relations appropriate to the problem
		are specified.
		
		\item The Helmholtz matrices for the staggered $u$- and $v$-momentum
		equations are assembled and LU-factorized. The predictor-stage velocity
		boundary-condition relations are incorporated into the Helmholtz systems
		during their construction.
		
		\item The pressure solver is initialized according to the selected
		projection method.
		
		\begin{enumerate}
			
			\item For the standard projection, the pressure Poisson operator
			$\Dop_f\Gop$ is assembled and LU-factorized.
			
			\item For the Perot projection, the second-order approximate
			inverse operators $\Bop_2^u$ and $\Bop_2^v$ are constructed from
			the Helmholtz operators. The generalized pressure operator
			$\Dop_f\Bop_2\Gop$ is then assembled and LU-factorized.
			
		\end{enumerate}
		
		\item The Adams--Bashforth state is initialized with an empty convection history.
		
	\end{enumerate}	
	
	\item \textit{Momentum predictor.}
	
	At each time step, a pressure-free provisional velocity
	$\mathbf{V}^{*}$ is calculated.
	
	\begin{enumerate}
		
		\item The velocity field $\mathbf{V}^{n}$ is copied to a working
		velocity field.
		
		\item The predictor-stage velocity boundary conditions are applied to the
		working copy of $\mathbf{V}^{n}$, updating the required physical
		boundary-face velocities and ghost-layer velocities.
		
		\item The convection operator $\mathcal{C}$ is evaluated using the current
		velocity field $\mathbf{V}^{n}$, giving the integrated convective flux
		imbalance $\mathcal{C}_f^{\,n}$.
		
		\item On the first time step, convection is advanced using
		\begin{equation}
			\mathcal{C}_f^{\mathrm{AB}}
			=
			\mathcal{C}_f^{\,n}.
		\end{equation}
		On subsequent time steps, the second-order Adams--Bashforth
		approximation is used,
		\begin{equation}
			\mathcal{C}_f^{\mathrm{AB}}
			=
			\frac{3}{2}\mathcal{C}_f^{\,n}
			-
			\frac{1}{2}\mathcal{C}_f^{\,n-1}.
		\end{equation}
		
		The result is divided by the corresponding momentum
		control-volume size as a consequence of the finite-volume discretization.
		
		\item The Helmholtz systems for the $u$- and $v$-momentum equations
		are solved using the previously LU-factorized matrices. This gives the
		pressure-free provisional velocity $\mathbf{V}^{*}$.
		
		\item The predictor-stage velocity boundary conditions are applied
		again to $\mathbf{V}^{*}$. The physical boundary and ghost values
		required by the subsequent pressure projection are therefore
		consistent with the provisional velocity field.
		
	\end{enumerate}

	\item \textit{Pressure projection.}
	
	The provisional velocity $\mathbf{V}^{*}$ is projected to obtain the
	divergence-free velocity $\mathbf{V}^{n+1}$. The procedure depends on the
	selected projection method.
	
	\begin{enumerate}
		
		\item For the standard projection:
		
		\begin{enumerate}
			
			\item The discrete divergence of $\mathbf{V}^{*}$ is used to
			construct the right-hand side of the pressure Poisson equation.
			
			\item The pressure Poisson equation is solved for $p^{n+1/2}$.
			
			\item The pressure boundary conditions are applied to
			$p^{n+1/2}$. For nonperiodic boundaries these conditions are imposed
			through the pressure ghost cells. Periodic pressure ghost cells are filled with the corresponding pressure values from the opposite side of the domain.
			
			\item The staggered pressure gradient $\Gop p^{n+1/2}$ is evaluated
			using the pressure field with its updated ghost values.
			
			\item The provisional velocity is corrected according to
			Equation~\ref{eq:velocity_projection} to obtain
			$\mathbf{V}^{n+1}$.
			
		\end{enumerate}
		
		\item For the Perot projection:
		
		\begin{enumerate}
			
			\item The discrete divergence of $\mathbf{V}^{*}$ is used to
			construct the right-hand side of the generalized Perot pressure
			equation.
			
			\item The generalized pressure equation is solved for
			$p^{n+1/2}$.
			
			\item The pressure boundary conditions are applied to
			$p^{n+1/2}$. For nonperiodic boundaries these conditions are imposed
			through the pressure ghost cells. Periodic pressure ghost cells are filled with the corresponding pressure values from the opposite side of the domain.
			
			\item The staggered pressure gradient $\Gop p^{n+1/2}$ is
			evaluated.
			
			\item The second-order approximate inverse $\Bop_2$ is applied
			to the pressure-gradient field.
			
			\item The resulting correction is applied to
			$\mathbf{V}^{*}$ according to
			Equation~\ref{eq:perot_velocity_correction} to obtain
			$\mathbf{V}^{n+1}$.
			
		\end{enumerate}
		
		\item For a doubly periodic domain, duplicate periodic velocity faces
		are synchronized and the associated ghost values are filled with the
		corresponding values from the opposite side of the domain after the
		velocity correction.
		
	\end{enumerate}

	\item \textit{Finalization.}
	
	The completed solution is prepared for the next time step.
	
	\begin{enumerate}
		
		\item The final velocity boundary conditions are applied to
		$\mathbf{V}^{n+1}$. One exception is the prescribed-pressure outlet,
		where the pressure-corrected boundary-normal velocity is retained.
		
		\item The current integrated convective flux imbalance
		$\mathcal{C}_f^{\,n}$ is stored as the previous convective flux
		imbalance for the next Adams--Bashforth step.
		
		\item The completed-step counter is incremented and
		$\mathbf{V}^{n+1}$, $p^{n+1/2}$, and the updated
		Adams--Bashforth state are returned.
		
		\item The run driver advances the physical time and evaluates or
		writes the requested diagnostic and output quantities. The procedure
		is then repeated for the next time step.
		
	\end{enumerate}
	
\end{enumerate}

The explicit-Euler implementation follows the same overall projection
structure but does not require convection history or Helmholtz solves.
After application of the predictor boundary conditions, the convective and
diffusive contributions are evaluated explicitly to form the pressure-free
momentum rate. This rate is advanced with explicit Euler to obtain
$\mathbf{V}^{*}$. The predictor boundary conditions are then reapplied,
followed by the standard pressure projection and the final velocity boundary
conditions. The present explicit-Euler implementation uses only the standard
projection.

\clearpage

\chapter{Verification}

Verification is carried out using a number of test problems with gradual increase in complexity These are arranged as follows:
\begin{enumerate}
	\item \textit{Stokes' first problem}. This isolates transient viscous diffusion. The
	velocity field remains divergence-free without a pressure gradient, allowing
	the diffusion and time-integration treatments to be examined independently
	of convection and pressure--velocity coupling.
	
	\item \textit{Advection of a sine-wave velocity profile}. This isolates transient
	convection. The velocity field remains divergence-free without a pressure
	gradient. The problem is also used to examine the numerical dissipation
	and dispersion of the convection discretization.
	
	\item \textit{The inviscid advecting Taylor--Green vortex}. This one introduces nonlinear
	convection and a nonzero pressure gradient in a periodic domain. It
	therefore brings pressure--velocity coupling into the verification while
	excluding physical diffusion.
	
	\item \textit{The viscous advecting Taylor--Green vortex}. This introduces viscous
	diffusion and therefore exercises the complete incompressible
	Navier--Stokes formulation in a periodic domain.
	
	\item \textit{The wall-bounded vortex from \citet{perot1993analysis}}. It replaces the
	periodic boundaries with solid walls. It is used to examine the behavior
	of the projection methods in a wall-bounded flow.
	
	\item \textit{The lid-driven cavity}. This test provides a canonical wall-bounded benchmark.
	Numerical velocity profiles are compared with established reference data.
	
	\item \textit{Plane channel flow}. This one introduces inlet and outlet boundary conditions.
	The resulting velocity field is compared with the exact fully developed
	solution.
\end{enumerate}
Spatial and temporal refinement studies are performed where appropriate
to establish the expected convergence behavior. Before considering these
problems, the error measures used
throughout the verification studies are defined. The chapter concludes
with further observations drawn from the verification results.

\section{Error measures}
\label{sec:error_measures}

For a discrete error field $e_k=\phi_k-\phi_k^{\mathrm{ref}}$, the
control-volume-weighted $L_2$ norm is
\begin{equation}
	E_\phi
	=
	\left(
	\frac{
		\displaystyle\sum_k V_k e_k^2
	}{
		\displaystyle\sum_k V_k
	}
	\right)^{1/2},
\end{equation}
where $\phi_k$ is the numerical value, $\phi_k^{\mathrm{ref}}$ is the
corresponding reference value, and $V_k$ is the control-volume area
associated with the discrete unknown at location $k$. The discrete
errors in the velocity components and pressure are defined as
\begin{equation}
	e_{u,k}=u_k-u_k^{\mathrm{ref}},
	\qquad
	e_{v,k}=v_k-v_k^{\mathrm{ref}},
	\qquad
	e_{p,k}=p_k-p_k^{\mathrm{ref}}.
\end{equation}
Here, $u$ and $v$ are stored on their respective staggered control
volumes. The two components are combined into the single error measure
\begin{equation}
	E_{\mathbf{V}}
	=
	\left[
	\frac{
		\displaystyle
		\sum_k V_k^{(u)}e_{u,k}^2
		+
		\sum_k V_k^{(v)}e_{v,k}^2
	}{
		\displaystyle
		\sum_k V_k^{(u)}
		+
		\sum_k V_k^{(v)}
	}
	\right]^{1/2},
\end{equation}
where $V_k^{(u)}$ and $V_k^{(v)}$ are the areas of the staggered
control volumes associated with $u_k$ and $v_k$, respectively.
The pressure error is defined as
\begin{equation}
	E_p
	=
	\left[
	\frac{
		\displaystyle
		\sum_k V_k^{(p)}e_{p,k}^2
	}{
		\displaystyle
		\sum_k V_k^{(p)}
	}
	\right]^{1/2}.
\end{equation}

The time levels of velocity and pressure must be distinguished when
these errors are evaluated. A calculation advanced to
$t^{n+1}=T$ provides the velocity $\mathbf{V}^{n+1}$ at $T$.
The pressure $p^{n+1/2}$ is associated with the midpoint of the last
time interval,
\begin{equation}
	t^{n+1/2}
	=
	T-\frac{\Delta t}{2}.
\end{equation}
The reference velocity is therefore evaluated at $T$, while the
reference pressure is evaluated at $T-\Delta t/2$. This convention is
used for all velocity and pressure error calculations in this chapter.

For spatial refinement, the observed order is denoted by $q$ and is
evaluated from
\begin{equation}
	q
	=
	\frac{
		\log(E_1/E_2)
	}{
		\log(\Delta_1/\Delta_2)
	},
\end{equation}
where $E_1$ and $E_2$ are errors on two successively refined grids and
$\Delta_1$ and $\Delta_2$ are their nominal grid spacings. The
corresponding velocity and pressure orders are written as
$q_{\mathbf{V}}$ and $q_p$.

For temporal refinement, the velocity errors are defined as
\begin{equation}
	e_{u,k}^{(\Delta t)}
	=
	u_k^{(\Delta t)}-u_k^{\mathrm{ref}}(T),
	\qquad
	e_{v,k}^{(\Delta t)}
	=
	v_k^{(\Delta t)}-v_k^{\mathrm{ref}}(T),
\end{equation}
where the numerical solution has been advanced to $T$. The pressure requires a different interpretation. For a calculation advanced to $T$, the pressure obtained from the last
time step should not be interpreted as $p(T)$. The last time step spans
the interval from $T-\Delta t$ to $T$. Since the pressure in the
time-centered momentum equation is located at $n+1/2$, the computed
pressure represents
\begin{equation}
	p^{n+1/2}
	=
	p\left(T-\frac{\Delta t}{2}\right).
\end{equation}
The pressure error is therefore defined as
\begin{equation}
	e_{p,k}^{(\Delta t)}
	=
	p_k^{(\Delta t)}
	-
	p_k^{\mathrm{ref}}
	\left(T-\frac{\Delta t}{2}\right).
\end{equation}
Thus, the reference pressure is evaluated at the time represented by
the numerical pressure.

This distinction is important when a fine-time-step numerical solution
is used as the reference. Suppose that both the test and reference
calculations are advanced to the same final velocity time $T$. If their
time steps are different, their final pressure fields do not represent
the same time. They are associated with
\begin{equation}
	T-\frac{\Delta t}{2}
	\qquad\text{and}\qquad
	T-\frac{\Delta t_{\mathrm{ref}}}{2},
\end{equation}
respectively. The two final pressure fields should therefore not be
compared directly.

For example, consider a test calculation with $\Delta t=0.1$ and a
reference calculation with $\Delta t_{\mathrm{ref}}=0.01$. Let both
calculations be advanced to $T=1$. The final pressure of the test
calculation represents
\begin{equation}
	1-\frac{0.1}{2}=0.95,
\end{equation}
whereas the final pressure of the reference calculation represents
\begin{equation}
	1-\frac{0.01}{2}=0.995.
\end{equation}
A direct comparison would therefore compare pressure at $0.95$ with
pressure at $0.995$. The difference between these time locations is
\begin{equation}
	\frac{\Delta t-\Delta t_{\mathrm{ref}}}{2}.
\end{equation}
This introduces an $O(\Delta t)$ error into the pressure comparison
\citep{BrownCortezMinion2001, ArmfieldStreet2003}.

The correct comparison uses the reference pressure at the time
represented by the test pressure. In the example above, the required
comparison is therefore between the test pressure at $0.95$ and the
reference pressure at $0.95$.

A single fine-time-step numerical solution is used as the reference
for a temporal convergence study. Its stored pressure fields are
available at its own midpoint time levels. The required time
$T-\Delta t/2$ will not, in general, coincide with one of these levels.
The reference pressure is then obtained by linear interpolation between
the two pressure fields that bracket the required time.

For example, with $\Delta t_{\mathrm{ref}}=0.01$, reference pressures
are available at
\begin{equation}
	\ldots,\;0.945,\;0.955,\;\ldots
\end{equation}
The reference pressure required at $0.95$ is therefore obtained by
linear interpolation between the pressure fields at $0.945$ and
$0.955$.

In general, if the required time $T-\Delta t/2$ lies between two
reference pressure time levels $t_a$ and $t_b$, the interpolated
reference pressure is
\begin{equation}
	p_k^{\mathrm{ref}}
	\left(T-\frac{\Delta t}{2}\right)
	=
	(1-\theta)p_k^{\mathrm{ref}}(t_a)
	+
	\theta p_k^{\mathrm{ref}}(t_b),
\end{equation}
where
\begin{equation}
	\theta
	=
	\frac{
		T-\Delta t/2-t_a
	}{
		t_b-t_a
	}.
\end{equation}
The test and reference pressures are then compared at the same time.
For a smooth pressure history, linear interpolation is second-order
accurate in time and is consistent with the temporal accuracy of the
reference calculation.

The temporal velocity error is
\begin{equation}
	E_{\mathbf{V}}
	=
	\left[
	\frac{
		\displaystyle
		\sum_k V_k^{(u)}
		\left(e_{u,k}^{(\Delta t)}\right)^2
		+
		\sum_k V_k^{(v)}
		\left(e_{v,k}^{(\Delta t)}\right)^2
	}{
		\displaystyle
		\sum_k V_k^{(u)}
		+
		\sum_k V_k^{(v)}
	}
	\right]^{1/2},
\end{equation}
and the temporal pressure error is
\begin{equation}
	E_p
	=
	\left[
	\frac{
		\displaystyle
		\sum_k V_k^{(p)}
		\left(e_{p,k}^{(\Delta t)}\right)^2
	}{
		\displaystyle
		\sum_k V_k^{(p)}
	}
	\right]^{1/2}.
\end{equation}

The observed temporal order is denoted by $r$ and is evaluated from
\begin{equation}
	r
	=
	\frac{
		\log(E_1/E_2)
	}{
		\log(\Delta t_1/\Delta t_2)
	},
\end{equation}
where $E_1$ and $E_2$ are the errors obtained with two successively
refined time steps $\Delta t_1$ and $\Delta t_2$. The corresponding
velocity and pressure orders are written as $r_{\mathbf{V}}$ and $r_p$.

\section{Stokes' first problem}
\label{sec:stokes_first}

Stokes' first problem is considered to isolate the spatial discretization
and temporal integration of the diffusion term.  The problem
consists of an initially quiescent fluid occupying the half-space above
an infinite flat plate.  At $t=0$, the plate is impulsively set into
motion with a constant velocity $U_w$ in the $x$ direction. The
resulting velocity field only has an $x$-component and that depends on the $y$-coordinate and time: $u(y,t)$. Therefore $\nabla \cdot\left(\mathbf{V}\mathbf{V}\right)$
vanishes. Furthermore, no pressure gradient is required. The governing momentum equation consequently reduces to
\begin{equation}
	\frac{\partial u}{\partial t}
	=
	\nu
	\frac{\partial^2 u}{\partial y^2}.
	\label{eq:stokes_first_diffusion}
\end{equation}
The analytical solution is
\begin{equation}
	\frac{u(y,t)}{U_w}
	=
	\operatorname{erfc}
	\left(
	\frac{y}{2\sqrt{\nu t}}
	\right),
	\label{eq:stokes_first_exact}
\end{equation}
where $\nu$ is the kinematic viscosity.  The viscous disturbance
shears the fluid progressively further from the wall as time
increases.

The problem therefore constitutes a pure viscous-diffusion test of the
finite-volume Laplacian, its moving-wall ghost-cell treatment, and the
Crank--Nicolson temporal discretization.  Neither the AB2 convection
scheme nor a pressure field contributes to the analytical
evolution.

The computational domain is a finite rectangular approximation to the
semi-infinite physical domain, with $L_x=0.125$ and $L_y=1$.  A uniform
Cartesian grid is employed.  The resolutions considered were
\begin{equation}
	N_x\times N_y
	=
	4 \times 32,\;\; 8 \times 64,\;\; 16 \times 128,\;\; 32 \times 256 .
\end{equation}
Periodic boundary conditions are imposed
at the left and right boundaries.  At the lower boundary, the moving-wall
conditions are
\begin{equation}
	u=U_w,
	\qquad
	v=0,
\end{equation}
while the upper boundary is treated as a stationary no-slip wall,
\begin{equation}
	u=v=0.
\end{equation}
The parameters are set to $U_w=1$ and $\nu=0.01$.  The height of the
domain is selected such that the viscous disturbance remains
negligible at the upper boundary over the time interval considered.
In this example, at $t=0.5$ s,
\begin{equation}
	\frac{u(L_y,t)}{U_w}
	=
	\operatorname{erfc}
	\left(
	\frac{L_y}{2\sqrt{\nu t}}
	\right)
	=
	\operatorname{erfc}(7.071),
\end{equation}
which is about zero.

The impulsive start at $t=0$ produces a singular wall-normal velocity
gradient.  Since the purpose of the present calculation is to examine
the formal discretization errors rather than the numerical treatment
of this initial singularity, the computation is initialized from the
analytical solution at the finite physical time $t_1=0.1$ s.  The
analytical profile given by Eq.~\eqref{eq:stokes_first_exact} is sampled
directly at the staggered $u$-velocity locations at $t=t_1$, and the
numerical solution is subsequently advanced to $t=0.5$.

The pressure reference used in the present solver is $p_{\mathrm{ref}}=0$.

\begin{figure}[!htbp]
	\centering
	
	\begin{subfigure}[t]{0.6\textwidth}
		\centering
		\includegraphics[width=\linewidth]{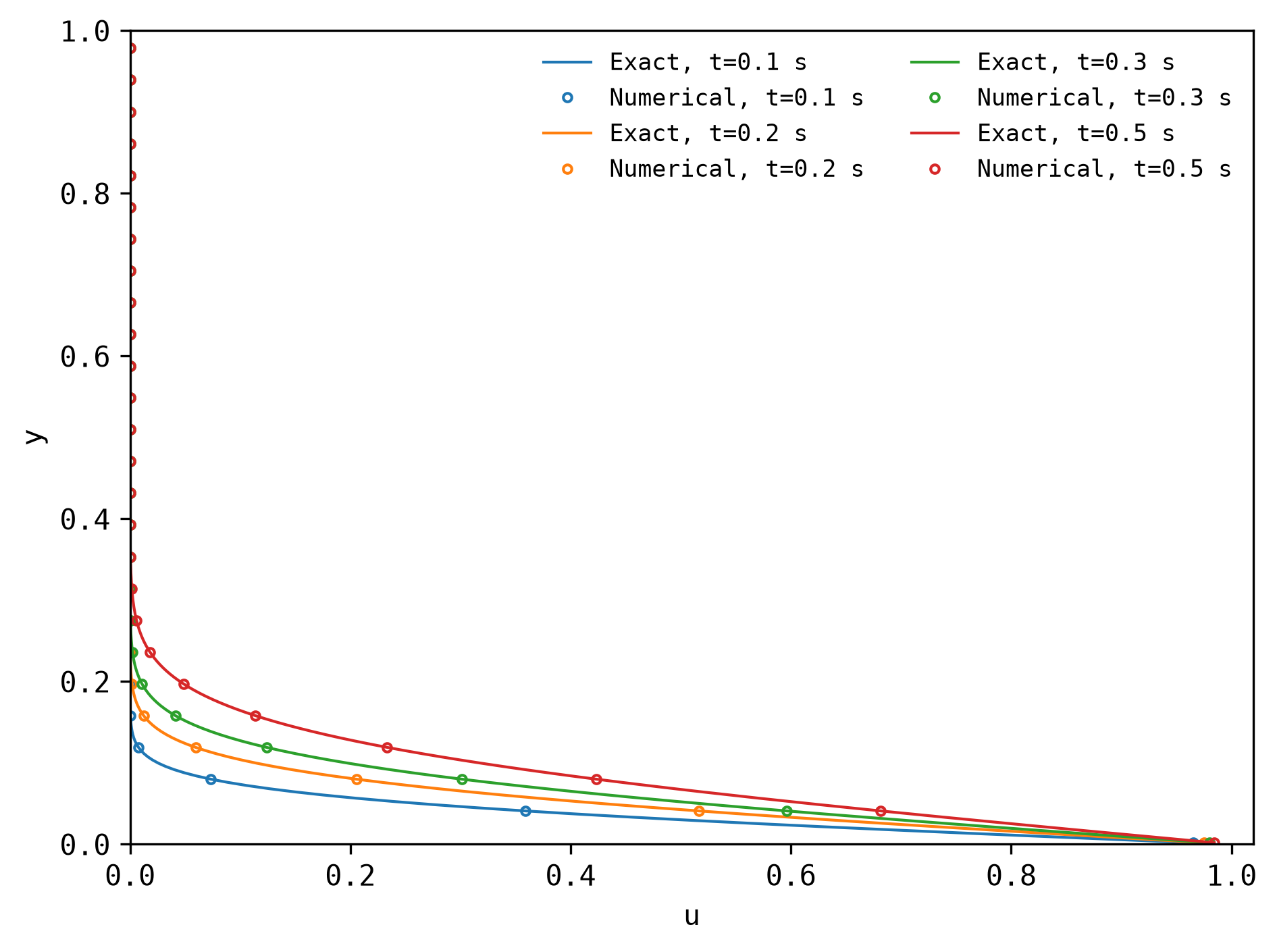}
	\end{subfigure}
	\begin{subfigure}[t]{0.24\textwidth}
		\centering
		\includegraphics[width=\linewidth]{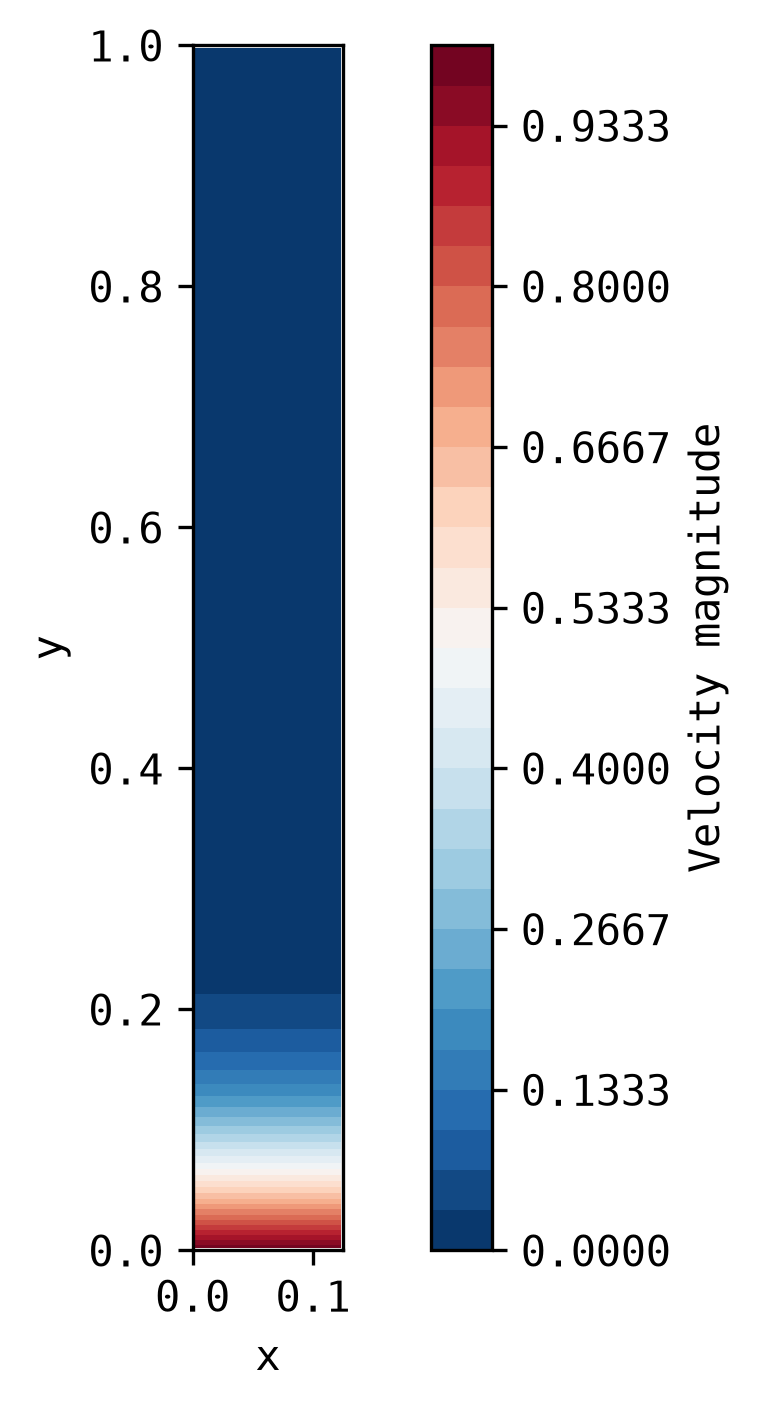}
	\end{subfigure}
	
	\caption{Stokes' first problem. The left panel compares the numerical
		velocity profiles with the analytical solution at selected instants,
		while the right panel shows the velocity-magnitude field at $t=0.5~\mathrm{s}$.}
	\label{fig:stokes_first_profiles}
\end{figure}

The intermediate velocity remains divergence-free because it retains
the form $(u(y,t),0)$. Hence the pressure correction has
a zero right-hand side.  Apart from round-off and discretization-level
residuals, the projection therefore produces no velocity correction.
For this reason, the Standard and Perot projections become
equivalent for this particular problem: the pressure correction is
zero, so the distinction between their pressure-correction operators
does not enter the velocity update.  Only the Standard projection is
therefore reported. Velocity profiles obtained from numerical solutions are compared with the exact solution in Figure \ref{fig:stokes_first_profiles}.

\subsection{Spatial convergence}

Spatial convergence is examined using four successively refined uniform
Cartesian grids. The physical domain is kept fixed, while both
directions are refined by a factor of two so that the cells remain
square. The time step is fixed at
$\Delta t=1.0\times10^{-3}~\mathrm{s}$, and the error is evaluated at
the physical time $t=0.5~\mathrm{s}$ using the analytical solution in
Eq.~\eqref{eq:stokes_first_exact}. The error measures are those defined
in Section~\ref{sec:error_measures}.

\begin{table}[!htbp]
	\centering
	\caption{Spatial convergence for Stokes' first problem at
		$t=0.5~\mathrm{s}$. The uniform grids contain square cells and
		$\Delta t=1.0\times10^{-3}~\mathrm{s}$.}
	\label{tab:stokes_first_spatial}
	
	\begin{tabular}{ccc}
		\hline
		$N_x\times N_y$ & $E_{\mathbf{V}}$ & $q_{\mathbf{V}}$ \\
		\hline
		$4\times32$
		& $1.0743\times10^{-3}$
		& 2.0160 \\
		$8\times64$
		& $2.6560\times10^{-4}$
		& 2.0045 \\
		$16\times128$
		& $6.6193\times10^{-5}$
		& 2.0034 \\
		$32\times256$
		& $1.6509\times10^{-5}$
		& -- \\
		\hline
	\end{tabular}
\end{table}

Table~\ref{tab:stokes_first_spatial} shows second-order spatial
convergence over the complete refinement range. The observed orders
remain very close to two, demonstrating the nominal second-order
accuracy of the finite-volume diffusion operator and its wall-boundary
treatment for this pure-diffusion problem.

\subsection{Temporal convergence}

Temporal convergence is examined on a fixed $128\times1024$ uniform
grid. The numerical solution is initialized from the analytical
profile at $t_1=0.1~\mathrm{s}$ and advanced to
$t=0.5~\mathrm{s}$. Since the exact transient solution is available,
the temporal error is evaluated directly against
Eq.~\eqref{eq:stokes_first_exact}, using the error measures introduced
previously.

The time step is successively halved according to
\begin{equation}
	\Delta t
	=
	8.0\times10^{-2},\;
	4.0\times10^{-2},\;
	2.0\times10^{-2}~\mathrm{s}.
\end{equation}

\begin{table}[!htbp]
	\centering
	\caption{Temporal convergence for Stokes' first problem on the
		$128\times1024$ uniform grid. The numerical solution is advanced
		from $t_1=0.1~\mathrm{s}$ to $t=0.5~\mathrm{s}$ and compared
		directly with the analytical solution.}
	\label{tab:stokes_first_temporal}
	
	\begin{tabular}{ccc}
		\hline
		$\Delta t$ (s)
		& $E_{\mathbf{V}}$
		& $r_{\mathbf{V}}$ \\
		\hline
		$8.0000\times10^{-2}$
		& $2.7600\times10^{-4}$
		& 2.0205 \\
		$4.0000\times10^{-2}$
		& $6.8028\times10^{-5}$
		& 2.0602 \\
		$2.0000\times10^{-2}$
		& $1.6312\times10^{-5}$
		& -- \\
		\hline
	\end{tabular}
\end{table}

Table~\ref{tab:stokes_first_temporal} shows second-order temporal
convergence over the reported refinement range. The observed orders,
2.0205 and 2.0602, are consistent with the nominal second-order
accuracy of the Crank--Nicolson discretization. Since convection
vanishes identically and the projection produces no appreciable
velocity correction for this problem, the observed temporal behavior
isolates the viscous time integration.

\section{Advection of a sine wave velocity profile}

Stokes' first problem isolates the viscous-diffusion term because convection
and the pressure gradient vanish analytically. A complementary problem is
considered here to isolate convection. Flow is inviscid and a
velocity field is chosen for which the pressure gradient also vanishes.

Consider the two-dimensional velocity field
\begin{equation}
	u=U_a,
	\qquad
	v=v(x,t),
\end{equation}
where $U_a$ is a constant. This field satisfies the incompressibility
condition $\nabla\cdot\mathbf{V}=0$.

For $\nu=0$ and a spatially uniform pressure, all terms in the
$x$-momentum equation vanish and no additional information is obtained
from this equation. The $y$-momentum equation reduces to
\begin{equation}
	\frac{\partial v}{\partial t}
	+
	U_a\frac{\partial v}{\partial x}
	=0.
	\label{eq:sine_advection_equation}
\end{equation}
The incompressible momentum equations therefore reduce to the classical
linear-convection equation for the transverse velocity component.

The exact
solution is
\begin{equation}
	u(x,y,t)=U_a,
	\qquad
	v(x,y,t)
	=
	A\sin\left[k(x-U_at)\right].
	\label{eq:sine_advection_exact}
\end{equation}
The corresponding initial condition is
\begin{equation}
	u(x,y,0)=U_a,
	\qquad
	v(x,y,0)=A\sin(kx),
\end{equation}
where $A$ is the wave amplitude and $k$ is the wavenumber.  
The wave therefore translates in the positive $x$-direction at the
constant speed $U_a$ without a change in amplitude or shape. Sinusoidal
and truncated-sinusoidal profiles are classical test functions for the
linear-advection equation \citep{Fletcher1991,HoffmannChiang2000}. Discontinuous profiles are particularly relevant to compressible-flow algorithms because shock waves may occur. Incompressible DNS algorithms are not designed to capture shocks. The
present problem therefore uses a complete periodic sine wave, so that the
advected velocity field remains smooth throughout the domain.

The problem is solved in a doubly periodic rectangular domain of lengths
$L_x=2\pi$~m and $L_y=\pi/2$~m using a uniform Cartesian grid. $x$-periodicity requires the wavenumber to satisfy
\begin{equation}
	k=\frac{2\pi n}{L_x},
	\qquad
	n=1,2,\ldots .
\end{equation}
The calculations reported here use
\begin{equation}
	U_a=1~\mathrm{m/s},
	\qquad
	A=1~\mathrm{m/s}.
\end{equation}
The selected wavenumbers are
\begin{equation}
	k=1,\;4,\;8,\;16~\mathrm{m}^{-1}.
\end{equation}
These values correspond to progressively shorter wavelengths ($\lambda = 2\pi/k$) and are used
to examine the convection of spatial structures of different sizes.

Motif executes the pressure solution as part of its projection algorithm.
The pressure reference used in the present solver is
$p_{\mathrm{ref}}=0$.

The convection step preserves the form $(U_a,v(x,t))$, so the intermediate
velocity is also discretely divergence-free. Hence the pressure-correction
equation has a zero right-hand side. Apart from round-off and solver
residuals, the projection produces no velocity correction. The Standard
and Perot projections are consequently equivalent for this particular
problem, and only the Standard projection is reported.

The calculation starts at $t=0$ and is advanced to $0.2$~s for the
spatial-convergence tests and to $0.02$~s for the temporal-convergence
tests.

\begin{figure}[htbp]
	\centering
	\includegraphics[width=0.9\textwidth]
	{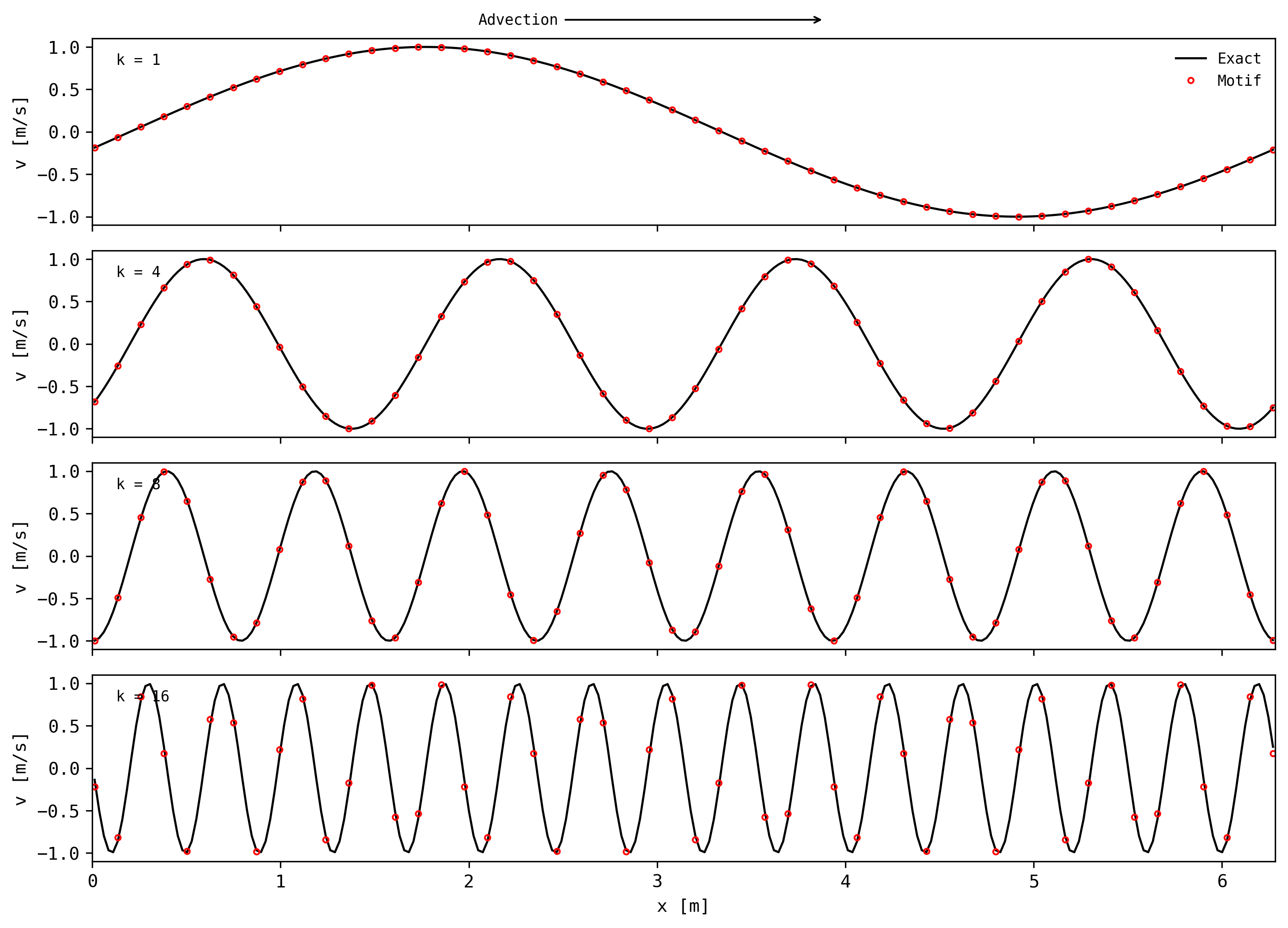}
	\caption{Comparison of the analytical and numerical transverse-velocity
		profiles for advection of a sine wave at $t=0.2$~s. All cases ran with $\Delta t = 0.0003125$ s.}
	\label{fig:sine_advection_profiles}
\end{figure}

Figure~\ref{fig:sine_advection_profiles} compares the analytical and
numerical transverse-velocity profiles for the selected wavenumbers.

\subsection{Spatial convergence}

Spatial convergence is examined using the uniform grids
\begin{equation}
	N_x\times N_y
	=
	32\times8,\;\;
	64\times16,\;\;
	128\times32,\;\;
	256\times64.
\end{equation}
The timestep is fixed at
\begin{equation}
	\Delta t=3.125\times10^{-4}~\mathrm{s},
\end{equation}
and is chosen sufficiently small that temporal error does not dominate the
reported spatial error. The case $k=4~\mathrm{m}^{-1}$ is used. The
numerical solution at $t=0.2$~s is compared with
Eq.~\eqref{eq:sine_advection_exact}. The velocity error
$E_{\mathbf V}$ is evaluated using the control-volume-weighted error
measure defined above.

Table~\ref{tab:sine_advection_spatial} gives the spatial-convergence results. The results clearly indicate a second-order spatial order of accuracy.

\begin{table}[htbp]
	\centering
	\caption{Spatial convergence for convection of a sine wave with
		$k=4~\mathrm{m}^{-1}$ and
		$\Delta t=3.125\times10^{-4}$~s.}
	\label{tab:sine_advection_spatial}
	\begin{tabular}{ccc}
		\hline
		$N_x\times N_y$ &
		$E_{\mathbf V}$ &
		$q_{\mathbf V}$ \\
		\hline
		$32\times8$   & $3.98627\times10^{-2}$ & -- \\
		$64\times16$  & $1.02014\times10^{-2}$ & $1.966$ \\
		$128\times32$ & $2.56500\times10^{-3}$ & $1.992$ \\
		$256\times64$ & $6.41984\times10^{-4}$ & $1.998$ \\
		\hline
	\end{tabular}
\end{table}

\subsection{Temporal convergence}

Temporal convergence is examined on the fixed uniform grid
\begin{equation}
	N_x\times N_y
	=
	512\times128.
\end{equation}
The timesteps are
\begin{equation}
	\Delta t
	=
	0.010,\;\;0.005,\;\;0.0025~\mathrm{s}.
\end{equation}
The case $k=1~\mathrm{m}^{-1}$ is used. The numerical solutions at
$t=0.02$~s are compared with Eq.~\eqref{eq:sine_advection_exact}.
The temporal velocity error $E_{\mathbf V}$ is evaluated using
the error measure defined above.

Table~\ref{tab:sine_advection_temporal} gives the temporal-convergence
results. Temporal order of accuracy is shown to be 2 which is expected from the AB2 integration.

\begin{table}[htbp]
	\centering
	\caption{Temporal convergence for convection of a sine wave on the
		$512\times128$ uniform grid with $k=1~\mathrm{m}^{-1}$.}
	\label{tab:sine_advection_temporal}
	\begin{tabular}{ccc}
		\hline
		$\Delta t$ &
		$E_{\mathbf V}$ &
		$r_{\mathbf{V}}$ \\
		\hline
		$0.0100$ & $2.50039\times10^{-5}$ & $2.000$ \\
		$0.0050$ & $6.25265\times10^{-6}$ & $1.985$ \\
		$0.0025$ & $1.57914\times10^{-6}$ & -- \\
		\hline
	\end{tabular}
\end{table}

\subsection{Numerical dispersion and dissipation}

The sine-wave problem also permits the numerical dispersion and
dissipation of the convection discretization to be examined directly.
For the exact solution, the prescribed harmonic signal is advected without
a change in amplitude or phase speed. The numerical solution may instead
develop errors in both amplitude and phase.

The numerical amplitude and phase could be estimated from the heights
and locations of individual peaks. Such estimates, however, depend on
the locations of the grid points and require corresponding peaks of the
numerical and exact profiles to be identified. A more systematic measure
is obtained by using the complete numerical profile.

At the prescribed wavenumber $k$, a sine wave with numerical amplitude
$A_{\mathrm{num}}$ and phase $\phi_{\mathrm{num}}$ can be written as
\begin{equation}
	v_{\mathrm{num}}(x)
	=
	A_{\mathrm{num}}
	\sin\left(kx+\phi_{\mathrm{num}}\right).
\end{equation}
For the exact solution in Eq.~\eqref{eq:sine_advection_exact}, phase at time $t$ is $\phi_{\mathrm{exact}} = -kU_at$. Using the angle-addition identity,
\begin{equation}
	\sin\left(kx+\phi_{\mathrm{num}}\right)
	=
	\sin(kx)\cos\phi_{\mathrm{num}}
	+
	\cos(kx)\sin\phi_{\mathrm{num}},
\end{equation}
gives
\begin{equation}
	v_{\mathrm{num}}(x)
	=
	a\sin(kx)+b\cos(kx),
\end{equation}
where
\begin{equation}
	a
	=
	A_{\mathrm{num}}\cos\phi_{\mathrm{num}},
	\qquad
	b
	=
	A_{\mathrm{num}}\sin\phi_{\mathrm{num}}.
\end{equation}
The coefficients $a$ and $b$ are determined by a least-squares fit to
the complete numerical profile over the periodic domain. The numerical
amplitude and phase then follow as
\begin{align}
	A_{\mathrm{num}}
	&=
	\sqrt{a^2+b^2},
	\\
	\phi_{\mathrm{num}}
	&=
	\operatorname{atan2}(b,a).
\end{align}

The amplitude ratio $A_{\mathrm{num}}/A$ therefore quantifies the change
in amplitude of the numerical mode. Numerical dissipation would appear
as a reduction of this ratio below unity. The numerical phase error is
defined as
\begin{equation}
	\Delta\phi
	=
	\phi_{\mathrm{num}}-\phi_{\mathrm{exact}}.
\end{equation}
This provides the phase difference between the numerical and exact
profiles within one wavelength. 

A longer propagation test is performed on the $256\times64$ grid using
$\Delta t=3.125\times10^{-4}$~s and a final time of $t=6.4$~s.
The small timestep is retained so that the contribution of the AB2 time
discretization is small compared with the spatial dispersion error.

Table~\ref{tab:sine_dispersion_dissipation} compares the measured
amplitude and phase errors with the modified-wavenumber prediction. The amplitude remains essentially unchanged for all four wavenumbers,
indicating negligible numerical dissipation in this test. In contrast,
the phase error increases strongly as the wavelength becomes less
resolved. For the higher wavenumbers considered here, doubling the
wavenumber increases the phase error by approximately a factor of eight.
The results therefore show that the principal error observed in this
long-time advection test is associated with phase propagation rather
than amplitude decay.

\begin{table}[htbp]
	\centering
	\caption{Numerical dispersion and dissipation for convection of a
		sine wave at $t=6.4$~s.}
	\label{tab:sine_dispersion_dissipation}
	\begin{tabular}{cccc}
		\hline
		$k$ ($\mathrm{m}^{-1}$) &
		Points/$\lambda$ &
		$A_{\mathrm{num}}/A$ &
		$|\Delta\phi|$ ($^\circ$) \\
		\hline
		1  & 256 & 1.000000 & 0.0368 \\
		4  & 64  & 1.000001 & 2.354  \\
		8  & 32  & 1.000003 & 18.806 \\
		16 & 16  & 1.000015 & 149.581 \\
		\hline
	\end{tabular}
\end{table}

This simple test therefore exposes an important limitation of the present second-order centered convection discretization: poorly resolved spatial structures can retain their amplitude while accumulating substantial phase error. Grid must be refined to reduce the phase error.

\section{Inviscid advecting Taylor--Green vortex}
\label{sec:inviscid_atgv}

The preceding tests isolate viscous diffusion and convection separately.
The inviscid advecting Taylor--Green vortex provides a more coupled
verification problem in which convection, a nonzero pressure gradient,
and the projection operation act simultaneously. The kinematic viscosity
is set to $\nu=0$, so the diffusion term vanishes identically and the
Crank--Nicolson discretization does not contribute to the numerical
evolution. Unlike the convection-of-a-sine-wave problem, the intermediate
velocity generally requires a pressure correction. The problem therefore
tests the interaction between the convection and pressure--velocity
coupling in the absence of viscous diffusion.

For the advection in the $x$-direction with a speed $U_a$, define
\begin{equation}
	X = x - U_a t
\end{equation}
The flow is considered in the square domain
\begin{equation}
	0 \leq x \leq 2\pi,
	\qquad
	0 \leq y \leq 2\pi,
\end{equation}
with periodic boundary conditions in both coordinate directions. The
exact solution of the incompressible Euler equations is
\begin{align}
	u(x,y,t)
	&=
	U_a
	+
	U_0\sin(X)\cos(y),
	\\
	v(x,y,t)
	&=
	-U_0\cos(X)\sin(y),
	\\
	p(x,y,t)
	&=
	\frac{\rho U_0^2}{4}
	\left[
	\cos(2X)+\cos(2y)
	\right]
	+C,
\end{align}
where $C$ is an arbitrary pressure constant. The analytical pressure is shifted to
satisfy the pressure reference used by the numerical solution.

The parameters are set to
\begin{equation}
	U_0=1,
	\qquad
	U_a=1,
	\qquad
	\rho=1,
	\qquad
	\nu=0.
\end{equation}
Periodic boundary conditions are imposed in both coordinate directions.
The numerical solution is initialized by sampling the exact velocity
field directly at the native staggered-grid locations. The sampled
initial field is subsequently projected onto the discrete
divergence-free space before time integration.

In contrast to Stokes' first problem, the pressure field is nonuniform
and the projection remains active. The present problem is therefore not
a test of the convection operator alone. Rather, it isolates the
combined action of the AB2 convection discretization and the pressure
projection after removal of viscous diffusion. Spatial and temporal
errors can be evaluated directly against the analytical Euler solution.

The inviscid limit also provides a direct test of the relation derived
for the second-order inverse approximation. When $\nu=0$, 
\begin{equation}
	\Bop_2
	=
	\Delta t I
	=
	\Bop_1.
\end{equation}
The distinction between the first- and second-order inverse
approximations therefore disappears. Accordingly, the Standard and
Perot projection formulations are expected to produce identical
velocity and pressure solutions, apart from round-off error. This
equivalence is examined explicitly below.

\begin{figure}[!htbp]
	\centering
	
	\begin{tabular}{ccc}
		\includegraphics[width=0.31\textwidth]{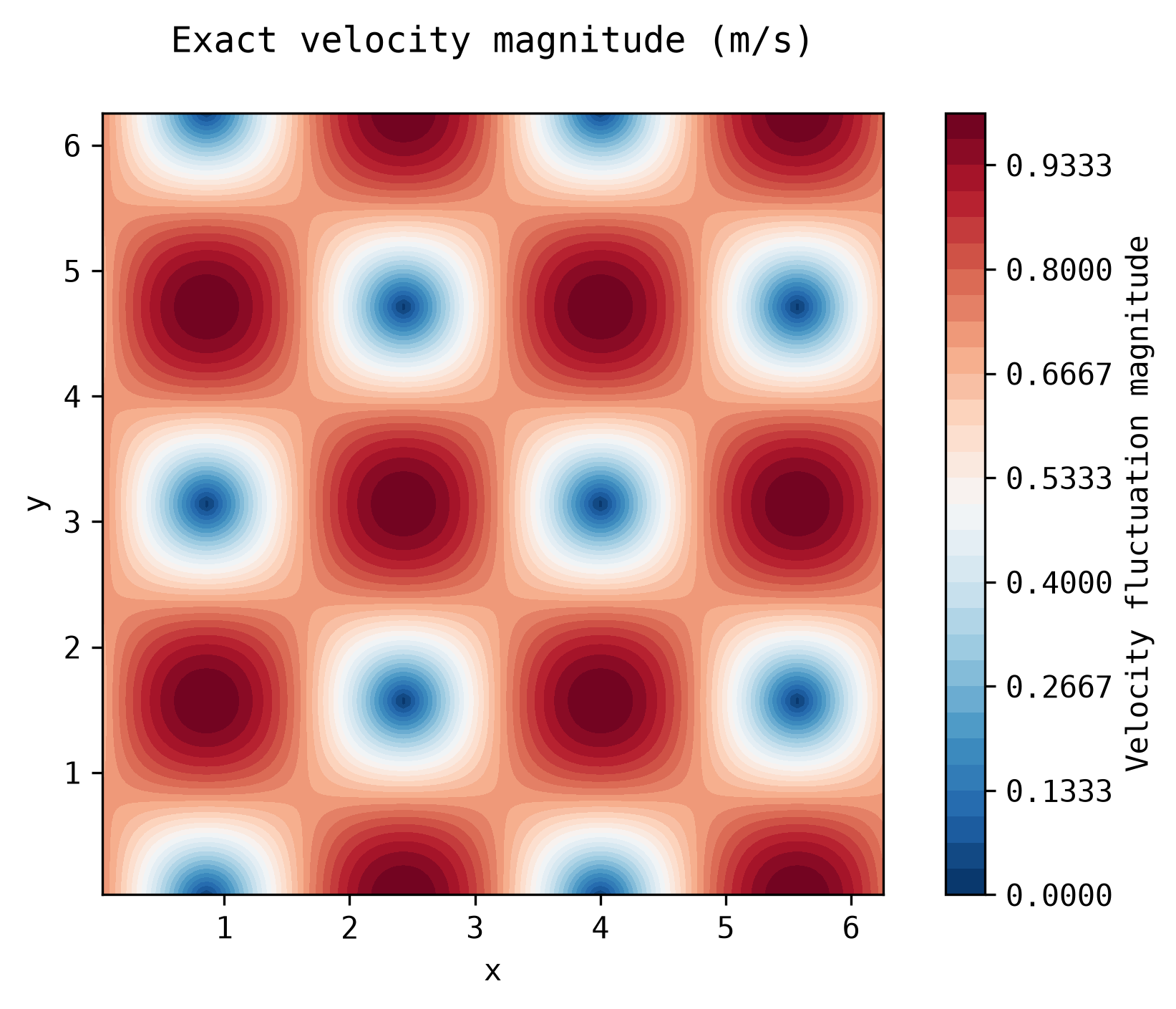}
		&
		\includegraphics[width=0.31\textwidth]{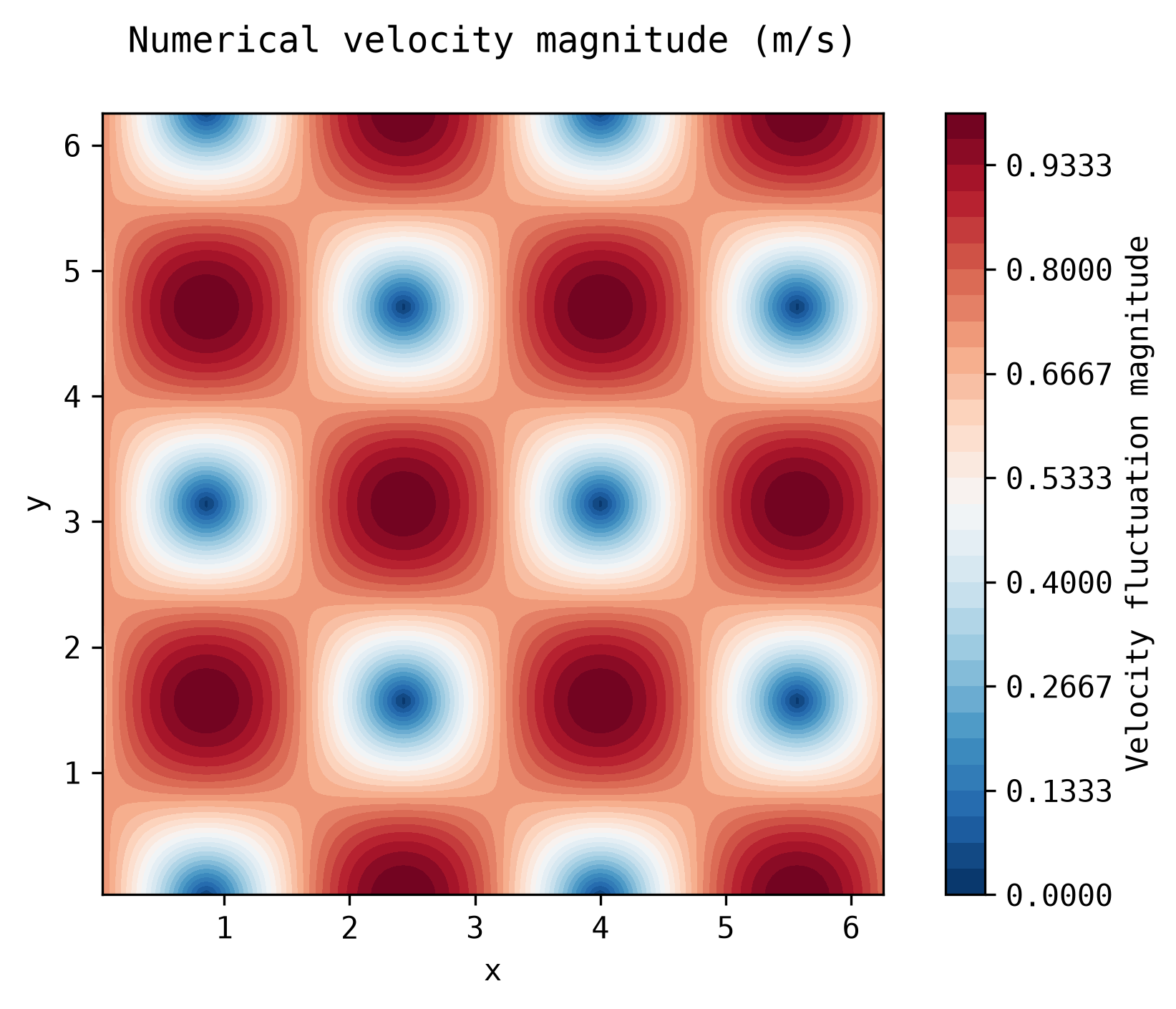}
		&
		\includegraphics[width=0.31\textwidth]{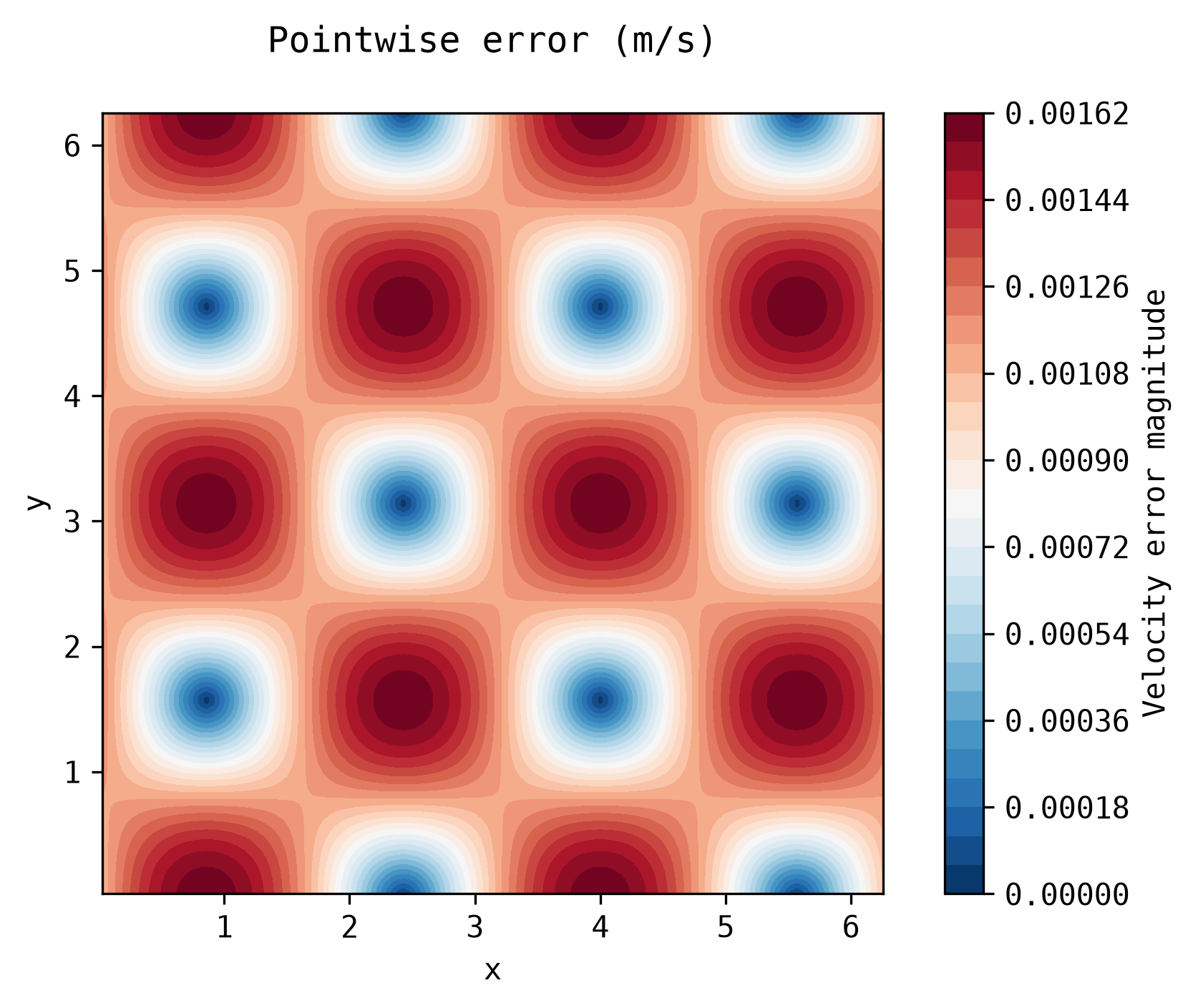}
		\\
		
		\includegraphics[width=0.31\textwidth]{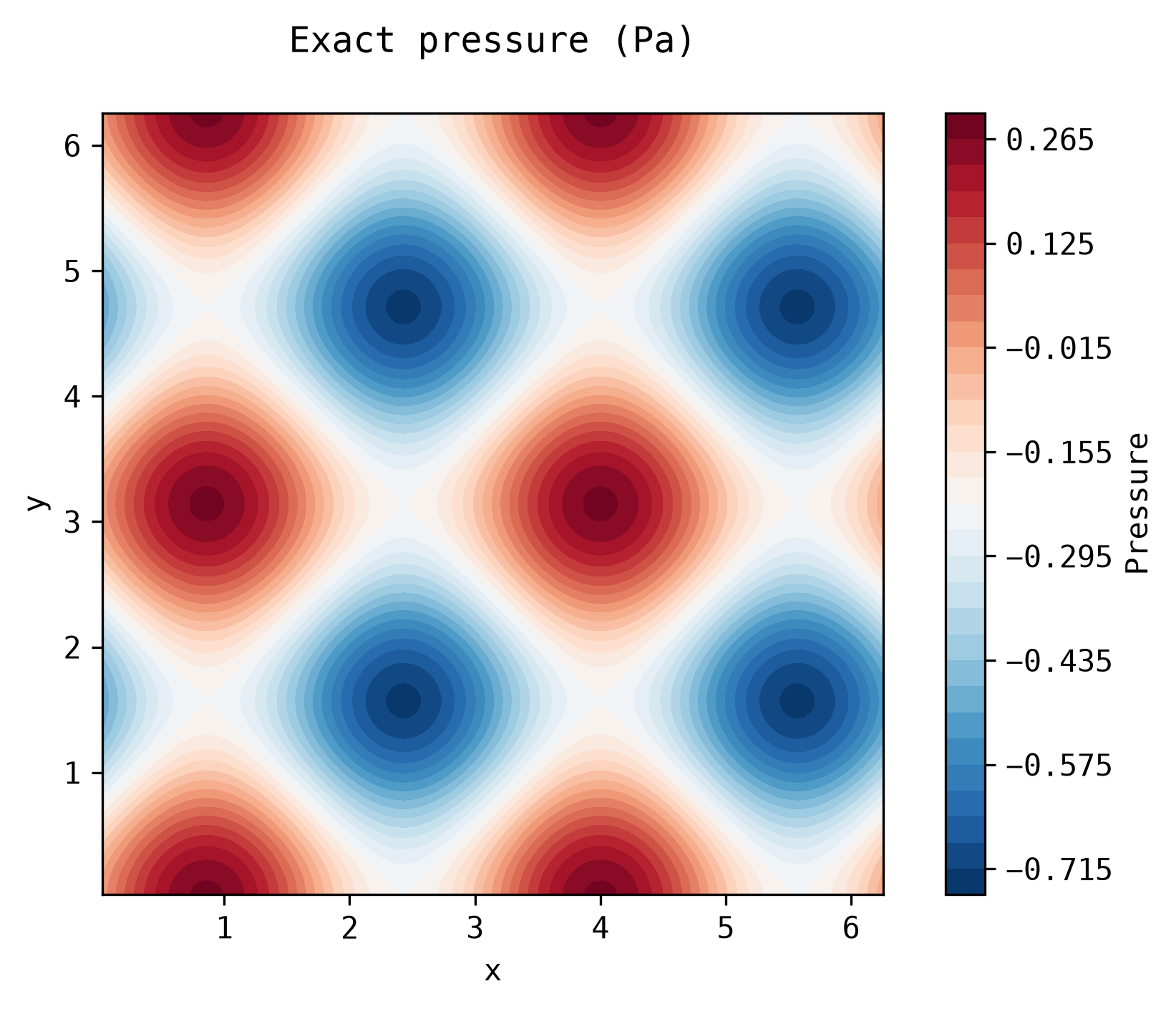}
		&
		\includegraphics[width=0.31\textwidth]{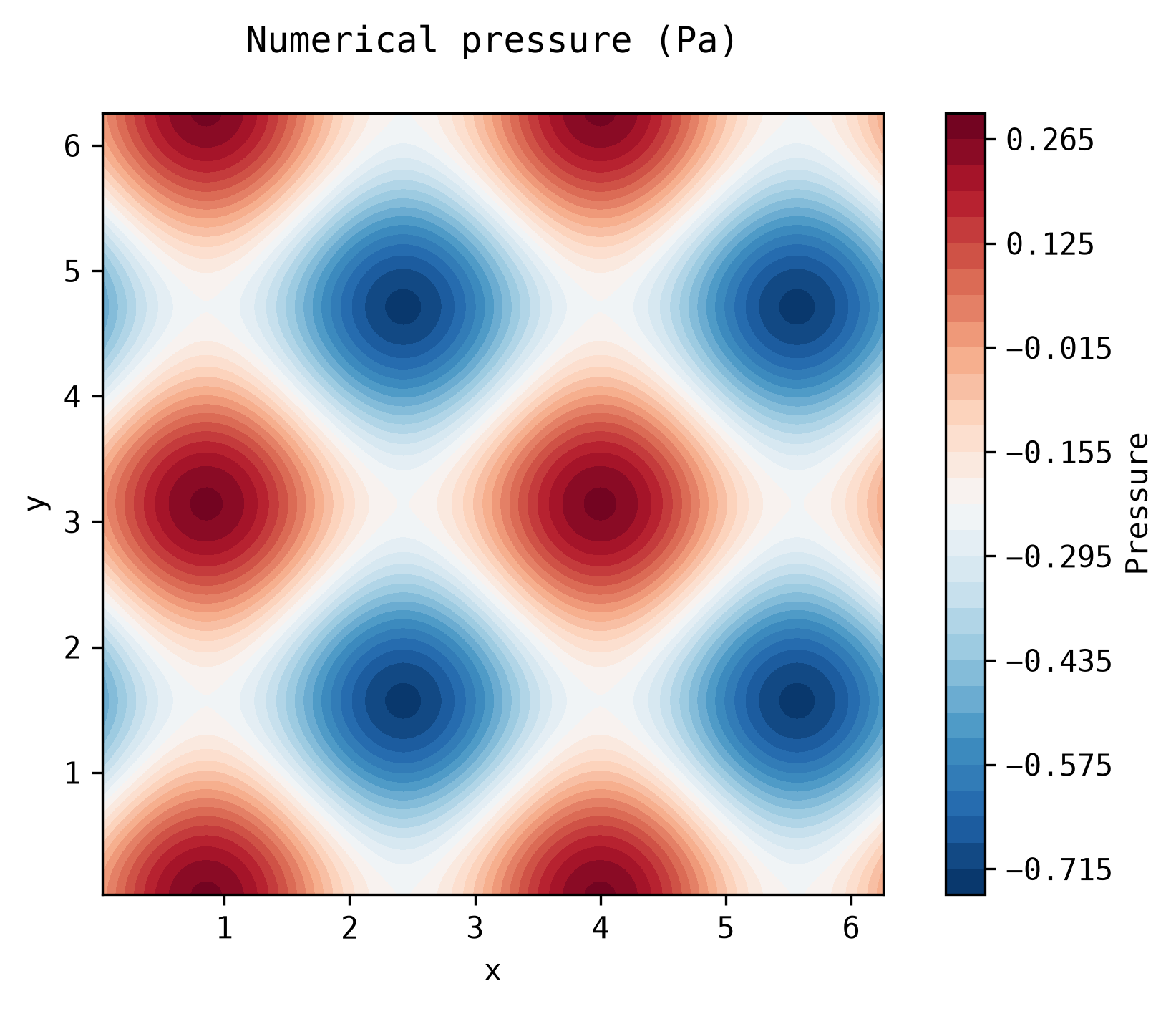}
		&
		\includegraphics[width=0.32\textwidth]{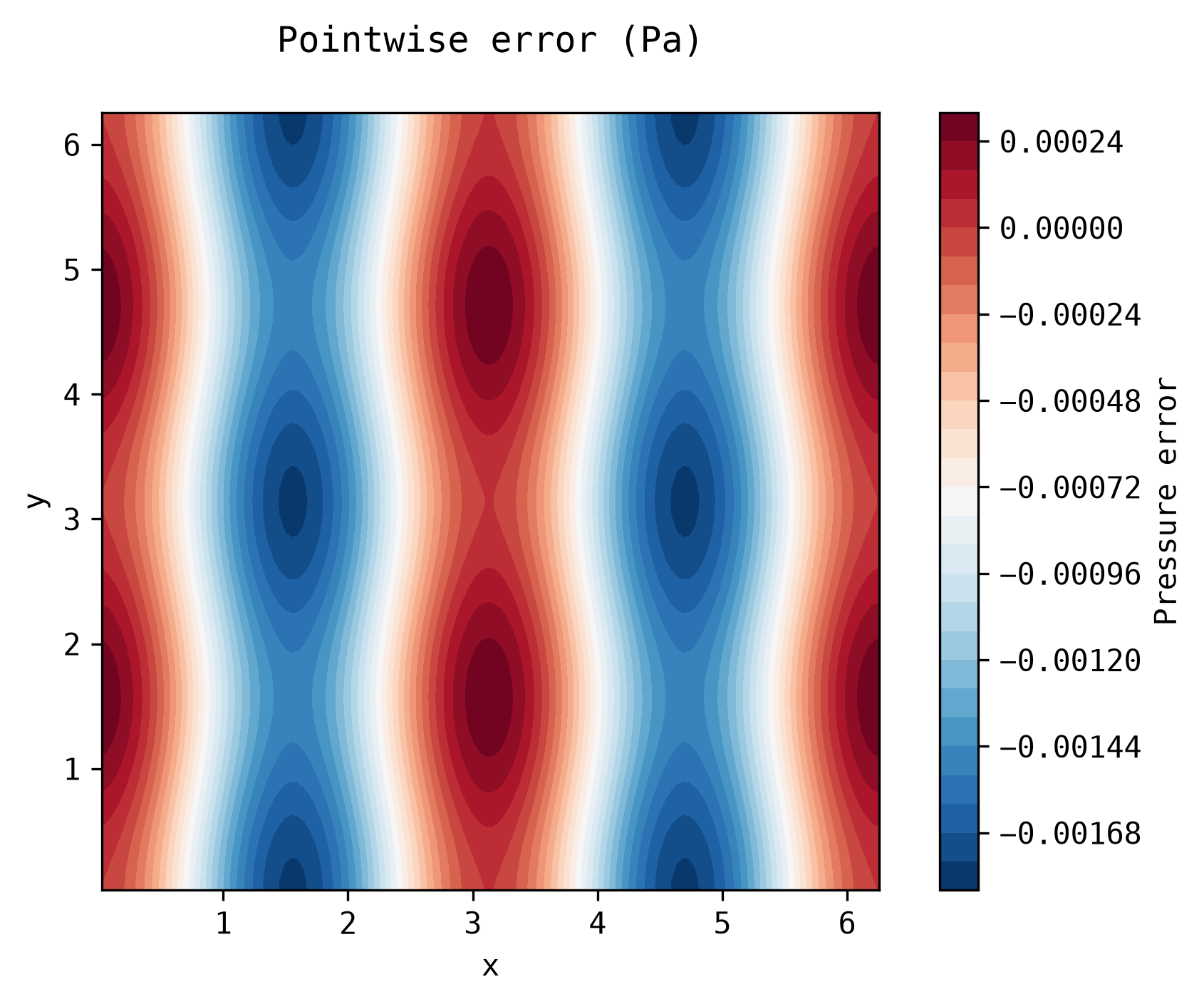}
		\\
		
	\end{tabular}
	
	\caption{Comparison of analytical and numerical fields for the inviscid
		advecting Taylor--Green vortex with a $128\times128$ uniform grid and $t_{\mathrm{final}}=4~\mathrm{s}$. Note that advection velocity is subtracted before velocity magnitude calculations.}
	\label{fig:invatg_fields}
\end{figure}

For visualization, the solution is advanced to $t=4$ s. Velocity and pressure errors are shown in Figure \ref{fig:invatg_fields}.

\subsection{Spatial convergence}

Spatial convergence is examined on the same sequence of successively
refined Cartesian grids used for the viscous advecting Taylor--Green
problem,
\begin{equation}
	16\times16,\;32\times32,\;64\times64,\;128\times128.
\end{equation}
Both uniform grids and smoothly stretched grids with
$s_x=s_y=0.6$ are considered. The calculations use
$\Delta t=3.125\times10^{-4}~\mathrm{s}$ and are advanced to
$t_{\mathrm{final}}=0.1~\mathrm{s}$. Numerical errors are evaluated
 against the analytical Euler solution using the measures
defined in Section~\ref{sec:error_measures}.

Tables~\ref{tab:inviscid_atg_spatial_uniform} and
\ref{tab:inviscid_atg_spatial_stretched} show second-order spatial
convergence in both velocity and pressure. The Standard and Perot
projection results are identical to the reported precision, consistent
with the inviscid relation $\Bop_2=\Bop_1=\Delta t I$. Grid stretching
increases the error magnitude at a given nominal resolution but does
not alter the asymptotic second-order convergence rate.

\begin{table}[!htbp]
	\centering
	\caption{Spatial convergence of the inviscid advecting
		Taylor--Green vortex on uniform grids. The calculations use
		$\Delta t=3.125\times10^{-4}$ s and
		$t_{\mathrm{final}}=0.1$ s.}
	\label{tab:inviscid_atg_spatial_uniform}
	
	\begin{tabular}{ccccc}
		\hline
		& \multicolumn{2}{c}{Standard} & \multicolumn{2}{c}{Perot} \\
		\cline{2-3}\cline{4-5}
		$N_x\times N_y$
		& $E_{\mathbf{V}}$ & $q_{\mathbf{V}}$
		& $E_{\mathbf{V}}$ & $q_{\mathbf{V}}$ \\
		\hline
		$16\times16$  & $1.2752\times10^{-3}$ & --     & $1.2752\times10^{-3}$ & --     \\
		$32\times32$  & $3.2066\times10^{-4}$ & 1.9917 & $3.2066\times10^{-4}$ & 1.9917 \\
		$64\times64$  & $8.0278\times10^{-5}$ & 1.9979 & $8.0278\times10^{-5}$ & 1.9979 \\
		$128\times128$ & $2.0075\times10^{-5}$ & 1.9996 & $2.0075\times10^{-5}$ & 1.9996 \\
		\hline
	\end{tabular}
	
	\vspace{0.8em}
	
\begin{tabular}{ccccc}
	\hline
	& \multicolumn{2}{c}{Standard} & \multicolumn{2}{c}{Perot} \\
	\cline{2-3}\cline{4-5}
	$N_x\times N_y$
	& $E_p$ & $q_p$
	& $E_p$ & $q_p$ \\
	\hline
	$16\times16$
	& $2.0704\times10^{-2}$ & --
	& $2.0704\times10^{-2}$ & -- \\
	$32\times32$
	& $5.3333\times10^{-3}$ & 1.9568
	& $5.3333\times10^{-3}$ & 1.9568 \\
	$64\times64$
	& $1.3344\times10^{-3}$ & 1.9988
	& $1.3344\times10^{-3}$ & 1.9988 \\
	$128\times128$
	& $3.3252\times10^{-4}$ & 2.0047
	& $3.3252\times10^{-4}$ & 2.0047 \\
	\hline
\end{tabular}
\end{table}

\begin{table}[!htbp]
	\centering
	\caption{Spatial convergence of the inviscid advecting
		Taylor--Green vortex on smoothly stretched grids with
		$s_x=s_y=0.6$. The calculations use
		$\Delta t=3.125\times10^{-4}$ s and
		$t_{\mathrm{final}}=0.1$ s.}
	\label{tab:inviscid_atg_spatial_stretched}
	
	\begin{tabular}{ccccc}
		\hline
		& \multicolumn{2}{c}{Standard} & \multicolumn{2}{c}{Perot} \\
		\cline{2-3}\cline{4-5}
		$N_x\times N_y$
		& $E_{\mathbf{V}}$ & $q_{\mathbf{V}}$
		& $E_{\mathbf{V}}$ & $q_{\mathbf{V}}$ \\
		\hline
		$16\times16$  & $4.3153\times10^{-3}$ & --     & $4.3153\times10^{-3}$ & --     \\
		$32\times32$  & $1.1010\times10^{-3}$ & 1.9706 & $1.1010\times10^{-3}$ & 1.9706 \\
		$64\times64$  & $2.7671\times10^{-4}$ & 1.9924 & $2.7671\times10^{-4}$ & 1.9924 \\
		$128\times128$ & $6.9268\times10^{-5}$ & 1.9981 & $6.9268\times10^{-5}$ & 1.9981 \\
		\hline
	\end{tabular}
	
	\vspace{0.8em}
	
	\begin{tabular}{ccccc}
		\hline
		& \multicolumn{2}{c}{Standard} & \multicolumn{2}{c}{Perot} \\
		\cline{2-3}\cline{4-5}
		$N_x\times N_y$
		& $E_p$ & $q_p$
		& $E_p$ & $q_p$ \\
		\hline
		$16\times16$
		& $3.6927\times10^{-2}$ & --
		& $3.6927\times10^{-2}$ & -- \\
		$32\times32$
		& $8.9151\times10^{-3}$ & 2.0503
		& $8.9151\times10^{-3}$ & 2.0503 \\
		$64\times64$
		& $2.1890\times10^{-3}$ & 2.0260
		& $2.1890\times10^{-3}$ & 2.0260 \\
		$128\times128$
		& $5.4216\times10^{-4}$ & 2.0135
		& $5.4216\times10^{-4}$ & 2.0135 \\
		\hline
	\end{tabular}
\end{table}

\subsection{Temporal convergence}

Temporal convergence is examined on the fixed $128\times128$ grid.
The calculations are advanced to $t_{\mathrm{final}}=0.1$ s using
\begin{equation}
	\Delta t
	=
	1.0\times10^{-2},\;
	5.0\times10^{-3},\;
	2.5\times10^{-3},\;
	1.25\times10^{-3}\ {\rm s}.
\end{equation}
A numerical reference solution is computed on the same grid using
$\Delta t_{\mathrm{ref}}=3.125\times10^{-4}$ s. The same reference
solution is used to evaluate both the velocity and pressure errors.

The numerical and reference velocities are compared directly at
$t_{\mathrm{final}}$. The pressure produced by a test calculation with
time step $\Delta t$ is instead associated with
\begin{equation}
	t_p
	=
	t_{\mathrm{final}}-\frac{\Delta t}{2}.
\end{equation}
The reference pressure at $t_p$ is obtained by linear interpolation
between the two reference pressure fields that bracket this time.
Thus, the numerical and reference pressures are compared at the same
physical time. This avoids the $O(\Delta t)$ time-location error
discussed in Section~\ref{sec:error_measures}.

The temporal errors $E_{\mathbf{V}}$ and $E_p$ and the corresponding
orders $r_{\mathbf{V}}$ and $r_p$ are evaluated using the measures
defined previously. Both the uniform grid and the smoothly stretched
grid with $s_x=s_y=0.6$ are considered. According to the results in
Tables~\ref{tab:inviscid_atg_temporal_uniform} and
\ref{tab:inviscid_atg_temporal_stretched}, we observe:
\begin{enumerate}
	\item
	The velocity is second-order accurate in time on both the uniform and
	stretched grids.
	
	\item
	The pressure is also second-order accurate in time on both grids.
	The observed pressure order remains close to two over the entire
	range of time steps.
	
	\item
	Grid stretching considered here has essentially no effect on the
	observed temporal orders.
	
	\item
	The Standard and Perot results are identical to the precision reported
	in the tables. This follows from the inviscid reduction
	$\Bop_2=\Bop_1$, and their differences are at the level of numerical
	round-off. This equivalence does not imply first-order temporal
	accuracy. It only establishes that the two projection formulations
	coincide when $\nu=0$. The complete AB2--projection scheme remains
	second-order accurate in time for both velocity and pressure in this
	periodic problem.
\end{enumerate}

\begin{table}[!htbp]
	\centering
	\caption{Temporal convergence of the inviscid advecting
		Taylor--Green vortex on the $128\times128$ uniform grid.
		The final time is $t_{\mathrm{final}}=0.1~\mathrm{s}$.}
	\label{tab:inviscid_atg_temporal_uniform}
	
	\begin{tabular}{ccccc}
		\hline
		& \multicolumn{2}{c}{Standard} & \multicolumn{2}{c}{Perot} \\
		\cline{2-3}\cline{4-5}
		$\Delta t$ (s)
		& $E_{\mathbf{V}}$ & $r_{\mathbf{V}}$
		& $E_{\mathbf{V}}$ & $r_{\mathbf{V}}$ \\
		\hline
		$1.0000\times10^{-2}$
		& $2.5084\times10^{-5}$ & 2.0041
		& $2.5084\times10^{-5}$ & 2.0041 \\
		$5.0000\times10^{-3}$
		& $6.2531\times10^{-6}$ & 2.0170
		& $6.2531\times10^{-6}$ & 2.0170 \\
		$2.5000\times10^{-3}$
		& $1.5450\times10^{-6}$ & 2.0703
		& $1.5450\times10^{-6}$ & 2.0703 \\
		$1.2500\times10^{-3}$
		& $3.6786\times10^{-7}$ & --
		& $3.6786\times10^{-7}$ & -- \\
		\hline
	\end{tabular}
	
	\vspace{0.8em}
	
	\begin{tabular}{ccccc}
		\hline
		& \multicolumn{2}{c}{Standard} & \multicolumn{2}{c}{Perot} \\
		\cline{2-3}\cline{4-5}
		$\Delta t$ (s)
		& $E_p$ & $r_p$
		& $E_p$ & $r_p$ \\
		\hline
		$1.0000\times10^{-2}$
		& $9.8828\times10^{-5}$ & 2.0053
		& $9.8828\times10^{-5}$ & 2.0053 \\
		$5.0000\times10^{-3}$
		& $2.4617\times10^{-5}$ & 2.0153
		& $2.4617\times10^{-5}$ & 2.0153 \\
		$2.5000\times10^{-3}$
		& $6.0890\times10^{-6}$ & 2.0599
		& $6.0890\times10^{-6}$ & 2.0599 \\
		$1.2500\times10^{-3}$
		& $1.4608\times10^{-6}$ & --
		& $1.4608\times10^{-6}$ & -- \\
		\hline
	\end{tabular}
\end{table}

\begin{table}[!htbp]
	\centering
	\caption{Temporal convergence of the inviscid advecting
		Taylor--Green vortex on the $128\times128$ smoothly stretched
		grid with $s_x=s_y=0.6$. The final time is
		$t_{\mathrm{final}}=0.1~\mathrm{s}$.}
	\label{tab:inviscid_atg_temporal_stretched}
	
	\begin{tabular}{ccccc}
		\hline
		& \multicolumn{2}{c}{Standard} & \multicolumn{2}{c}{Perot} \\
		\cline{2-3}\cline{4-5}
		$\Delta t$ (s)
		& $E_{\mathbf{V}}$ & $r_{\mathbf{V}}$
		& $E_{\mathbf{V}}$ & $r_{\mathbf{V}}$ \\
		\hline
		$1.0000\times10^{-2}$
		& $2.5045\times10^{-5}$ & 2.0039
		& $2.5045\times10^{-5}$ & 2.0039 \\
		$5.0000\times10^{-3}$
		& $6.2445\times10^{-6}$ & 2.0162
		& $6.2445\times10^{-6}$ & 2.0162 \\
		$2.5000\times10^{-3}$
		& $1.5437\times10^{-6}$ & 2.0679
		& $1.5437\times10^{-6}$ & 2.0679 \\
		$1.2500\times10^{-3}$
		& $3.6818\times10^{-7}$ & --
		& $3.6818\times10^{-7}$ & -- \\
		\hline
	\end{tabular}
	
	\vspace{0.8em}
	
	\begin{tabular}{ccccc}
		\hline
		& \multicolumn{2}{c}{Standard} & \multicolumn{2}{c}{Perot} \\
		\cline{2-3}\cline{4-5}
		$\Delta t$ (s)
		& $E_p$ & $r_p$
		& $E_p$ & $r_p$ \\
		\hline
		$1.0000\times10^{-2}$
		& $9.8494\times10^{-5}$ & 2.0058
		& $9.8494\times10^{-5}$ & 2.0058 \\
		$5.0000\times10^{-3}$
		& $2.4525\times10^{-5}$ & 2.0158
		& $2.4525\times10^{-5}$ & 2.0158 \\
		$2.5000\times10^{-3}$
		& $6.0645\times10^{-6}$ & 2.0604
		& $6.0645\times10^{-6}$ & 2.0604 \\
		$1.2500\times10^{-3}$
		& $1.4552\times10^{-6}$ & --
		& $1.4552\times10^{-6}$ & -- \\
		\hline
	\end{tabular}
\end{table}

\section{Viscous advecting Taylor--Green vortex}
\label{sec:taylor_green}

\subsection{Problem definition}

 Next we consider the viscous variant of the advecting Taylor--Green vortex problem. As the name implies, it is a viscous Taylor--Green vortex viewed in a frame translating with a uniform background velocity. Advection velocity, problem domain and boundary conditions are identical with those of the inviscid problem. The
exact velocity field is
\begin{align}
	u(x,y,t) &= U_a + U_0 \sin X \cos y \, e^{-2\nu t}, \\
	v(x,y,t) &= - U_0 \cos X \sin y \, e^{-2\nu t},
\end{align}
and the corresponding pressure field is
\begin{equation}
	p(x,y,t)
	=
	\frac{\rho U_0^2}{4}
	\left(
	\cos 2X
	+
	\cos 2y
	\right)
	e^{-4\nu t}
	+
	C,
\end{equation}
where $U_0$ is the initial velocity amplitude and $C$ denotes the arbitrary pressure datum. In the numerical
comparisons, the analytical pressure field is shifted at each
evaluation time so that its value at the pressure-reference cell
matches the prescribed numerical reference value, $p_{\mathrm{ref}}=0$. In the calculations reported below,
\begin{equation}
	U_a=1,
	\qquad
	U_0=1,
	\qquad
	\rho=1,
	\qquad
	\nu=0.01.
\end{equation}

The analytical fields are evaluated directly at the locations of the
corresponding discrete variables. Thus, $u$ is evaluated at the
vertical velocity faces, $v$ at the horizontal velocity faces, and
$p$ at the pressure-cell centers. No interpolation between the
staggered locations is therefore required when the numerical and
analytical solutions are compared. The periodic duplicate $u$ and
$v$ faces are counted only once in the error evaluation.

To remove the discrete divergence that may arise when the analytical solution is sampled on a stretched grid, an initial projection is carried out. This is later disccused in Section \ref{ssec:violation}.

\begin{figure}[!htbp]
	\centering
	
	\begin{tabular}{ccc}
		\includegraphics[width=0.31\textwidth]{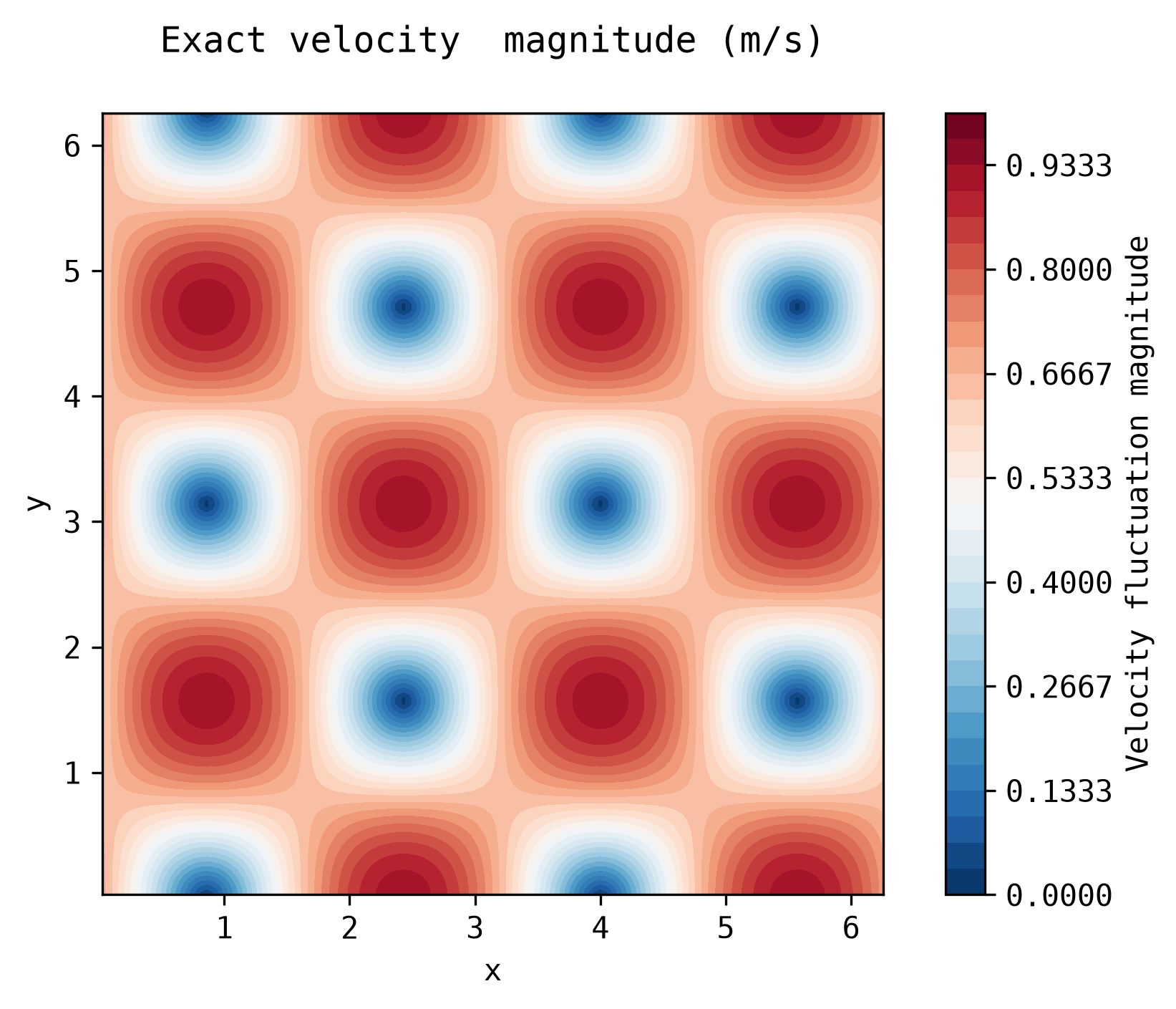}
		&
		\includegraphics[width=0.31\textwidth]{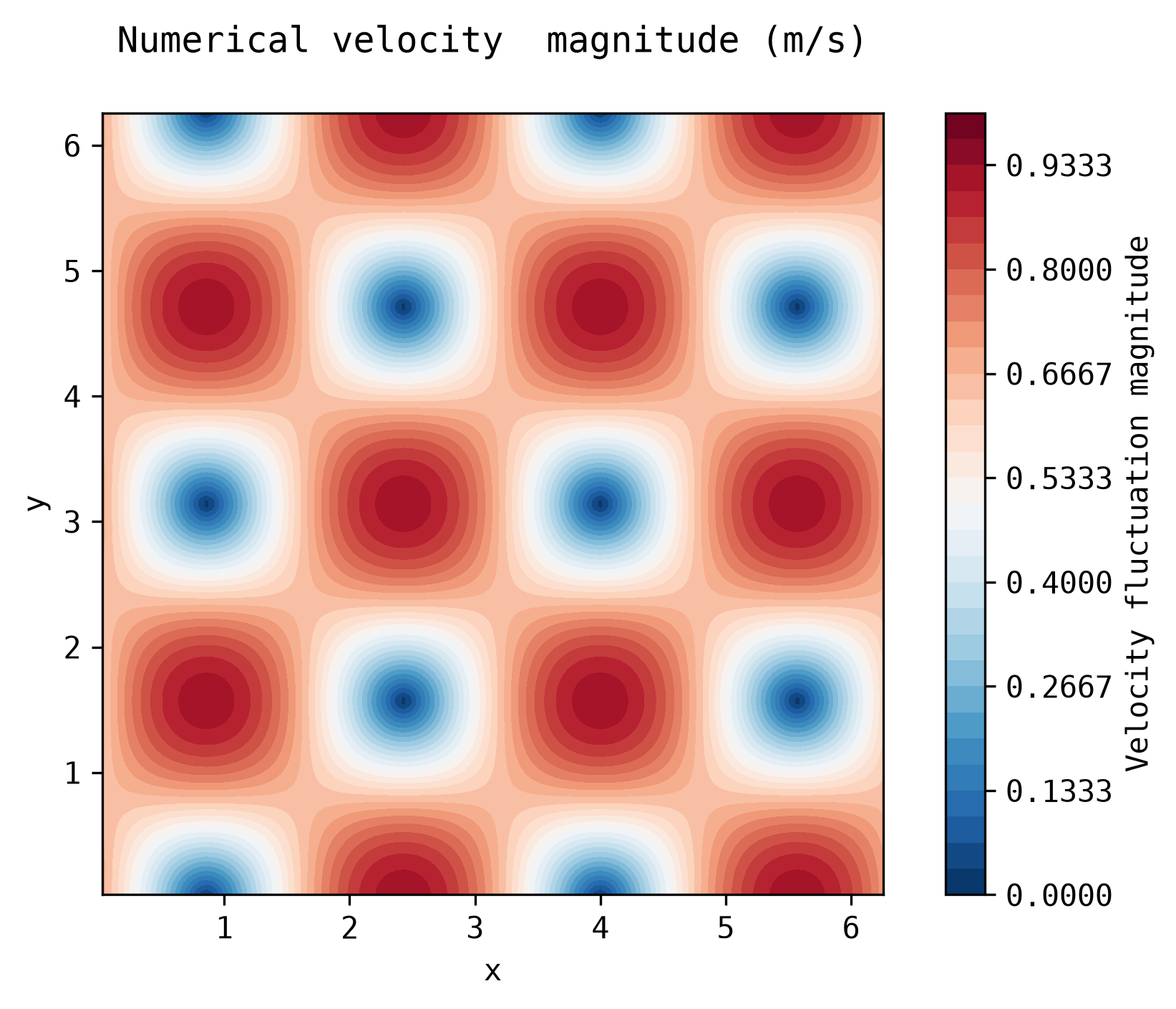}
		&
		\includegraphics[width=0.31\textwidth]{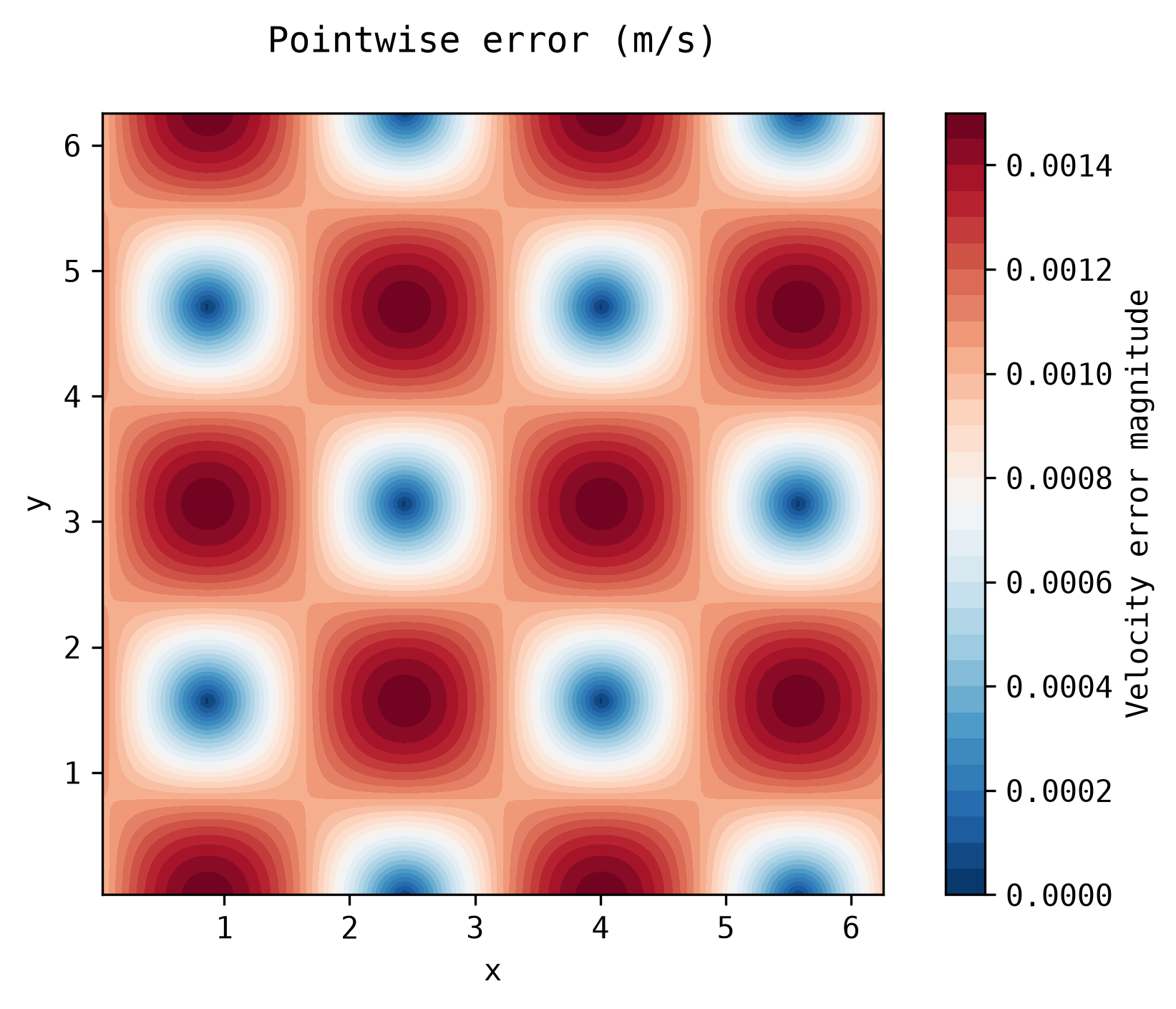}
		\\
		
		\includegraphics[width=0.31\textwidth]{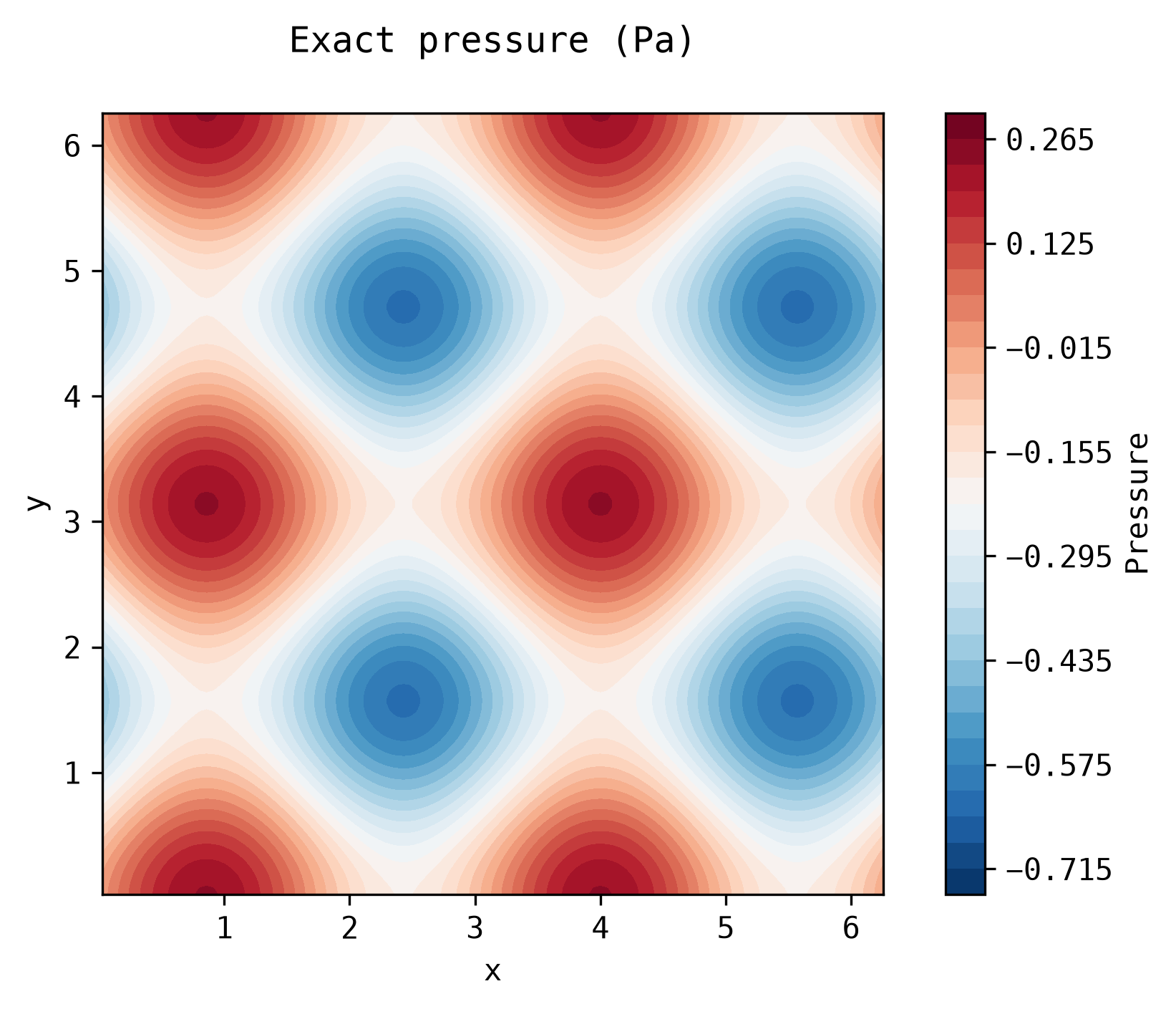}
		&
		\includegraphics[width=0.31\textwidth]{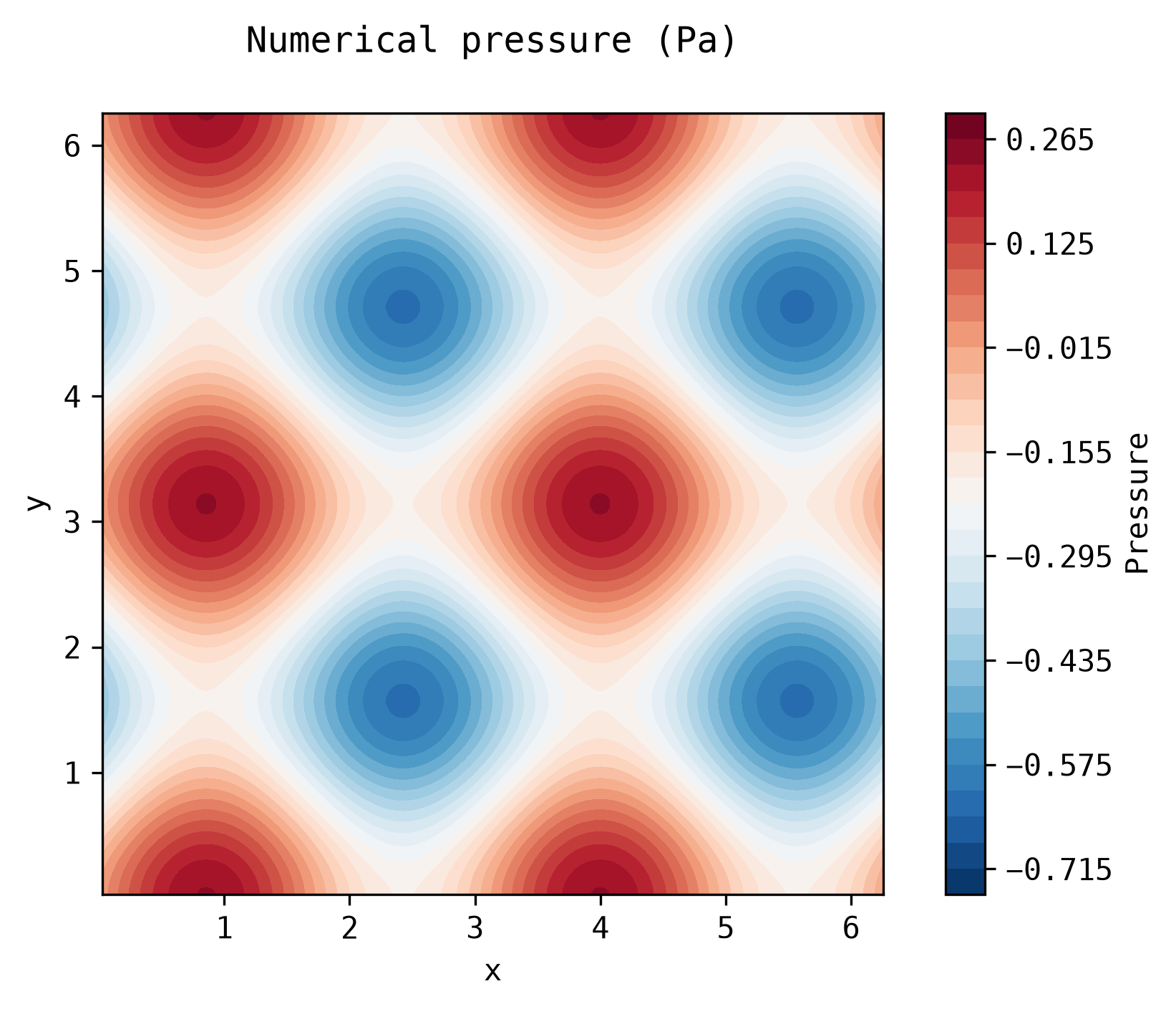}
		&
		\includegraphics[width=0.32\textwidth]{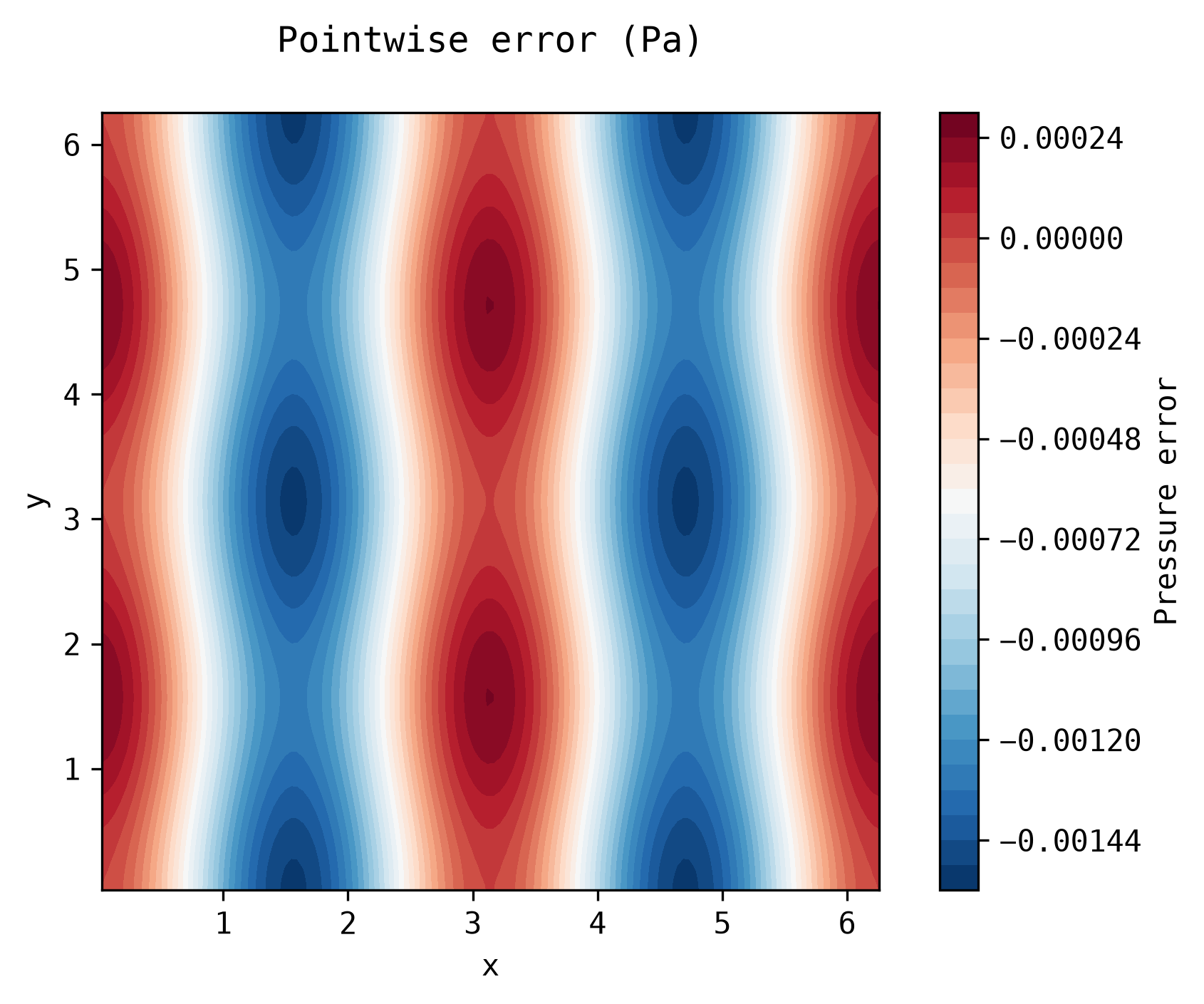}
		\\

	\end{tabular}
	
	\caption{Comparison of analytical and numerical fields for the viscous
		advecting Taylor--Green vortex with a $128\times128$ uniform grid and $t_{\mathrm{final}}=4~\mathrm{s}$. Note that advection velocity is subtracted before velocity magnitude calculations. }
	\label{fig:atg_fields}
\end{figure}

For visualization, the solution is advanced to $t=4$ s. Velocity and pressure errors are shown in Figure \ref{fig:atg_fields}. Velocity contours show slight decay compared to the inviscid case as expected.

\subsection{Spatial convergence}

Spatial convergence was examined by comparing the numerical solution with
the exact Taylor--Green solution on successively refined Cartesian grids.
The resolutions considered were
\begin{equation}
	N_x\times N_y
	=
	16 \times 16,\;32 \times 32,\;64 \times 64,\;128 \times 128 .
\end{equation}

Both uniform and smoothly stretched grids were considered. For the
stretched grids, the common mapping defined in
Section~\ref{sec:error_measures} was used with
$s_x=s_y=0.6$.

Tables~\ref{tab:atg_spatial_uniform} and
\ref{tab:atg_spatial_stretched} report the spatial-convergence results
for the uniform and smoothly stretched grids, respectively. Results are
given for both the standard and Perot projection methods. The time step
was selected sufficiently small that temporal discretization errors do
not substantially affect the spatial convergence rates reported here.
Both methods exhibit second-order spatial convergence in velocity. The
pressure errors likewise approach second-order convergence. The standard
and Perot methods produce essentially identical velocity errors at every
resolution, whereas small differences are observed in the pressure
errors.

Grid stretching increases the magnitude of the velocity error at a given
nominal resolution, but does not alter its asymptotic convergence rate.
Second-order behavior is also recovered for pressure on the stretched
grids. These results demonstrate that the spatial discretization retains
its nominal second-order accuracy on the smoothly nonuniform Cartesian
meshes considered here.

\begin{table}[!htbp]
	\centering
	\caption{Spatial convergence of the viscous advecting Taylor--Green
		vortex on uniform grids. The calculations use
		$\Delta t=3.125\times10^{-4}$ s and
		$t_{\mathrm{final}}=0.1$ s.}
	\label{tab:atg_spatial_uniform}
	
	\begin{tabular}{ccccc}
		\hline
		& \multicolumn{2}{c}{Standard} & \multicolumn{2}{c}{Perot} \\
		\cline{2-3}\cline{4-5}
		$N_x\times N_y$
		& $E_{\mathbf{V}}$ & $q_{\mathbf{V}}$
		& $E_{\mathbf{V}}$ & $q_{\mathbf{V}}$ \\
		\hline
		$16\times16$
		& $1.2728\times10^{-3}$ & --
		& $1.2728\times10^{-3}$ & -- \\
		$32\times32$
		& $3.2003\times10^{-4}$ & 1.9917
		& $3.2003\times10^{-4}$ & 1.9917 \\
		$64\times64$
		& $8.0122\times10^{-5}$ & 1.9979
		& $8.0122\times10^{-5}$ & 1.9979 \\
		$128\times128$
		& $2.0036\times10^{-5}$ & 1.9996
		& $2.0036\times10^{-5}$ & 1.9996 \\
		\hline
	\end{tabular}
	
	\vspace{0.8em}
	
	\begin{tabular}{ccccc}
		\hline
		& \multicolumn{2}{c}{Standard} & \multicolumn{2}{c}{Perot} \\
		\cline{2-3}\cline{4-5}
		$N_x\times N_y$
		& $E_p$ & $q_p$
		& $E_p$ & $q_p$ \\
		\hline
		$16\times16$
		& $2.0598\times10^{-2}$ & --
		& $2.0595\times10^{-2}$ & -- \\
		$32\times32$
		& $5.3084\times10^{-3}$ & 1.9562
		& $5.3051\times10^{-3}$ & 1.9569 \\
		$64\times64$
		& $1.3308\times10^{-3}$ & 1.9960
		& $1.3273\times10^{-3}$ & 1.9988 \\
		$128\times128$
		& $3.3421\times10^{-4}$ & 1.9935
		& $3.3075\times10^{-4}$ & 2.0047 \\
		\hline
	\end{tabular}
\end{table}

\begin{table}[!htbp]
	\centering
	\caption{Spatial convergence of the viscous advecting Taylor--Green
		vortex on smoothly stretched grids with $s_x=s_y=0.6$.
		The calculations use $\Delta t=3.125\times10^{-4}$ s and
		$t_{\mathrm{final}}=0.1$ s.}
	\label{tab:atg_spatial_stretched}
	
	\begin{tabular}{ccccc}
		\hline
		& \multicolumn{2}{c}{Standard} & \multicolumn{2}{c}{Perot} \\
		\cline{2-3}\cline{4-5}
		$N_x\times N_y$
		& $E_{\mathbf{V}}$ & $q_{\mathbf{V}}$
		& $E_{\mathbf{V}}$ & $q_{\mathbf{V}}$ \\
		\hline
		$16\times16$
		& $4.3038\times10^{-3}$ & --
		& $4.3038\times10^{-3}$ & -- \\
		$32\times32$
		& $1.0979\times10^{-3}$ & 1.9708
		& $1.0979\times10^{-3}$ & 1.9708 \\
		$64\times64$
		& $2.7593\times10^{-4}$ & 1.9925
		& $2.7593\times10^{-4}$ & 1.9925 \\
		$128\times128$
		& $6.9072\times10^{-5}$ & 1.9981
		& $6.9072\times10^{-5}$ & 1.9981 \\
		\hline
	\end{tabular}
	
	\vspace{0.8em}
	
	\begin{tabular}{ccccc}
		\hline
		& \multicolumn{2}{c}{Standard} & \multicolumn{2}{c}{Perot} \\
		\cline{2-3}\cline{4-5}
		$N_x\times N_y$
		& $E_p$ & $q_p$
		& $E_p$ & $q_p$ \\
		\hline
		$16\times16$
		& $3.6786\times10^{-2}$ & --
		& $3.6783\times10^{-2}$ & -- \\
		$32\times32$
		& $8.8829\times10^{-3}$ & 2.0501
		& $8.8796\times10^{-3}$ & 2.0505 \\
		$64\times64$
		& $2.1834\times10^{-3}$ & 2.0245
		& $2.1801\times10^{-3}$ & 2.0261 \\
		$128\times128$
		& $5.4320\times10^{-4}$ & 2.0070
		& $5.3996\times10^{-4}$ & 2.0135 \\
		\hline
	\end{tabular}
\end{table}

\subsection{Temporal convergence}

Temporal convergence is examined on a fixed $256\times256$ grid.
The calculations are advanced to $t_{\mathrm{final}}=0.1$ s using
\begin{equation}
	\Delta t
	=
	1.25\times10^{-3},\;
	6.25\times10^{-4},\;
	3.125\times10^{-4},\;
	1.5625\times10^{-4}\ {\rm s}.
\end{equation}
A single numerical reference solution is computed on the same spatial
grid using
\begin{equation}
	\Delta t_{\mathrm{ref}}
	=
	4.8828125\times10^{-6}\ {\rm s}.
\end{equation}
A separate reference solution is computed for each projection method
and grid.

The velocity fields are compared at $t_{\mathrm{final}}$. The pressure
produced by a calculation with time step $\Delta t$ is associated with
the midpoint of the final time interval,
\begin{equation}
	t_{\mathrm{final}}-\frac{\Delta t}{2}.
\end{equation}
The reference pressure is therefore evaluated at this same physical
time. When the required time lies between two pressure time levels of
the reference solution, the reference pressure is obtained by linear
interpolation, as described in Section~\ref{sec:error_measures}.

Using a numerical reference solution on the same spatial grid largely
removes the common spatial discretization error from the comparison.
This allows the temporal behavior to be examined separately. Both the
uniform grid and the smoothly stretched grid with $s_x=s_y=0.6$ are
considered.

According to the results in
Tables~\ref{tab:atg_temporal_uniform} and
\ref{tab:atg_temporal_stretched}, we observe:
\begin{enumerate}
	\item
	The velocity is second-order accurate in time on both the uniform and
	stretched grids. The Standard and Perot velocity errors are identical
	to the precision reported in the tables.
	
	\item
	The pressure obtained with the Standard projection approaches
	first-order temporal convergence. On the uniform grid, the observed
	orders are $0.9210$, $0.9708$, and $1.0032$. Nearly identical values
	are obtained on the stretched grid.
	
	\item
	The pressure obtained with the Perot projection is second-order
	accurate in time. The observed pressure orders remain very close to
	two over the complete range of time steps on both grids.
	
	\item
	Grid stretching has little effect on the observed temporal behavior.
	The velocity remains approximately second-order accurate, while the
	pressure retains the order associated with the selected projection
	method.
\end{enumerate}

For the periodic problem considered here, the Standard projection also
retains second-order temporal accuracy for velocity. The exceptional
behavior of the Standard projection with periodic boundary conditions
is discussed in Section~\ref{ssec:periodicproj}.

\begin{table}[!htbp]
	\centering
	\caption{Temporal convergence of the viscous advecting
		Taylor--Green vortex on the $256\times256$ uniform grid.
		The final time is $t_{\mathrm{final}}=0.1~\mathrm{s}$.}
	\label{tab:atg_temporal_uniform}
	
	\begin{tabular}{ccccc}
		\hline
		& \multicolumn{2}{c}{Standard} & \multicolumn{2}{c}{Perot} \\
		\cline{2-3}\cline{4-5}
		$\Delta t$ (s)
		& $E_{\mathbf{V}}$ & $r_{\mathbf{V}}$
		& $E_{\mathbf{V}}$ & $r_{\mathbf{V}}$ \\
		\hline
		$1.2500\times10^{-3}$
		& $3.9103\times10^{-7}$ & 2.0001
		& $3.9103\times10^{-7}$ & 2.0001 \\
		$6.2500\times10^{-4}$
		& $9.7753\times10^{-8}$ & 2.0003
		& $9.7753\times10^{-8}$ & 2.0003 \\
		$3.1250\times10^{-4}$
		& $2.4433\times10^{-8}$ & 2.0013
		& $2.4433\times10^{-8}$ & 2.0013 \\
		$1.5625\times10^{-4}$
		& $6.1030\times10^{-9}$ & --
		& $6.1030\times10^{-9}$ & -- \\
		\hline
	\end{tabular}
	
	\vspace{0.8em}
	
	\begin{tabular}{ccccc}
		\hline
		& \multicolumn{2}{c}{Standard} & \multicolumn{2}{c}{Perot} \\
		\cline{2-3}\cline{4-5}
		$\Delta t$ (s)
		& $E_p$ & $r_p$
		& $E_p$ & $r_p$ \\
		\hline
		$1.2500\times10^{-3}$
		& $1.2282\times10^{-5}$ & 0.9210
		& $1.5253\times10^{-6}$ & 2.0004 \\
		$6.2500\times10^{-4}$
		& $6.4866\times10^{-6}$ & 0.9708
		& $3.8123\times10^{-7}$ & 2.0004 \\
		$3.1250\times10^{-4}$
		& $3.3096\times10^{-6}$ & 1.0032
		& $9.5282\times10^{-8}$ & 2.0014 \\
		$1.5625\times10^{-4}$
		& $1.6511\times10^{-6}$ & --
		& $2.3797\times10^{-8}$ & -- \\
		\hline
	\end{tabular}
\end{table}

\begin{table}[!htbp]
	\centering
	\caption{Temporal convergence of the viscous advecting
		Taylor--Green vortex on the $256\times256$ smoothly stretched
		grid with $s_x=s_y=0.6$. The final time is
		$t_{\mathrm{final}}=0.1~\mathrm{s}$.}
	\label{tab:atg_temporal_stretched}
	
	\begin{tabular}{ccccc}
		\hline
		& \multicolumn{2}{c}{Standard} & \multicolumn{2}{c}{Perot} \\
		\cline{2-3}\cline{4-5}
		$\Delta t$ (s)
		& $E_{\mathbf{V}}$ & $r_{\mathbf{V}}$
		& $E_{\mathbf{V}}$ & $r_{\mathbf{V}}$ \\
		\hline
		$1.2500\times10^{-3}$
		& $3.9086\times10^{-7}$ & 1.9991
		& $3.9086\times10^{-7}$ & 1.9991 \\
		$6.2500\times10^{-4}$
		& $9.7779\times10^{-8}$ & 1.9963
		& $9.7779\times10^{-8}$ & 1.9963 \\
		$3.1250\times10^{-4}$
		& $2.4507\times10^{-8}$ & 1.9861
		& $2.4507\times10^{-8}$ & 1.9861 \\
		$1.5625\times10^{-4}$
		& $6.1858\times10^{-9}$ & --
		& $6.1858\times10^{-9}$ & -- \\
		\hline
	\end{tabular}
	
	\vspace{0.8em}
	
	\begin{tabular}{ccccc}
		\hline
		& \multicolumn{2}{c}{Standard} & \multicolumn{2}{c}{Perot} \\
		\cline{2-3}\cline{4-5}
		$\Delta t$ (s)
		& $E_p$ & $r_p$
		& $E_p$ & $r_p$ \\
		\hline
		$1.2500\times10^{-3}$
		& $1.2280\times10^{-5}$ & 0.9212
		& $1.5220\times10^{-6}$ & 2.0008 \\
		$6.2500\times10^{-4}$
		& $6.4849\times10^{-6}$ & 0.9709
		& $3.8030\times10^{-7}$ & 2.0008 \\
		$3.1250\times10^{-4}$
		& $3.3085\times10^{-6}$ & 1.0032
		& $9.5023\times10^{-8}$ & 2.0022 \\
		$1.5625\times10^{-4}$
		& $1.6506\times10^{-6}$ & --
		& $2.3720\times10^{-8}$ & -- \\
		\hline
	\end{tabular}
\end{table}

\section{Wall-bounded vortex}
\label{ssec:wall_bounded_vortex}

\subsection{Problem definition}

To examine temporal accuracy in the presence of non-periodic no-slip
boundaries, we next consider a wall-bounded vortex initialization problem taken from \cite{perot1993analysis}. Here a smooth divergence-free velocity field is considered on the unit square,
$0\leq x,y\leq1$:
\begin{align}
	u(x,y,0)
	&=
	\left[1-\cos(2\pi x)\right]\sin(2\pi y),
	\\
	v(x,y,0)
	&=
	\left[\cos(2\pi y)-1\right]\sin(2\pi x).
\end{align}
The velocity field satisfies the no-slip condition on all four
boundaries and is divergence-free in a continuous sense. It therefore provides
a wall-bounded alternative to the Taylor--Green problem for examining
the temporal behavior of the projection methods. It is also free of any velocity
discontinuities such as those occurring at the upper corners of the cavity. This is important, because such discontinuities make the temporal accuracy trend difficult to identify with the current metrics.

The flow is advanced by the unforced incompressible Navier--Stokes
equations with
\begin{equation}
	\rho=1,
	\qquad
	\nu=0.01.
\end{equation}
Because a closed-form transient solution is not used, temporal errors
are evaluated relative to a finely time-resolved numerical solution
computed on the same spatial grid.

The sampled initial velocity field is projected once onto the
discretely solenoidal velocity space before time integration. 

\begin{figure*}[!htbp]
	\centering
	
	\begin{subfigure}[t]{0.37\textwidth}
		\centering
		\includegraphics[width=\linewidth]{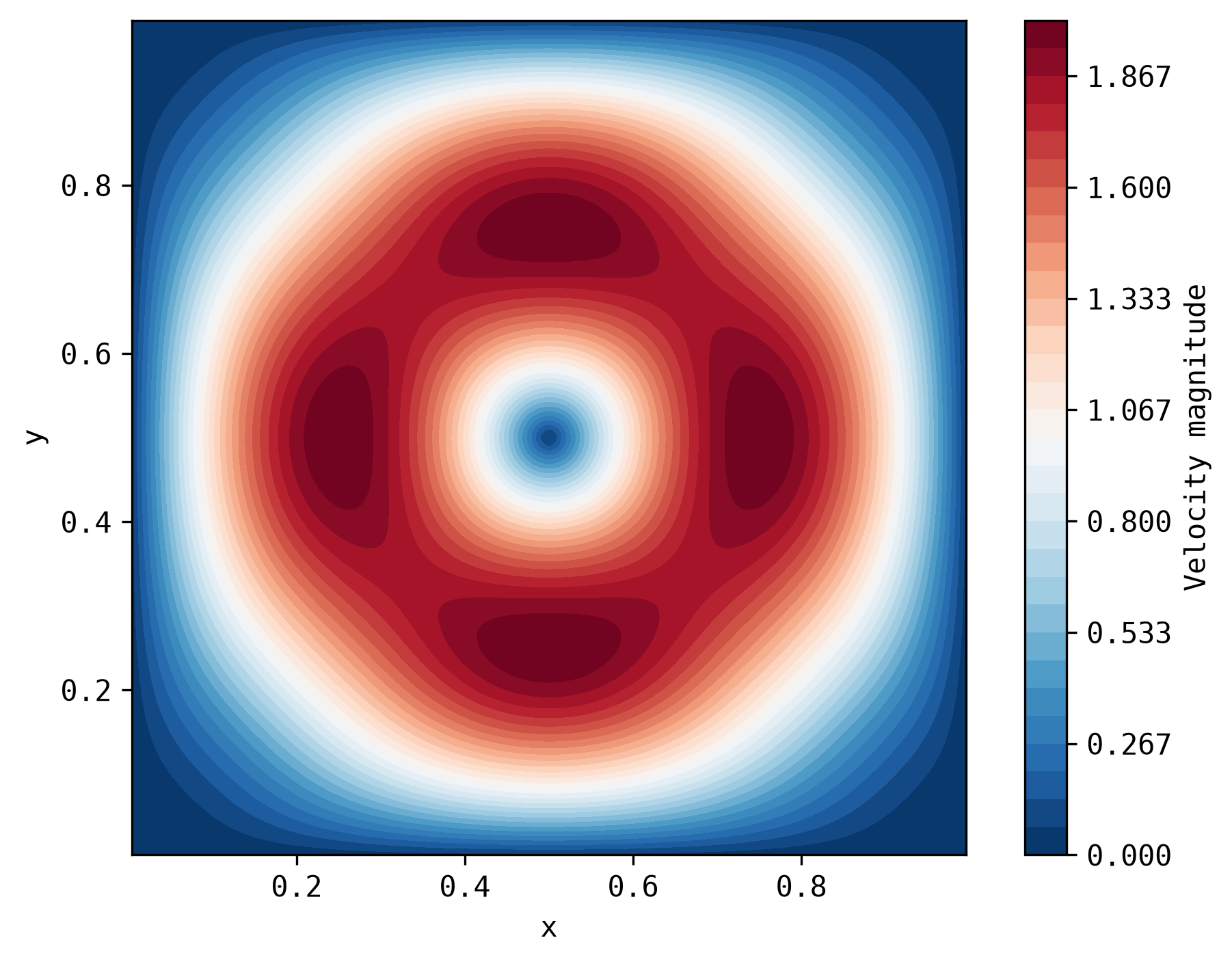}
	\end{subfigure}
	\begin{subfigure}[t]{0.3\textwidth}
		\centering
		\includegraphics[width=\linewidth]{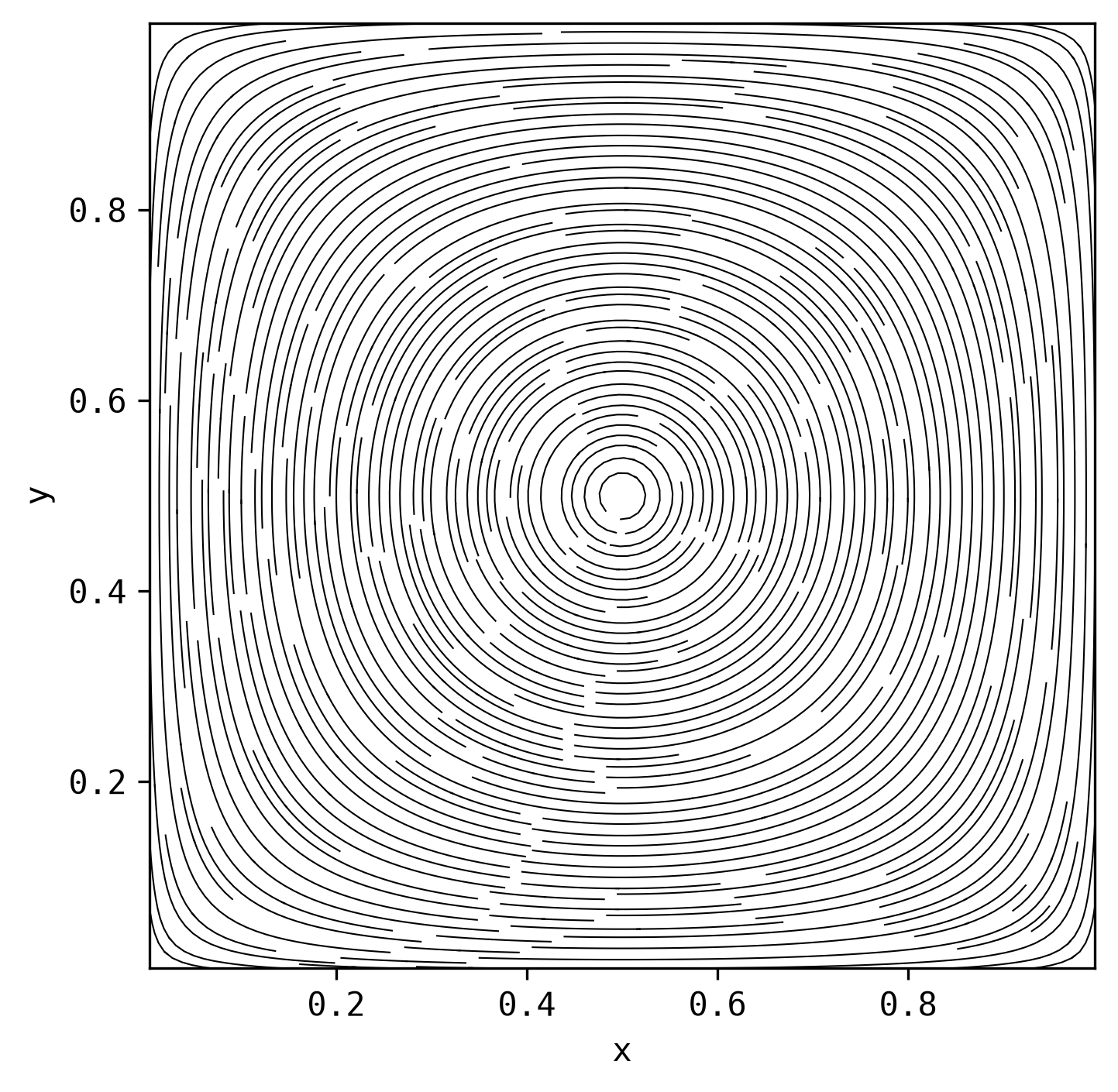}
	\end{subfigure}
	
	
	\begin{subfigure}[t]{0.37\textwidth}
		\centering
		\includegraphics[width=\linewidth]{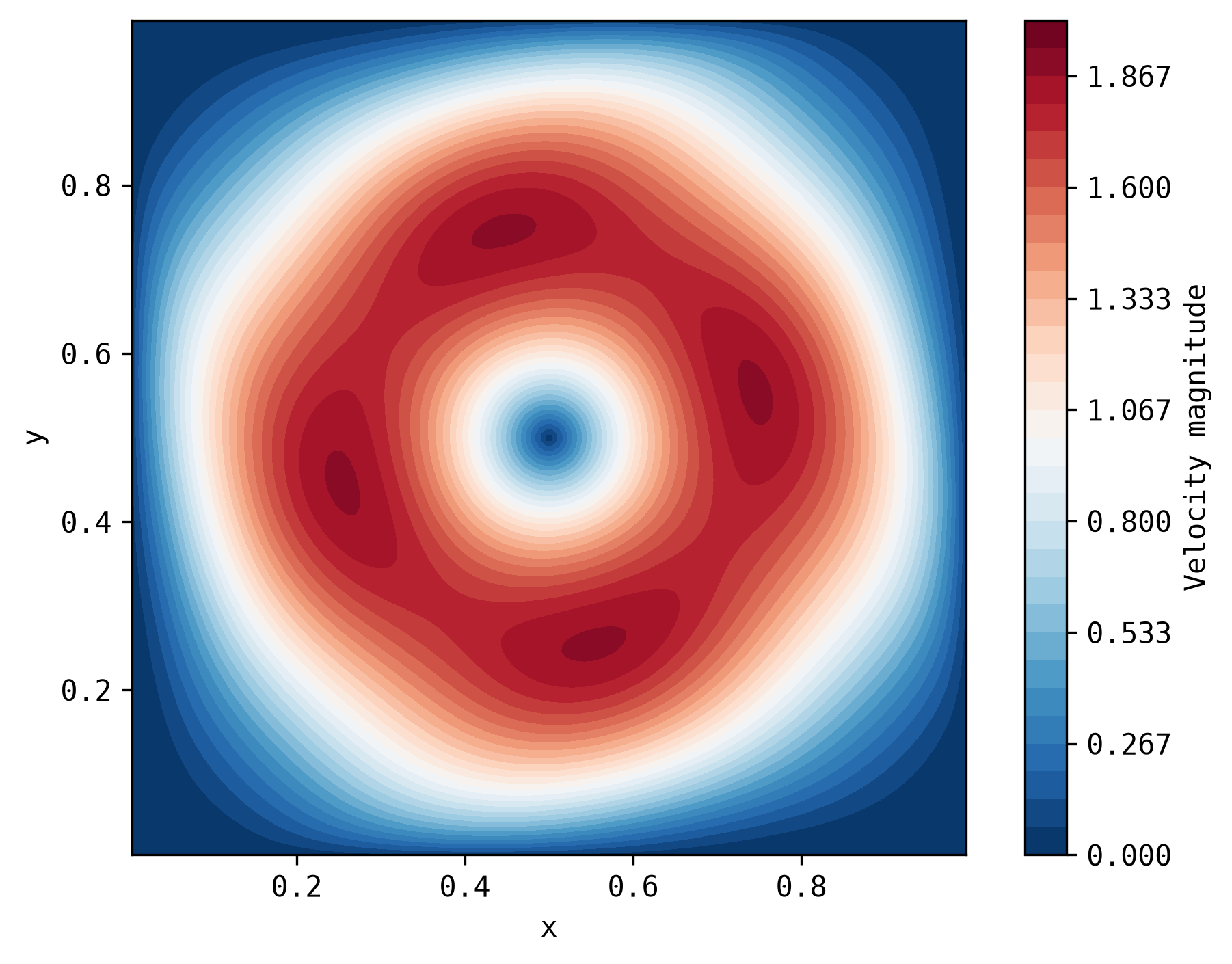}
	\end{subfigure}
	\begin{subfigure}[t]{0.3\textwidth}
		\centering
		\includegraphics[width=\linewidth]{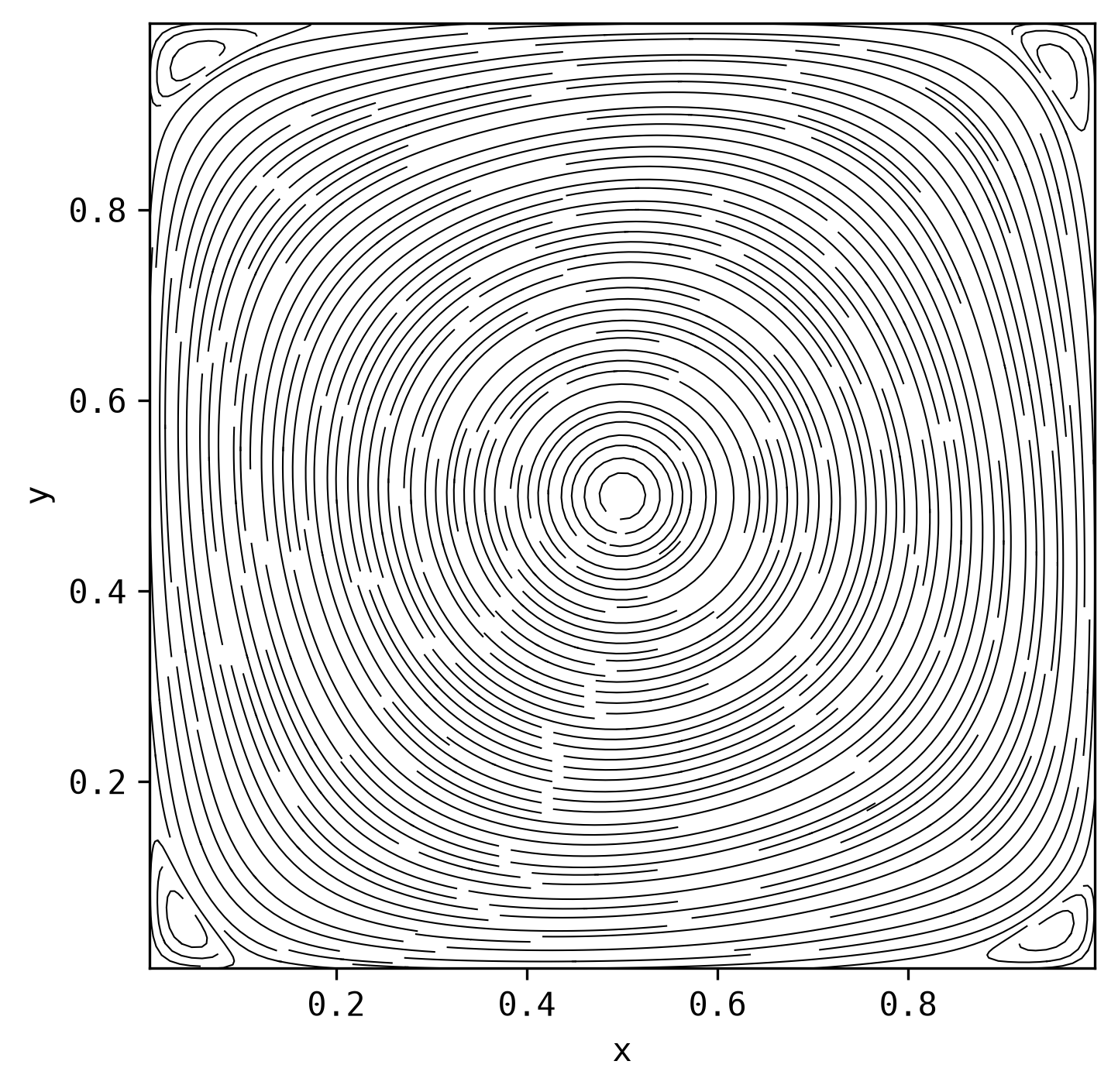}
	\end{subfigure}
	
	
	\begin{subfigure}[t]{0.37\textwidth}
		\centering
		\includegraphics[width=\linewidth]{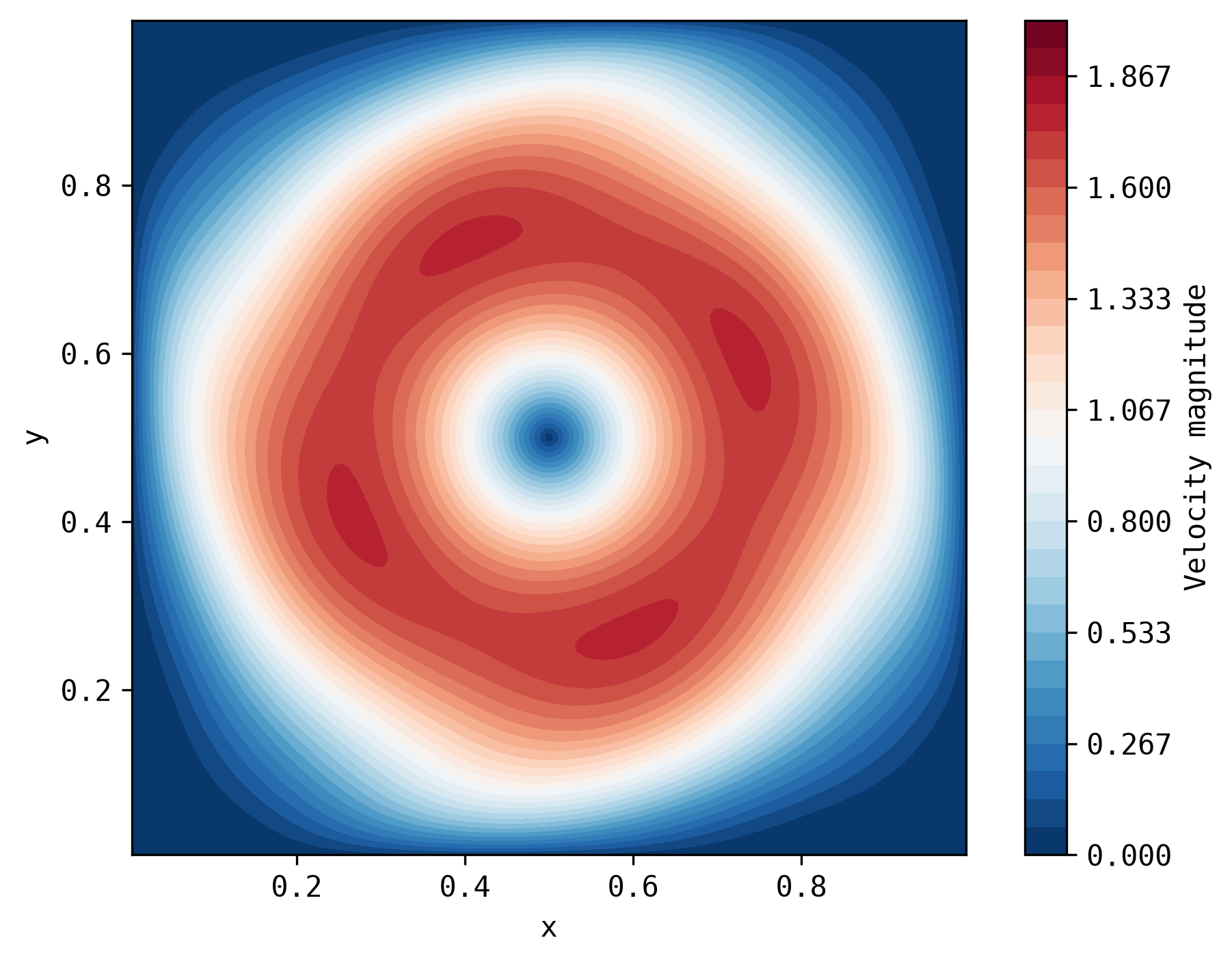}
	\end{subfigure}
	\begin{subfigure}[t]{0.3\textwidth}
		\centering
		\includegraphics[width=\linewidth]{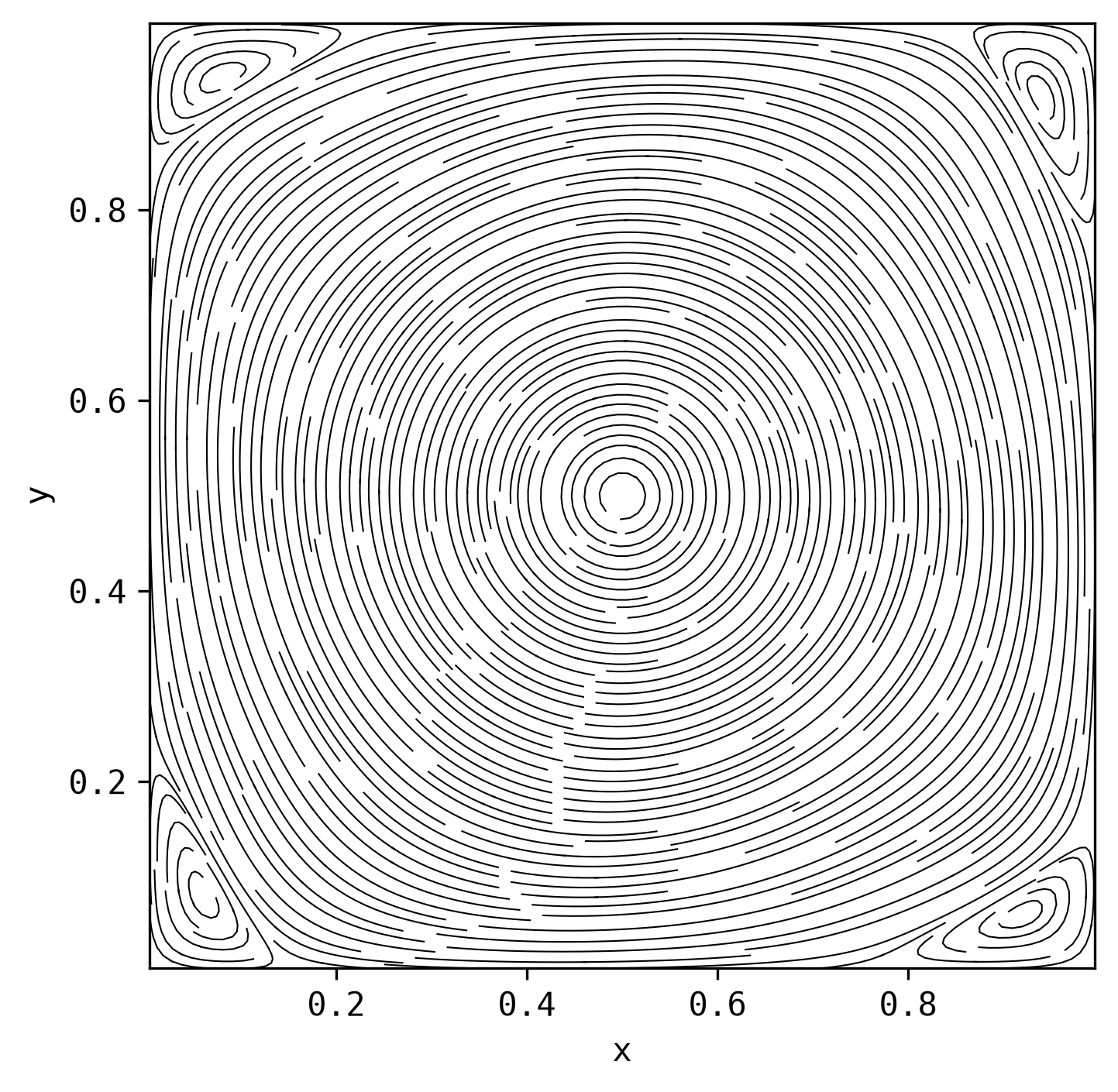}
	\end{subfigure}
	
	
	\begin{subfigure}[t]{0.37\textwidth}
		\centering
		\includegraphics[width=\linewidth]{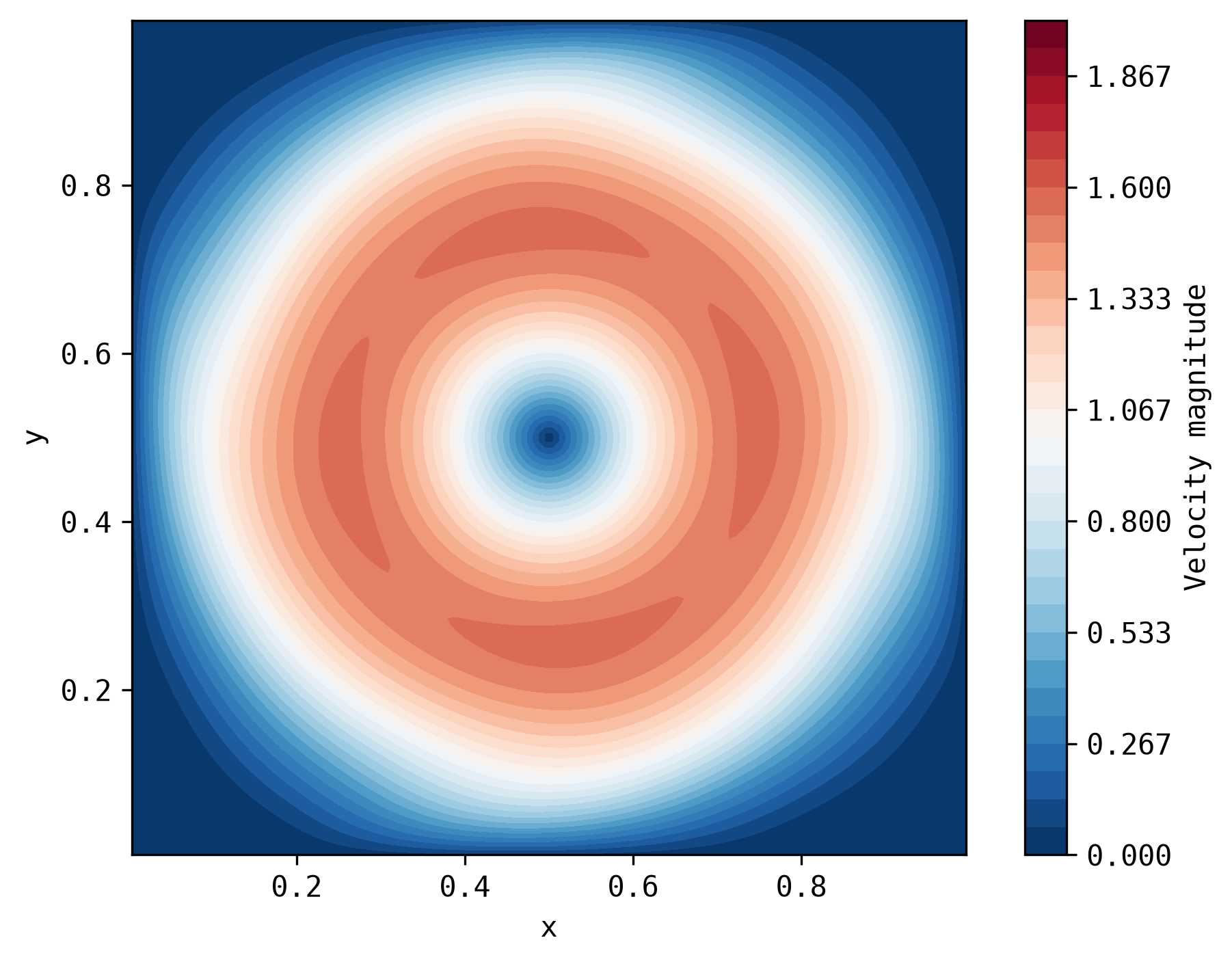}
	\end{subfigure}
	\begin{subfigure}[t]{0.3\textwidth}
		\centering
		\includegraphics[width=\linewidth]{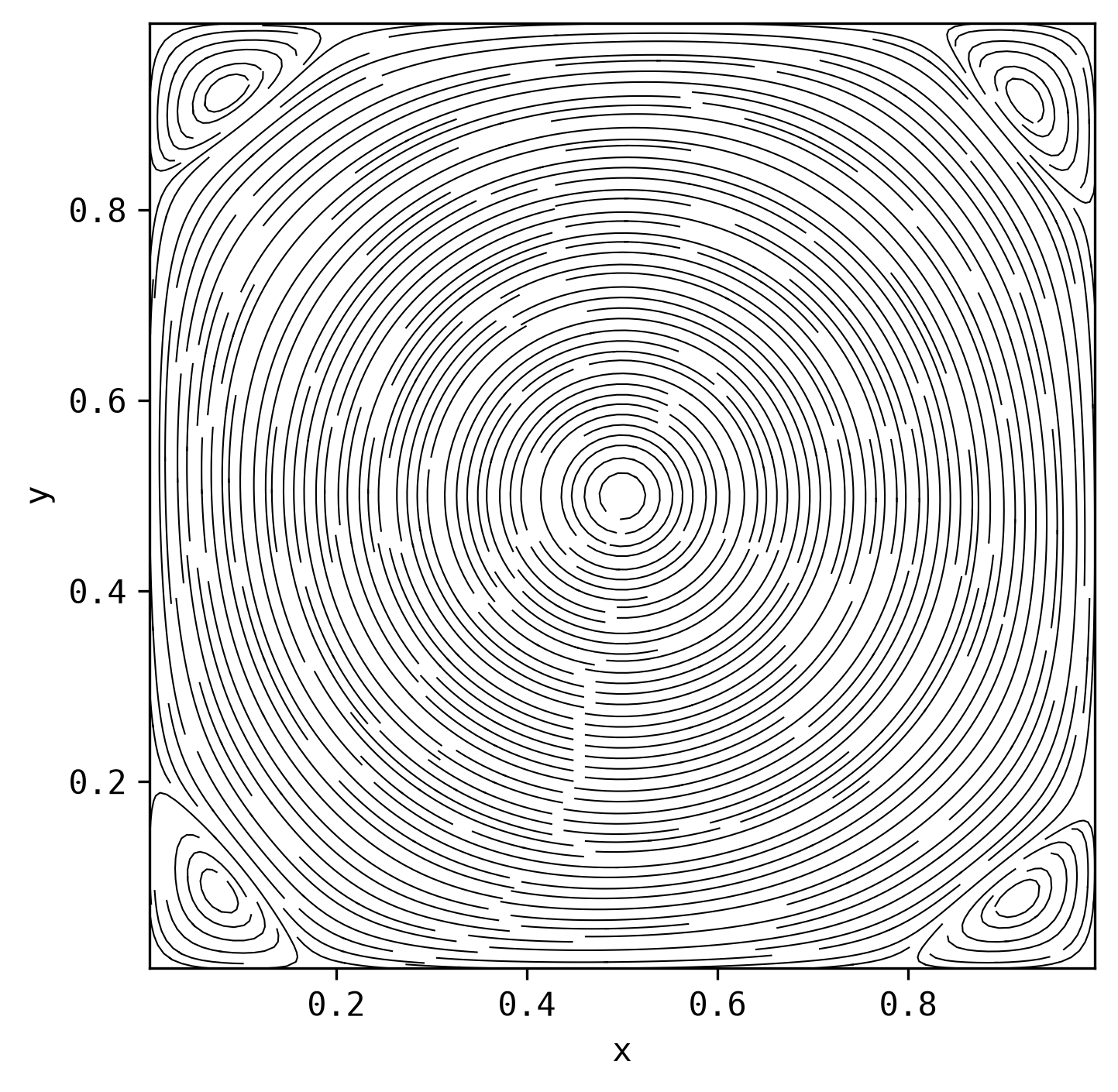}
	\end{subfigure}
	
	\caption{Evolution of the wall-bounded vortex. Rotation is counter-clockwise. The left column shows
		the velocity magnitude and the right column shows the corresponding
		streamlines at $t=0$, $0.1$, $0.2$, and $0.4~\mathrm{s}$,
		respectively. $128\times128$ uniform grid is used.}
	\label{fig:wbv_evolution}
\end{figure*}

The evolution of the wall-bounded vortex is illustrated in
Fig.~\ref{fig:wbv_evolution}. These images are generated using a $128\times128$ uniform grid and the second-order projection approach. Velocity-magnitude contours and streamlines
are shown at selected instants from $t=0$ to $0.4~\mathrm{s}$. A common
contour scale is used at all times to permit direct comparison of the
velocity magnitude and its decay, while the streamlines illustrate the
corresponding evolution of the vortex structure.

\subsection{Temporal convergence}

Temporal convergence is examined on a fixed $64\times64$ grid for both
the uniform mesh and the smoothly stretched mesh with $s_x=s_y=0.6$.
The grid size allows the temporal convergence trends to be identified
without reaching the positive-definiteness limits discussed in
Section~\ref{sec:verification_perot_positive_definiteness}.
The calculations are advanced to
$t_{\mathrm{final}}=0.1~\mathrm{s}$ using
\begin{equation}
	\Delta t
	=
	1.25\times10^{-3},\;
	6.25\times10^{-4},\;
	3.125\times10^{-4}\ {\rm s}.
\end{equation}
For each projection method, a numerical reference solution is computed
on the same spatial grid using
\begin{equation}
	\Delta t_{\mathrm{ref}}
	=
	1.953125\times10^{-5}\ {\rm s}.
\end{equation}

Temporal errors are evaluated using the measures and pressure-time
alignment procedure introduced in Section~\ref{sec:error_measures}.
Thus, velocity is compared at $t_{\mathrm{final}}$, while pressure is
compared at its corresponding midpoint time using the interpolated
fine-step reference solution. The numerical and reference pressure
fields use the same prescribed pressure-reference cell and reference
value.

Tables~\ref{tab:wbv_temporal_uniform} and
\ref{tab:wbv_temporal_stretched} report the temporal-convergence results.
According to these results, we observe:
\begin{enumerate}
	\item
	The Standard projection is approximately first-order accurate in time
	for velocity on both the uniform and stretched grids.
	
	\item
	The Perot projection is second-order accurate in time for velocity on
	both grids.
	
	\item
	The pressure obtained with the Standard projection is approximately
	first-order accurate in time. The observed pressure orders closely
	follow the corresponding velocity orders on both grids.
	
	\item
	The pressure obtained with the Perot projection approaches
	second-order temporal convergence. On the uniform grid, the observed
	orders are $1.9890$ and $1.9977$. On the stretched grid, the observed
	orders decrease from $2.1848$ to $2.0409$ as the time step is refined.
	
	\item
	Grid stretching does not change the asymptotic temporal behavior of
	either projection method.
\end{enumerate}

These results complement those of the advecting Taylor--Green vortex.
Under periodic boundary conditions, both projection methods exhibit
second-order temporal convergence in velocity. With non-periodic
no-slip boundaries, the present smooth wall-bounded problem reveals the
distinction between the two projection methods. The Standard projection
is first-order accurate in both velocity and pressure, whereas the
Perot projection is second-order accurate in both quantities when the
pressure is interpreted at its midpoint time level.

\begin{table}[!htbp]
	\centering
	\caption{Temporal convergence of the wall-bounded vortex on the
		$64\times64$ uniform grid. The final time is
		$t_{\mathrm{final}}=0.1~\mathrm{s}$. For each projection method,
		the reference solution is computed on the same grid using
		$\Delta t_{\mathrm{ref}}=1.953125\times10^{-5}~\mathrm{s}$.}
	\label{tab:wbv_temporal_uniform}
	
	\begin{tabular}{ccccc}
		\hline
		& \multicolumn{2}{c}{Standard} & \multicolumn{2}{c}{Perot} \\
		\cline{2-3}\cline{4-5}
		$\Delta t$ (s)
		& $E_{\mathbf{V}}$ & $r_{\mathbf{V}}$
		& $E_{\mathbf{V}}$ & $r_{\mathbf{V}}$ \\
		\hline
		$1.2500\times10^{-3}$
		& $1.3691\times10^{-4}$ & 1.0291
		& $8.8857\times10^{-6}$ & 2.0012 \\
		$6.2500\times10^{-4}$
		& $6.7086\times10^{-5}$ & 1.0500
		& $2.2195\times10^{-6}$ & 2.0043 \\
		$3.1250\times10^{-4}$
		& $3.2400\times10^{-5}$ & --
		& $5.5324\times10^{-7}$ & -- \\
		\hline
	\end{tabular}
	
	\vspace{0.8em}
	
	\begin{tabular}{ccccc}
		\hline
		& \multicolumn{2}{c}{Standard} & \multicolumn{2}{c}{Perot} \\
		\cline{2-3}\cline{4-5}
		$\Delta t$ (s)
		& $E_p$ & $r_p$
		& $E_p$ & $r_p$ \\
		\hline
		$1.2500\times10^{-3}$
		& $4.6717\times10^{-4}$ & 1.0278
		& $7.2452\times10^{-6}$ & 1.9890 \\
		$6.2500\times10^{-4}$
		& $2.2912\times10^{-4}$ & 1.0497
		& $1.8252\times10^{-6}$ & 1.9977 \\
		$3.1250\times10^{-4}$
		& $1.1068\times10^{-4}$ & --
		& $4.5702\times10^{-7}$ & -- \\
		\hline
	\end{tabular}
\end{table}

\begin{table}[!htbp]
	\centering
	\caption{Temporal convergence of the wall-bounded vortex on the
		$64\times64$ smoothly stretched grid with $s_x=s_y=0.6$.
		The final time is $t_{\mathrm{final}}=0.1~\mathrm{s}$.
		For each projection method, the reference solution is computed
		on the same grid using
		$\Delta t_{\mathrm{ref}}=1.953125\times10^{-5}~\mathrm{s}$.}
	\label{tab:wbv_temporal_stretched}
	
	\begin{tabular}{ccccc}
		\hline
		& \multicolumn{2}{c}{Standard} & \multicolumn{2}{c}{Perot} \\
		\cline{2-3}\cline{4-5}
		$\Delta t$ (s)
		& $E_{\mathbf{V}}$ & $r_{\mathbf{V}}$
		& $E_{\mathbf{V}}$ & $r_{\mathbf{V}}$ \\
		\hline
		$1.2500\times10^{-3}$
		& $1.4255\times10^{-4}$ & 1.0277
		& $2.6668\times10^{-5}$ & 2.0025 \\
		$6.2500\times10^{-4}$
		& $6.9919\times10^{-5}$ & 1.0493
		& $6.6555\times10^{-6}$ & 2.0045 \\
		$3.1250\times10^{-4}$
		& $3.3785\times10^{-5}$ & --
		& $1.6587\times10^{-6}$ & -- \\
		\hline
	\end{tabular}
	
	\vspace{0.8em}
	
	\begin{tabular}{ccccc}
		\hline
		& \multicolumn{2}{c}{Standard} & \multicolumn{2}{c}{Perot} \\
		\cline{2-3}\cline{4-5}
		$\Delta t$ (s)
		& $E_p$ & $r_p$
		& $E_p$ & $r_p$ \\
		\hline
		$1.2500\times10^{-3}$
		& $4.6739\times10^{-4}$ & 1.0265
		& $2.2965\times10^{-5}$ & 2.1848 \\
		$6.2500\times10^{-4}$
		& $2.2945\times10^{-4}$ & 1.0488
		& $5.0512\times10^{-6}$ & 2.0409 \\
		$3.1250\times10^{-4}$
		& $1.1091\times10^{-4}$ & --
		& $1.2275\times10^{-6}$ & -- \\
		\hline
	\end{tabular}
\end{table}

\section{Lid-driven cavity}
\label{sec:lid_driven_cavity}

\subsection{Problem definition}

Preceding tests highlighted the spatial and temporal orders of accuracy. The lid-driven cavity problem is now considered as a classical benchmark for
comparing the numerical solution against established reference data. In addition to nonlinear convection and
viscous diffusion, the problem exercises the implementation of
no-slip and moving-wall boundary conditions on the staggered Cartesian
grid. The numerical results are compared with the benchmark data of \cite{ghia1982high}.

The flow is considered in the square domain
\begin{equation}
	0 \leq x \leq L,
	\qquad
	0 \leq y \leq L,
\end{equation}
with $L=1$. The upper wall translates in the positive $x$ direction
with a constant velocity $U_{\mathrm{lid}}$, while the remaining three
walls are stationary. Thus,
\begin{align}
	u &= U_{\mathrm{lid}}, \qquad v=0,
	&& y=L,\\
	u &= 0, \qquad v=0,
	&& y=0,\;x=0,\;x=L .
\end{align}
In the present calculations,
\begin{equation}
	U_{\mathrm{lid}}=1,
	\qquad
	\rho=1,
\end{equation}
and the Reynolds number is defined as
\begin{equation}
	Re
	=
	\frac{U_{\mathrm{lid}}L}{\nu}.
\end{equation}

The benchmark comparison is based on the horizontal velocity along the
vertical cavity centerline, $u(x=L/2,y)$, and the vertical velocity
along the horizontal centerline, $v(x,y=L/2)$. 

\subsection{Comparison with reference data}

The steady cavity solution was assessed and will be reported against the reference benchmark data at $Re=1000$. $Re=100$ and 400 also showed comparable agreement however they will not be reported here.

The horizontal velocity along the vertical
centerline and the vertical velocity along the horizontal
centerline were compared with the corresponding tabulated
values of \cite{ghia1982high}. The numerical velocities were evaluated at the
sampling coordinates of this reference by interpolation from their native staggered
locations.

The comparison was performed using the AB2--Crank--Nicolson time
integration scheme. Calculations were continued until a steady state
was obtained, after which the centerline velocity profiles were
extracted for comparison with the reference data.

\begin{table}[!htbp]
	\centering
	\caption{Comparison of the normalized centerline velocities with the
		benchmark data of Ghia et al.~\citep{ghia1982high} for $Re=1000$.
		The Motif calculations use a $128\times128$ uniform grid,
		$\Delta t=0.001~\mathrm{s}$, and $t=50~\mathrm{s}$.}
	\label{tab:cavity_ghia_re1000}
	
	\begin{tabular}{cccc@{\qquad}cccc}
		\hline
		\multicolumn{4}{c}{$u/U_{\mathrm{lid}}$ at $x/L=0.5$}
		&
		\multicolumn{4}{c}{$v/U_{\mathrm{lid}}$ at $y/L=0.5$}
		\\
		\cline{1-4}\cline{5-8}
		$y/L$ & Ghia & Standard & Perot
		&
		$x/L$ & Ghia & Standard & Perot
		\\
		\hline
		1.0000 &  1.00000 &  1.00000 &  1.00000 &
		1.0000 &  0.00000 &  0.00000 &  0.00000 \\
		0.9766 &  0.65928 &  0.66136 &  0.66134 &
		0.9688 & -0.21388 & -0.22313 & -0.22326 \\
		0.9688 &  0.57492 &  0.57723 &  0.57723 &
		0.9609 & -0.27669 & -0.28806 & -0.28820 \\
		0.9609 &  0.51117 &  0.51273 &  0.51275 &
		0.9531 & -0.33714 & -0.34895 & -0.34910 \\
		0.9531 &  0.46604 &  0.46749 &  0.46752 &
		0.9453 & -0.39188 & -0.40329 & -0.40344 \\
		0.8516 &  0.33304 &  0.33159 &  0.33162 &
		0.9063 & -0.51550 & -0.51763 & -0.51768 \\
		0.7344 &  0.18719 &  0.18545 &  0.18547 &
		0.8594 & -0.42665 & -0.42219 & -0.42218 \\
		0.6172 &  0.05702 &  0.05573 &  0.05575 &
		0.8047 & -0.31966 & -0.31631 & -0.31631 \\
		0.5000 & -0.06080 & -0.06136 & -0.06134 &
		0.5000 &  0.02526 &  0.02573 &  0.02574 \\
		0.4531 & -0.10648 & -0.10674 & -0.10672 &
		0.2344 &  0.32235 &  0.32123 &  0.32125 \\
		0.2813 & -0.27805 & -0.27721 & -0.27719 &
		0.2266 &  0.33075 &  0.32963 &  0.32966 \\
		0.1719 & -0.38289 & -0.38179 & -0.38183 &
		0.1563 &  0.37095 &  0.37038 &  0.37041 \\
		0.1016 & -0.29730 & -0.29446 & -0.29456 &
		0.0938 &  0.32627 &  0.32669 &  0.32672 \\
		0.0703 & -0.22220 & -0.21897 & -0.21907 &
		0.0781 &  0.30353 &  0.30393 &  0.30395 \\
		0.0625 & -0.20196 & -0.19889 & -0.19899 &
		0.0703 &  0.29012 &  0.29049 &  0.29050 \\
		0.0547 & -0.18109 & -0.17832 & -0.17841 &
		0.0625 &  0.27485 &  0.27513 &  0.27514 \\
		0.0000 &  0.00000 &  0.00000 &  0.00000 &
		0.0000 &  0.00000 &  0.00000 &  0.00000 \\
		\hline
	\end{tabular}
\end{table}

Table~\ref{tab:cavity_ghia_re1000}
shows good agreement between the Motif solutions and the benchmark
centerline velocities of the reference. The variations of both velocity components, including the steep velocity gradients near
the cavity boundaries, are reproduced
by the $128\times128$ calculations. The standard and Perot projections produce nearly indistinguishable centerline velocities.
These results provide an independent verification of the spatially
discretized momentum equations and the Cartesian wall-boundary
treatment under strongly recirculating flow conditions.

\begin{figure}[!htbp]
	\centering
	
	\begin{subfigure}[t]{0.32\textwidth}
		\centering
		\includegraphics[width=\linewidth]{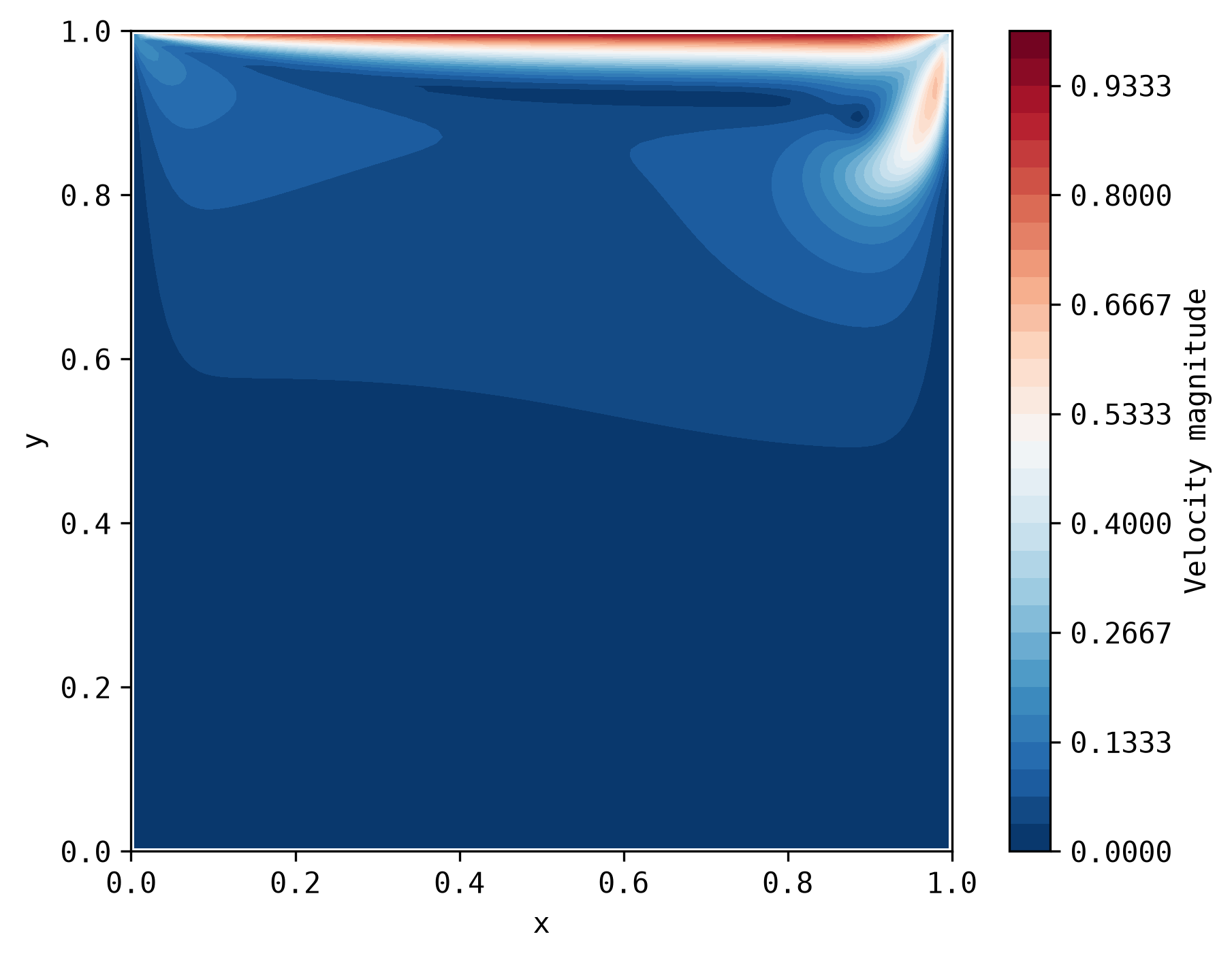}
	\end{subfigure}
	\begin{subfigure}[t]{0.26\textwidth}
		\centering
		\includegraphics[width=\linewidth]{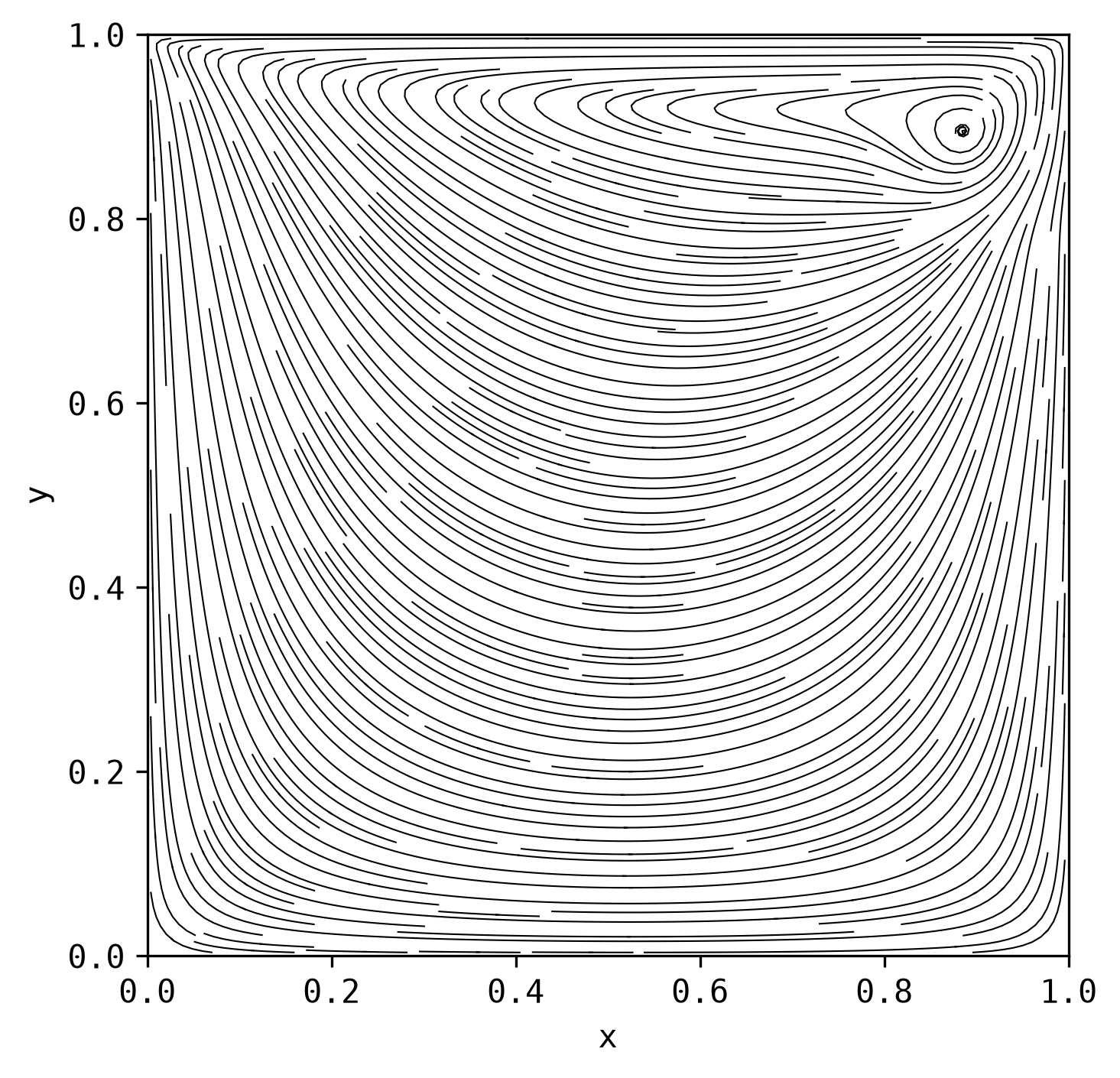}
	\end{subfigure}	
	\begin{subfigure}[t]{0.32\textwidth}
		\centering
		\includegraphics[width=\linewidth]{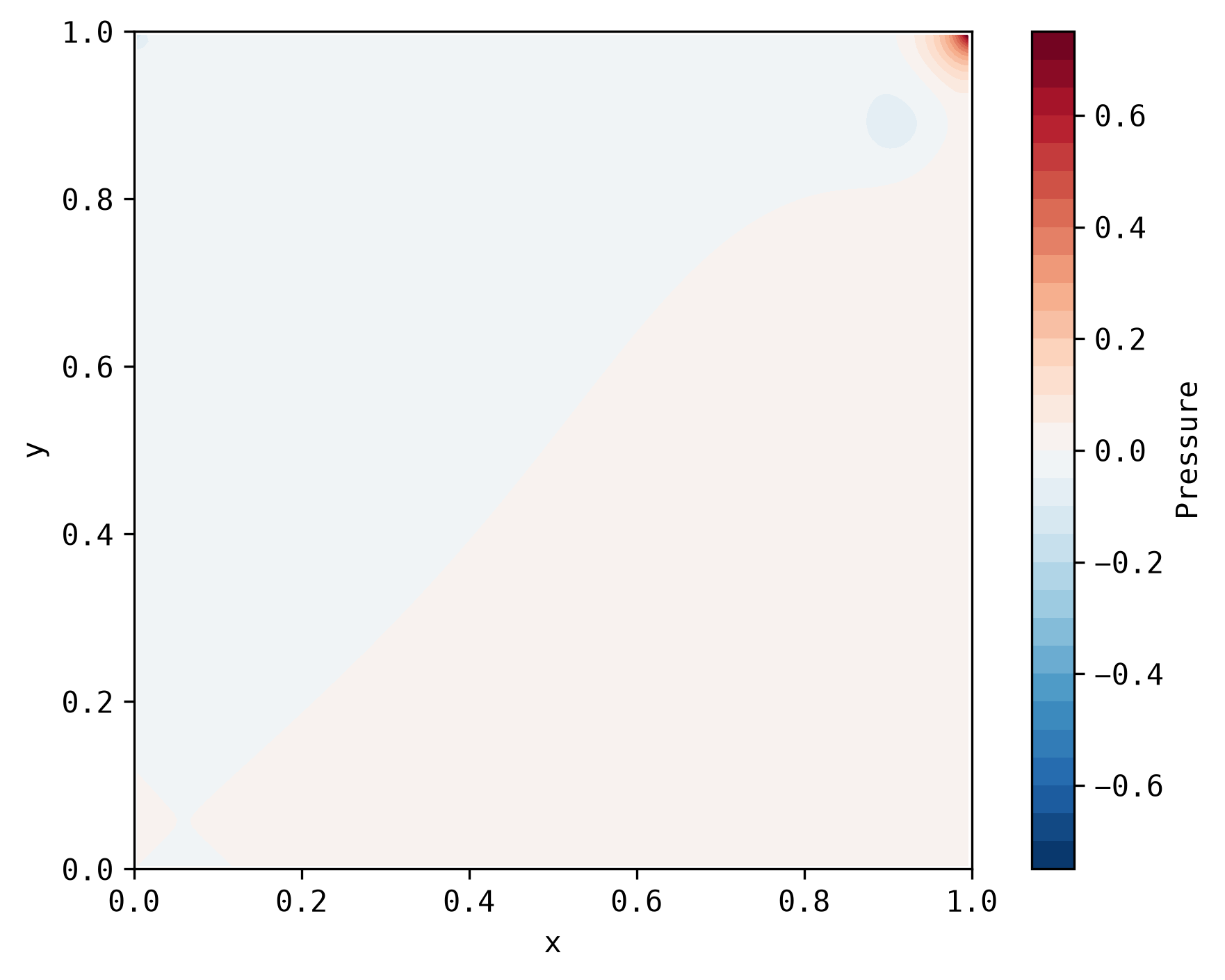}
	\end{subfigure}
	
	
	\begin{subfigure}[t]{0.32\textwidth}
		\centering
		\includegraphics[width=\linewidth]{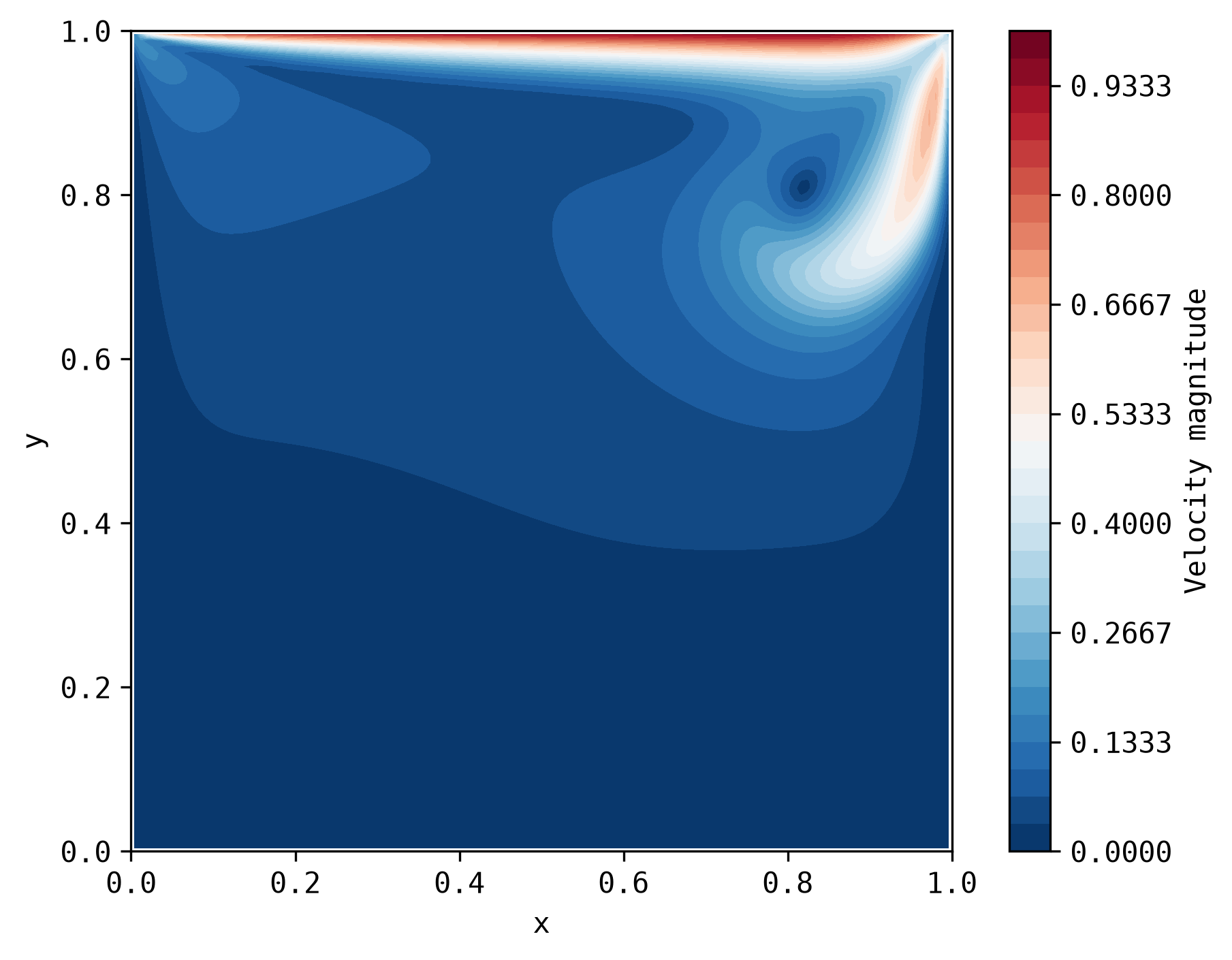}
	\end{subfigure}
	\begin{subfigure}[t]{0.26\textwidth}
		\centering
		\includegraphics[width=\linewidth]{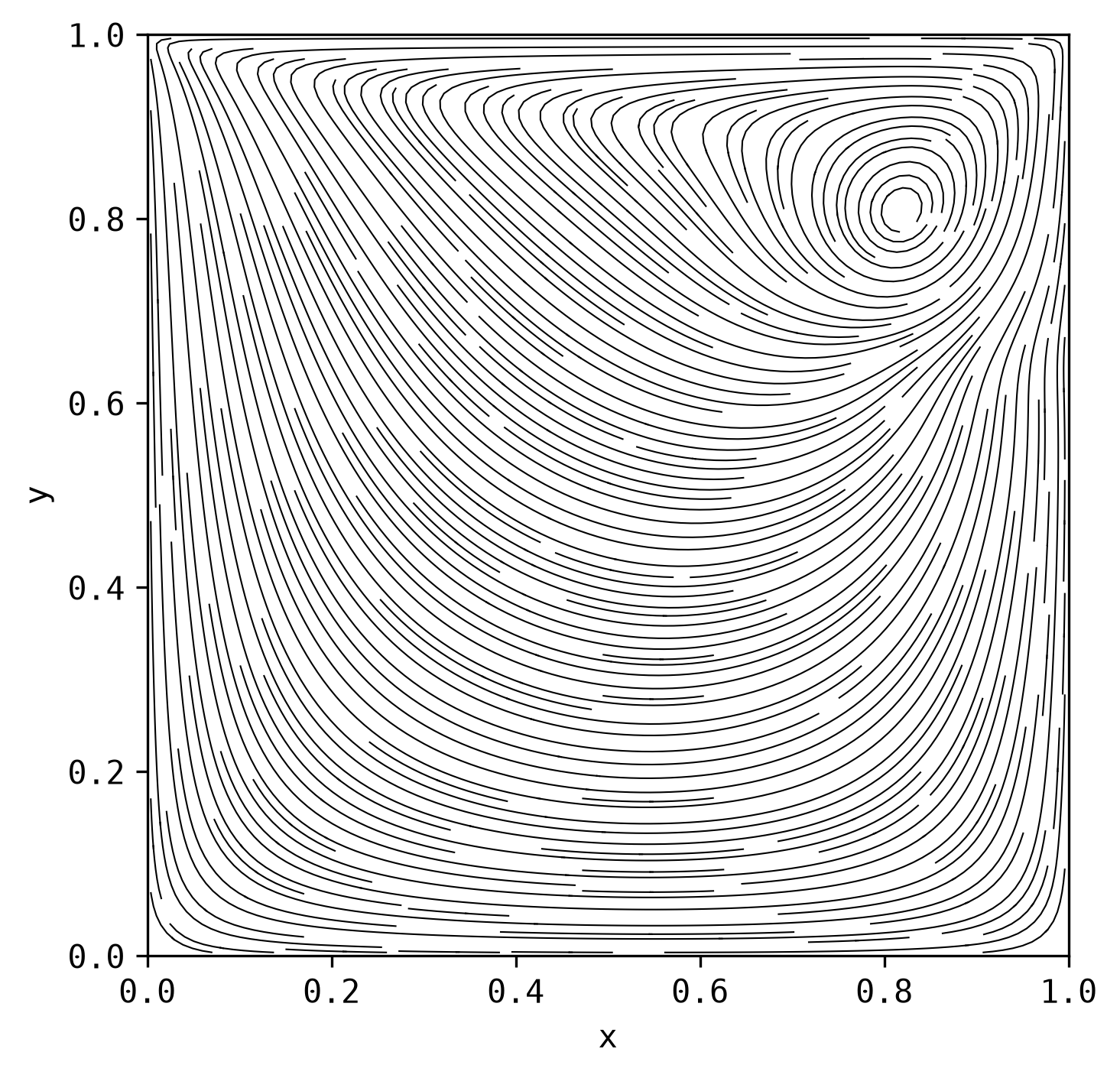}
	\end{subfigure}	
	\begin{subfigure}[t]{0.32\textwidth}
		\centering
		\includegraphics[width=\linewidth]{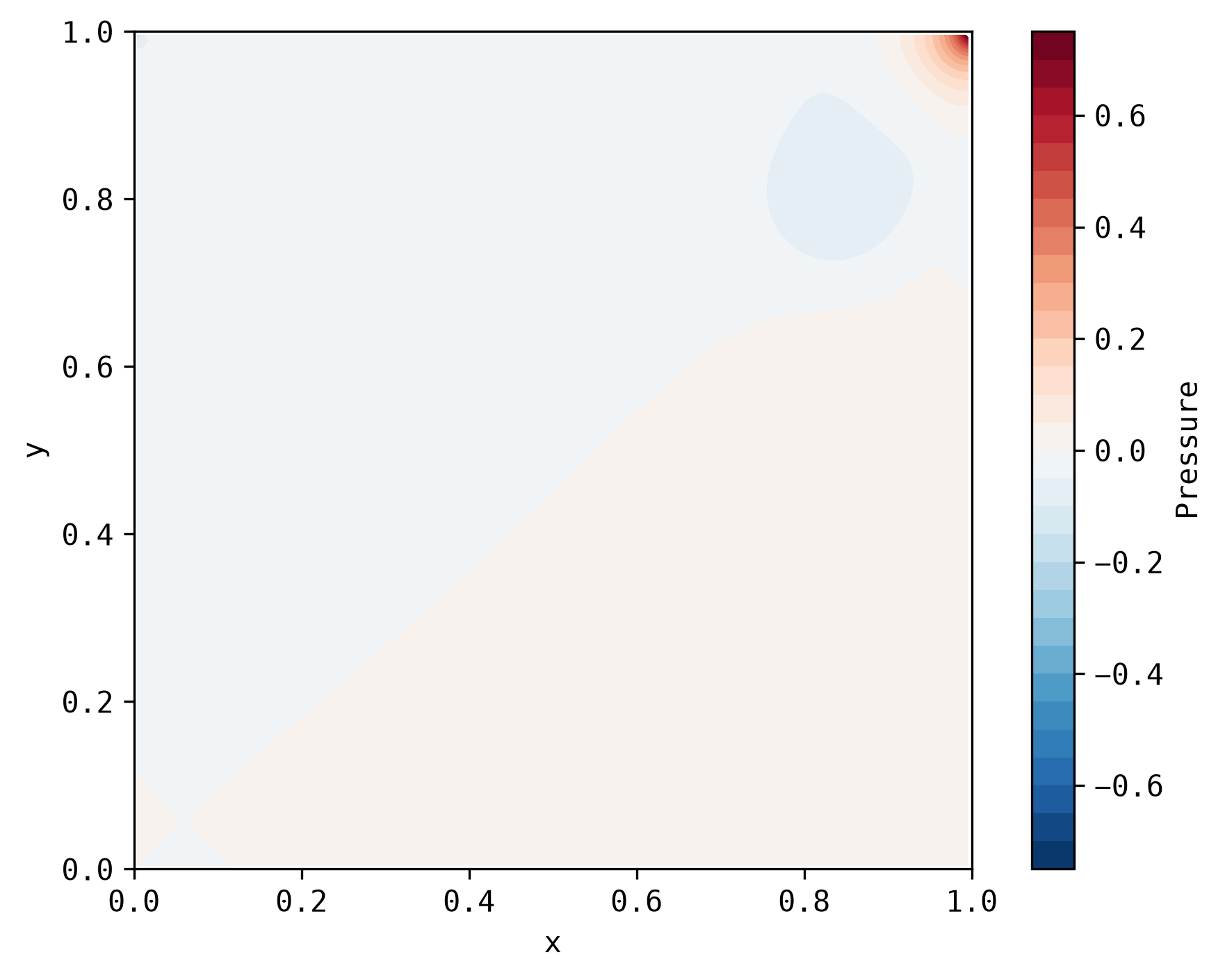}
	\end{subfigure}
	
	
	\begin{subfigure}[t]{0.32\textwidth}
		\centering
		\includegraphics[width=\linewidth]{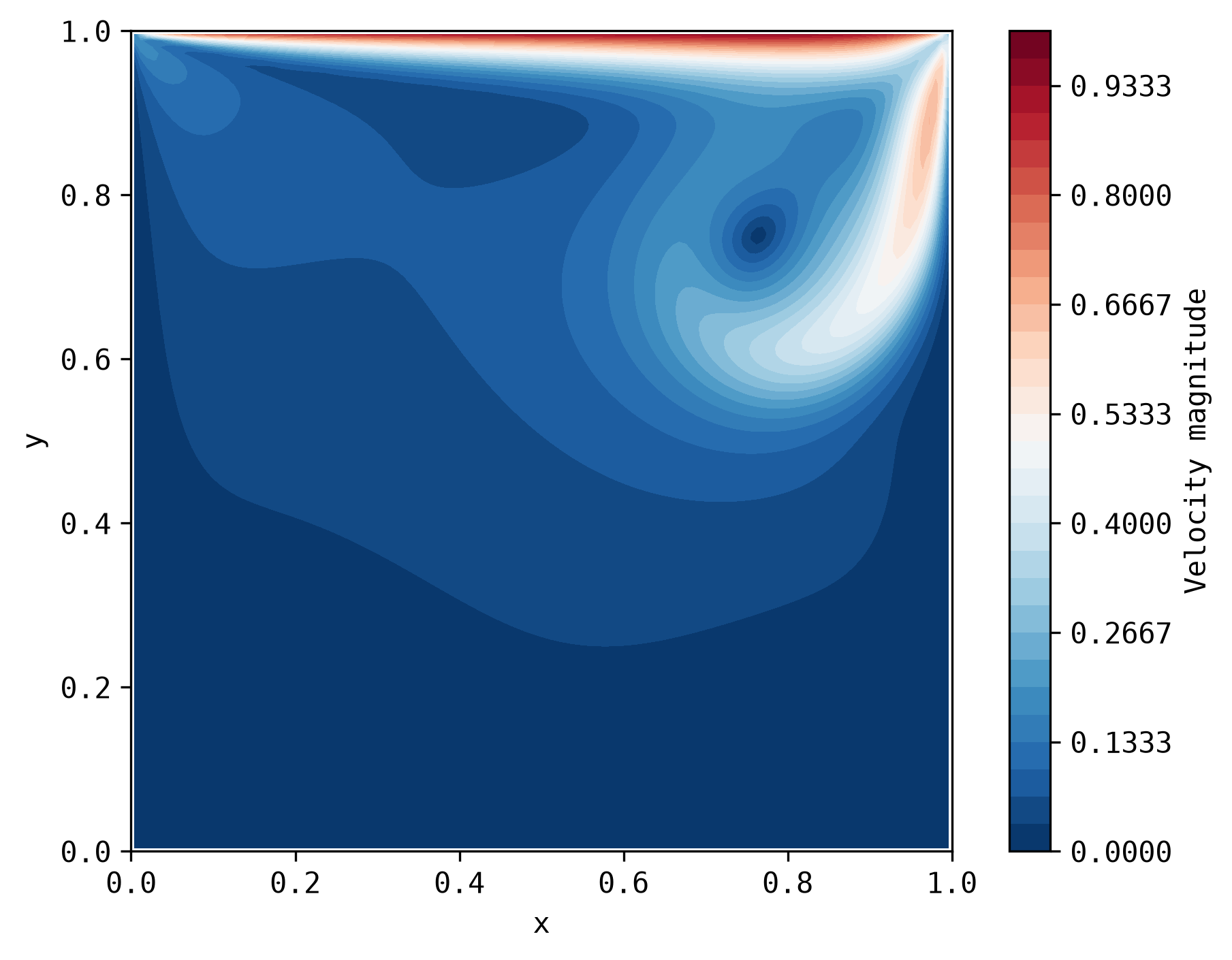}
	\end{subfigure}
	\begin{subfigure}[t]{0.26\textwidth}
		\centering
		\includegraphics[width=\linewidth]{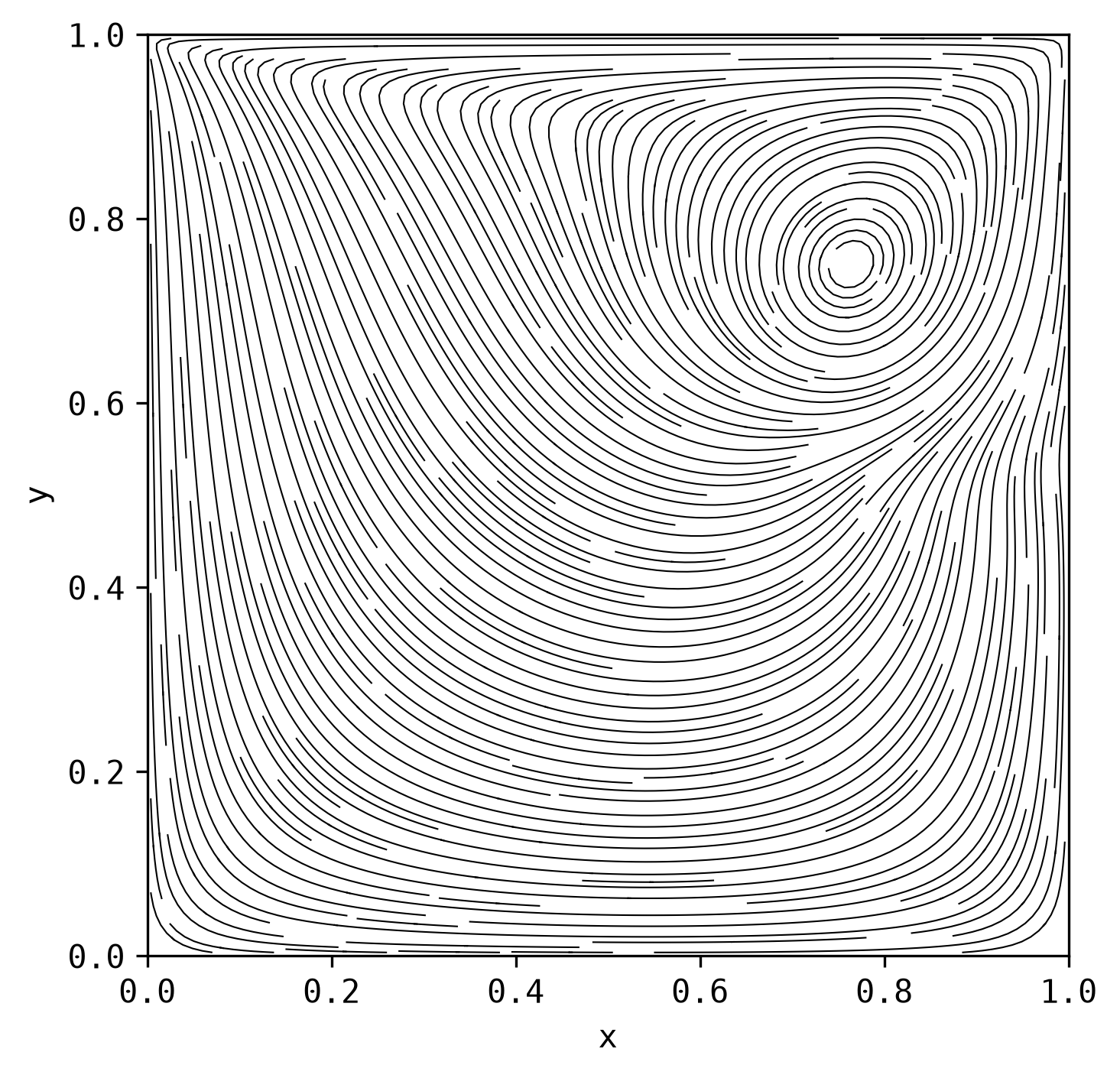}
	\end{subfigure}	
	\begin{subfigure}[t]{0.32\textwidth}
		\centering
		\includegraphics[width=\linewidth]{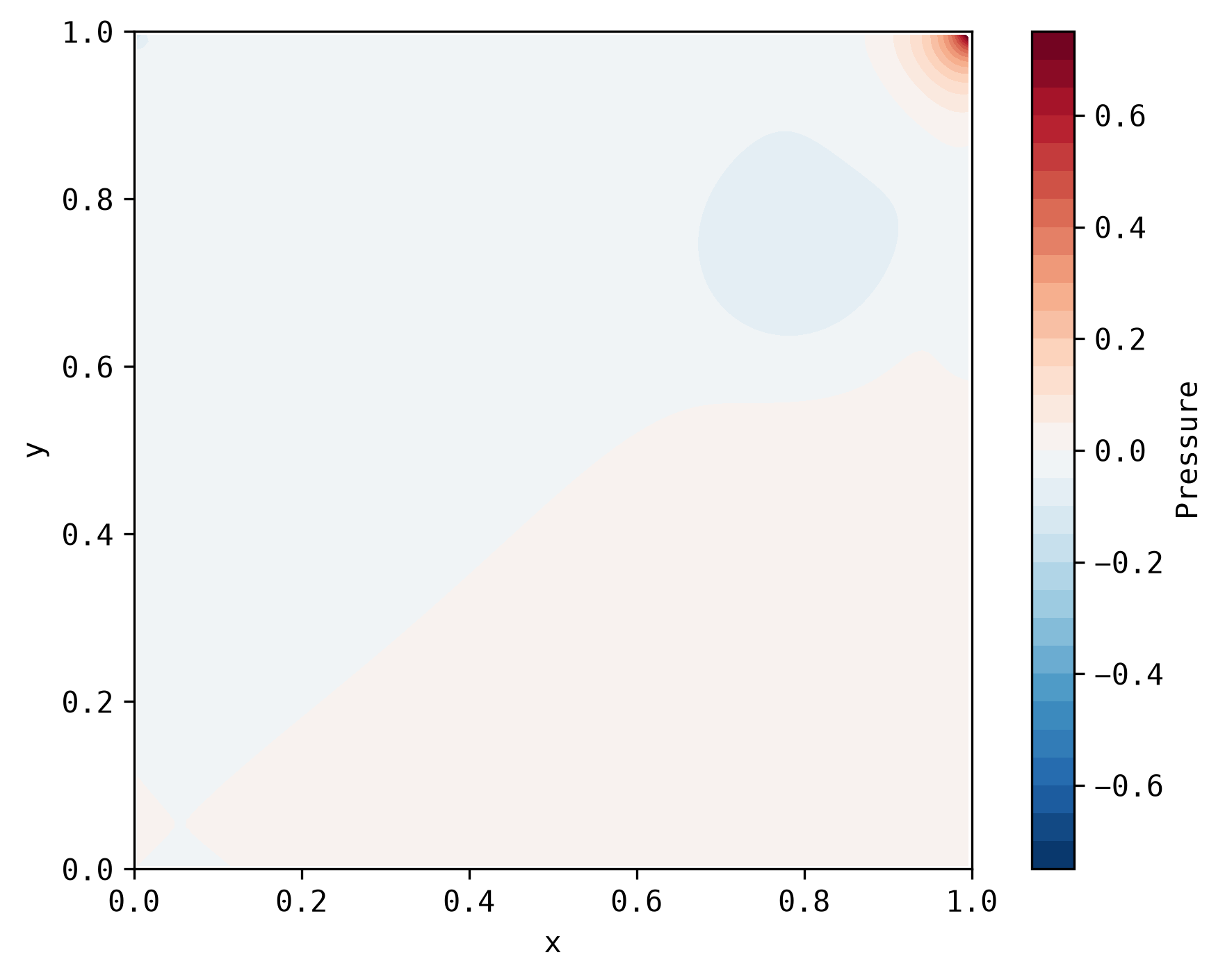}
	\end{subfigure}
	
	
	\begin{subfigure}[t]{0.32\textwidth}
		\centering
		\includegraphics[width=\linewidth]{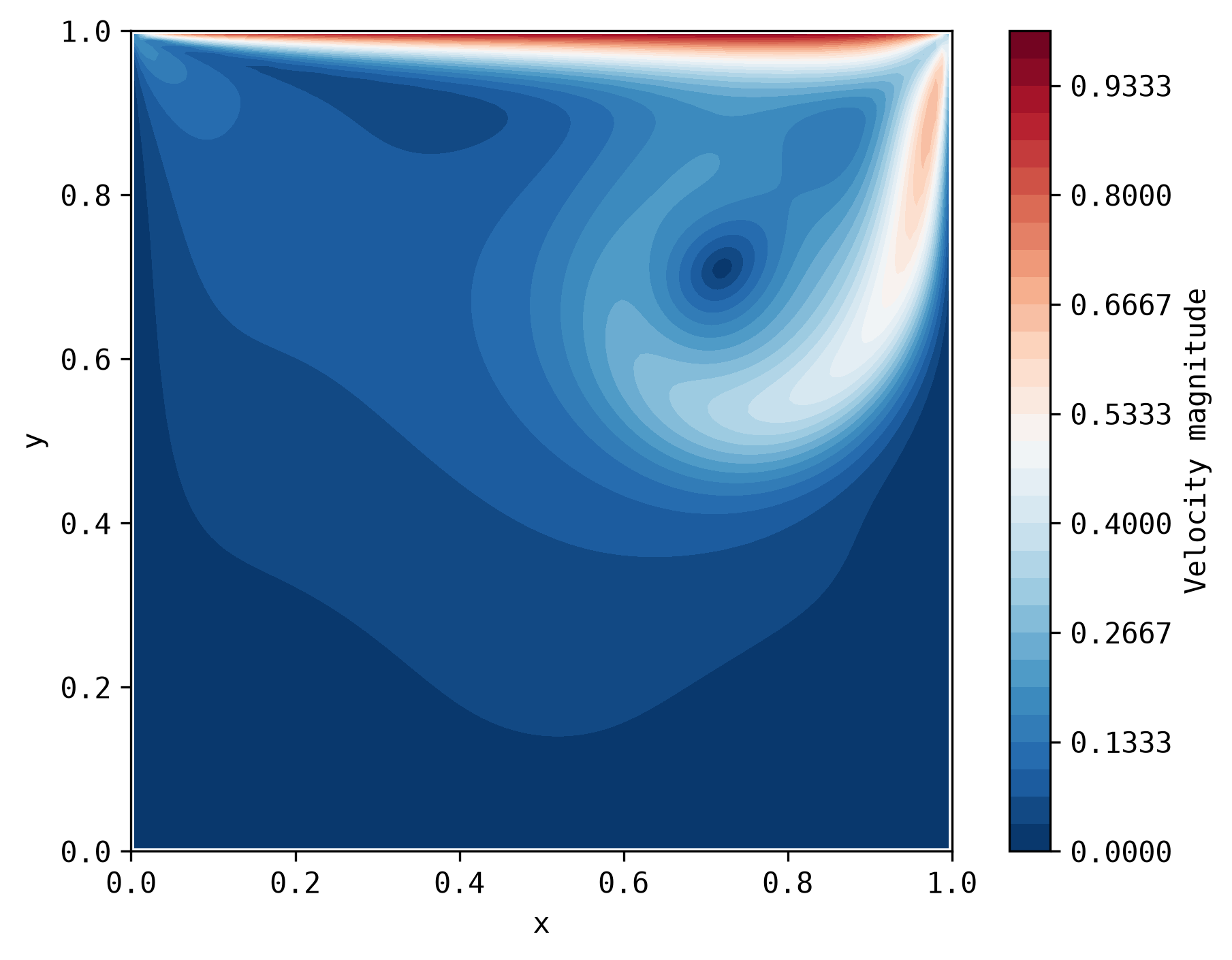}
	\end{subfigure}
	\begin{subfigure}[t]{0.26\textwidth}
		\centering
		\includegraphics[width=\linewidth]{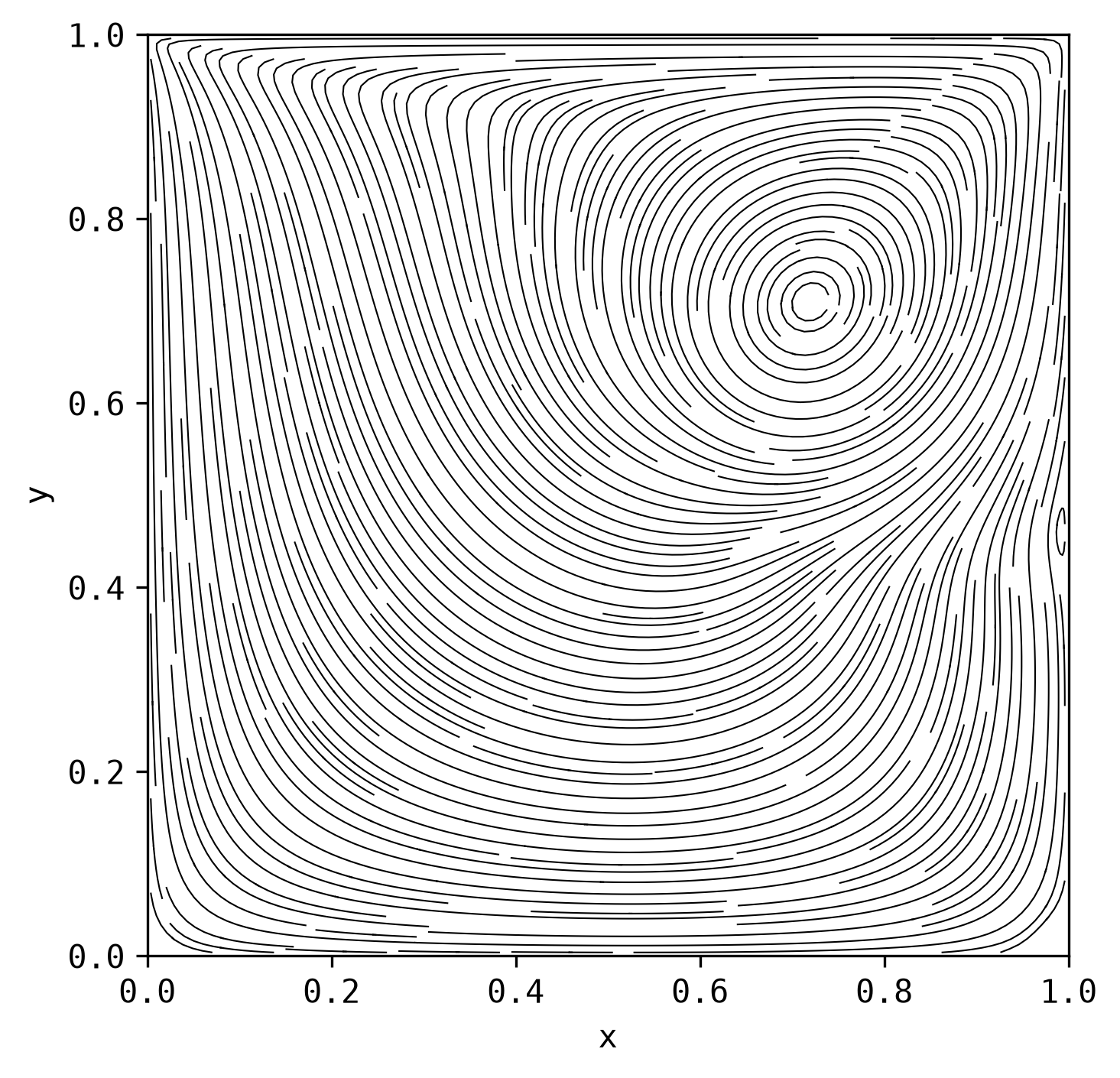}
	\end{subfigure}	
	\begin{subfigure}[t]{0.32\textwidth}
		\centering
		\includegraphics[width=\linewidth]{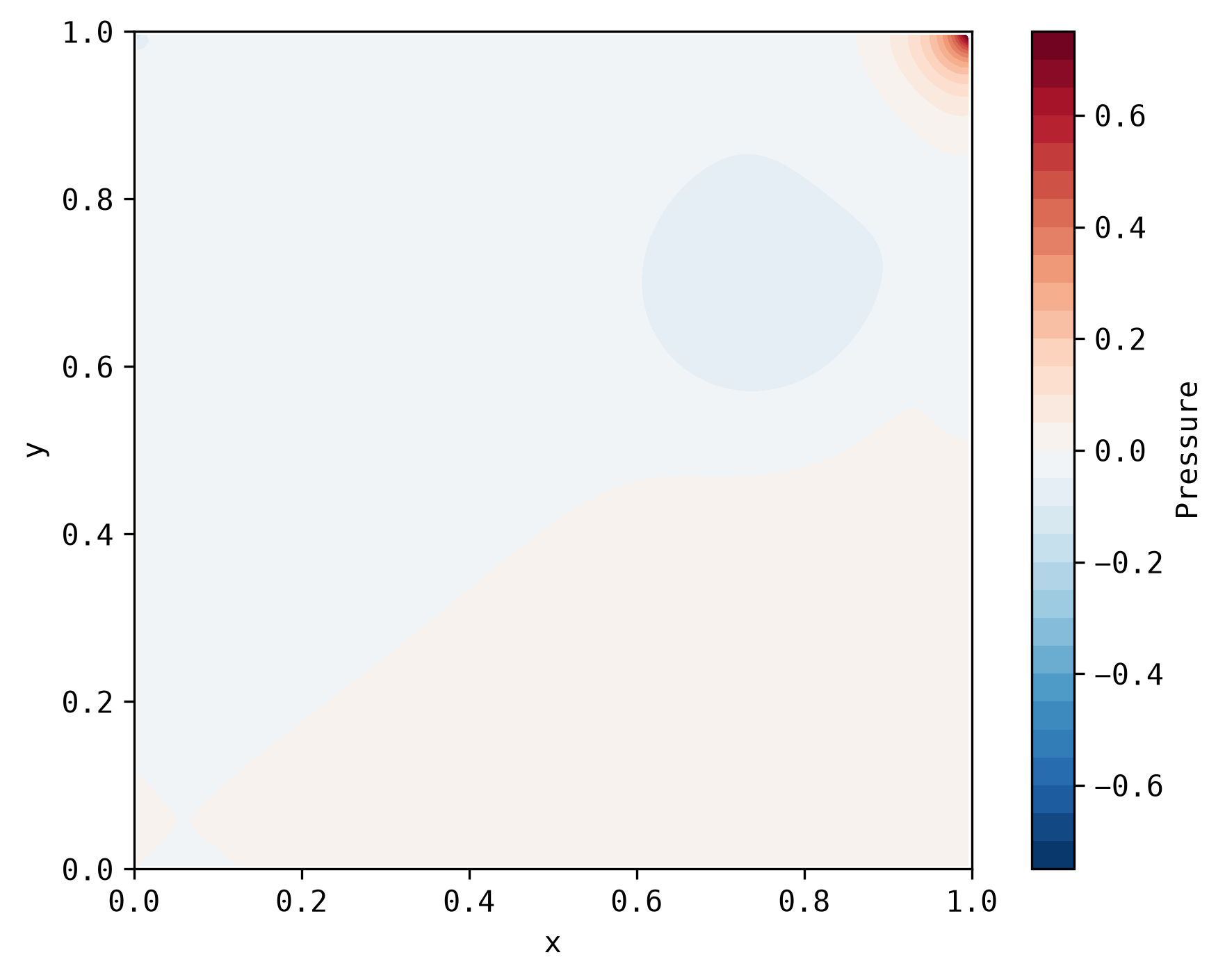}
	\end{subfigure}
	
	\caption{Evolution of the lid-driven cavity flow at $Re=1000$.
		The rows correspond to $t=1$, $2$, $3$, and $4~\mathrm{s}$.
		The columns show velocity magnitude, streamlines and pressure,
		respectively. $128\times128$ uniform grid is used.}
	\label{fig:cavity_evolution}
\end{figure}

The evolution of the cavity flow at $Re=1000$ over the first
$4~\mathrm{s}$ is illustrated in Fig.~\ref{fig:cavity_evolution}.
The primary recirculation progressively develops across the cavity within this interval.

\section{Plane channel flow}
\label{sec:channel_flow}

\subsection{Problem definition}

Developing laminar flow between two parallel stationary plates is
considered to assess the treatment of inlet, outlet, and no-slip wall
boundary conditions. The computational domain has length $L$ and
height $2h$. In terms of the wall-normal coordinate centered between
the plates, the walls are located at
\begin{equation}
	y=-h,
	\qquad
	y=+h .
\end{equation}
A uniform streamwise velocity is prescribed at the inlet,
\begin{equation}
	u(0,y,t)=U_0,
	\qquad
	v(0,y,t)=0,
\end{equation}
while the outlet pressure is prescribed as
\begin{equation}
	p(L,y,t)=p_{\mathrm{out}} .
\end{equation}
No-slip conditions are imposed at the upper and lower walls.

Sufficiently far downstream, the solution approaches fully developed
plane Poiseuille flow,
\begin{equation}
	u=u(y),
	\qquad
	v=0,
\end{equation}
for which the streamwise momentum equation reduces to
\begin{equation}
	\mu\frac{\mathrm{d}^2u}{\mathrm{d}y^2}
	=
	\frac{\mathrm{d}p}{\mathrm{d}x}.
\end{equation}
For a prescribed mean velocity $U_0$, the analytical velocity profile
is
\begin{equation}
	u(y)
	=
	\frac{3}{2}U_0
	\left[
	1-\left(\frac{y}{h}\right)^2
	\right],
\end{equation}
and the corresponding pressure gradient is
\begin{equation}
	\frac{\mathrm{d}p}{\mathrm{d}x}
	=
	-\frac{3\mu U_0}{h^2}.
\end{equation}
The maximum velocity is therefore
\begin{equation}
	u_{\max}
	=
	\frac{3}{2}U_0 .
\end{equation}

The analytical shear-stress distribution follows directly from the
velocity gradient,
\begin{equation}
	\tau_{xy}(y)
	=
	\mu\frac{\mathrm{d}u}{\mathrm{d}y}
	=
	-\frac{3\mu U_0}{h^2}y .
\end{equation}
Consequently, the shear stresses at the upper and lower walls are
\begin{equation}
	\tau_{xy}\big|_{y=+h}
	=
	-\frac{3\mu U_0}{h},
	\qquad
	\tau_{xy}\big|_{y=-h}
	=
	+\frac{3\mu U_0}{h},
\end{equation}
and the wall-shear magnitude is
\begin{equation}
	\tau_w
	=
	\frac{3\mu U_0}{h}.
\end{equation}

The Reynolds number is defined using the channel height,
\begin{equation}
	Re
	=
	\frac{U_0(2h)}{\nu}.
\end{equation}

\subsection{Inlet--outlet boundary treatment}

At the pressure outlet, the projection is allowed to correct the
outlet-normal velocity. The corrected outlet velocity is therefore
retained after the projection rather than being overwritten by
reapplication of the zero-gradient condition used for the intermediate velocity.

The consistency of the inlet--outlet treatment is assessed through
global volume conservation. Since the staggered $u$ velocities are located
directly on the inlet and outlet faces, the discrete volume flow rate
per unit depth on the outlet is
\begin{equation}
	Q_{\mathrm{out}}
	=
	\sum_{j=1}^{N_y}
	u_{N_x+1/2,j}\,\Delta y_j .
\end{equation}
For an incompressible flow in the fixed channel,
\begin{equation}
	Q_{\mathrm{out}}
	=
	Q_{\mathrm{in}},
\end{equation}
while the prescribed uniform inlet gives $Q_{\mathrm{in}} = 2hU_0$.

The relative inlet--outlet flow-rate imbalance is therefore monitored
using
\begin{equation}
	\varepsilon_Q
	=
	\frac{
		\left|
		Q_{\mathrm{out}}-Q_{\mathrm{in}}
		\right|
	}{
		\left|
		Q_{\mathrm{in}}
		\right|
	}.
	\label{eq:channel_mass_imbalance}
\end{equation}
The maximum discrete divergence-flux imbalance is monitored
by default in Motif.

\subsection{Comparison with the fully developed solution}

The developed-flow solution was assessed using the streamwise velocity
profile at $x/H=10$, the streamwise pressure gradient, and the wall
shear. The pressure gradient was obtained from a linear least-squares
fit to the cross-section-averaged pressure over
$8\leq x/H\leq9.5$ in order to avoid direct influence of the prescribed
outlet pressure.

The numerical results in Tables \ref{tab:channel_profile} and \ref{tab:channel_pressure_shear} show close agreement with the fully developed
plane-Poiseuille solution for both projection methods and both grid
types. At $x/H=10$, the computed velocity profiles closely reproduce
the analytical parabolic distribution, including the centerline
maximum and the no-slip behavior at the walls. The pressure gradients
obtained from the fully developed interior region are also close to
the analytical value, as are the wall-shear magnitudes evaluated at
$x/H=10$. The small differences between the standard and Perot
projections are comparable to the remaining spatial-discretization
errors. These results additionally verify
the treatment of the inlet, outlet, and no-slip wall boundary
conditions on both uniform and smoothly stretched grids.

\begin{table}[!htbp]
	\centering
	\caption{Comparison of the outlet streamwise velocity profile with
		the analytical plane-Poiseuille solution at $x/H=10$ for $Re=100$.
		The calculations use a $64\times32$ grid,
		$\Delta t=0.005~\mathrm{s}$, and $t=8~\mathrm{s}$.
		The stretched grid uses $s_y=0.6$ and $s_x=0$.}
	\label{tab:channel_profile}
	
	\begin{tabular}{c c cc cc}
		\hline
		&
		&
		\multicolumn{2}{c}{Uniform grid}
		&
		\multicolumn{2}{c}{Stretched grid}
		\\
		\cline{3-4}\cline{5-6}
		$y/h$
		&
		Exact
		&
		Standard & Perot
		&
		Standard & Perot
		\\
		\hline
		$-1.00$ & 0.00000 & 0.00000 & 0.00000 & 0.00000 & 0.00000 \\
		$-0.75$ & 0.65625 & 0.65564 & 0.65553 & 0.65496 & 0.65486 \\
		$-0.50$ & 1.12500 & 1.12307 & 1.12311 & 1.12314 & 1.12318 \\
		$-0.25$ & 1.40625 & 1.40290 & 1.40302 & 1.40427 & 1.40439 \\
		$ 0.00$ & 1.50000 & 1.49602 & 1.49617 & 1.49444 & 1.49460 \\
		$ 0.25$ & 1.40625 & 1.40290 & 1.40302 & 1.40427 & 1.40439 \\
		$ 0.50$ & 1.12500 & 1.12307 & 1.12311 & 1.12314 & 1.12318 \\
		$ 0.75$ & 0.65625 & 0.65564 & 0.65553 & 0.65496 & 0.65486 \\
		$ 1.00$ & 0.00000 & 0.00000 & 0.00000 & 0.00000 & 0.00000 \\
		\hline
	\end{tabular}
\end{table}

\begin{table}[!htbp]
	\centering
	\caption{Comparison of the streamwise pressure gradient, wall-shear
		magnitude, and inlet--outlet volume-flow imbalance with the
		analytical plane-Poiseuille solution for $Re=100$. The pressure
		gradient is obtained from a linear least-squares fit to the
		cross-section-averaged pressure over $8\leq x/H\leq9.5$, while
		the wall shear is evaluated at $x/H=10$. The calculations use a
		$64\times32$ grid, $\Delta t=0.005~\mathrm{s}$, and
		$t=8~\mathrm{s}$. The stretched grid uses $s_y=0.6$ and $s_x=0$.}
	\label{tab:channel_pressure_shear}
	
	\begin{tabular}{ccccc}
		\hline
		Grid & Projection
		& $\mathrm{d}p/\mathrm{d}x$
		& $|\tau_w|$
		& $\varepsilon_Q$ \\
		\hline
		Uniform
		& Standard
		& $-0.119953$
		& $0.060123$
		& $2.00\times10^{-15}$ \\
		
		Uniform
		& Perot
		& $-0.119989$
		& $0.059939$
		& $8.10\times10^{-15}$ \\
		
		Stretched
		& Standard
		& $-0.119839$
		& $0.060343$
		& $1.33\times10^{-15}$ \\
		
		Stretched
		& Perot
		& $-0.119875$
		& $0.059738$
		& $5.33\times10^{-15}$ \\
		\hline
		Exact
		& --
		& $-0.120000$
		& $0.060000$
		& $0$ \\
		\hline
	\end{tabular}
\end{table}

\begin{figure}[!htbp]
	\centering
	
	\includegraphics[width=0.92\textwidth]
	{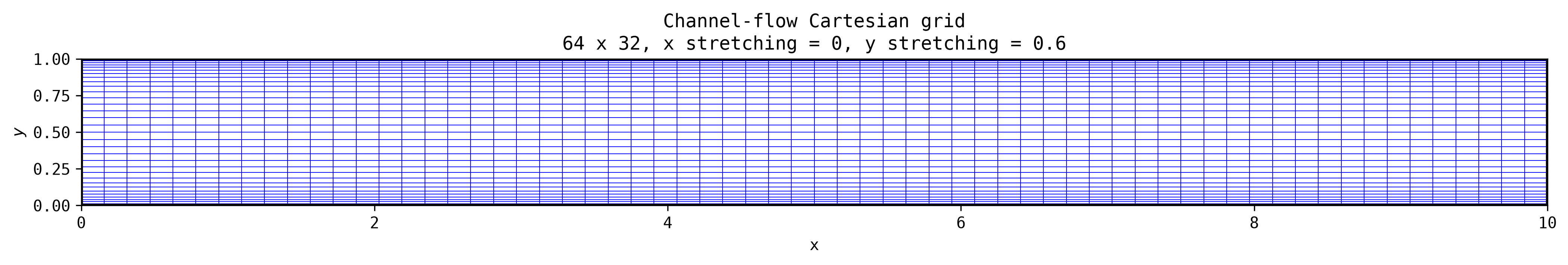}
	
	\vspace{0.5em}
	
	\includegraphics[width=0.95\textwidth]
	{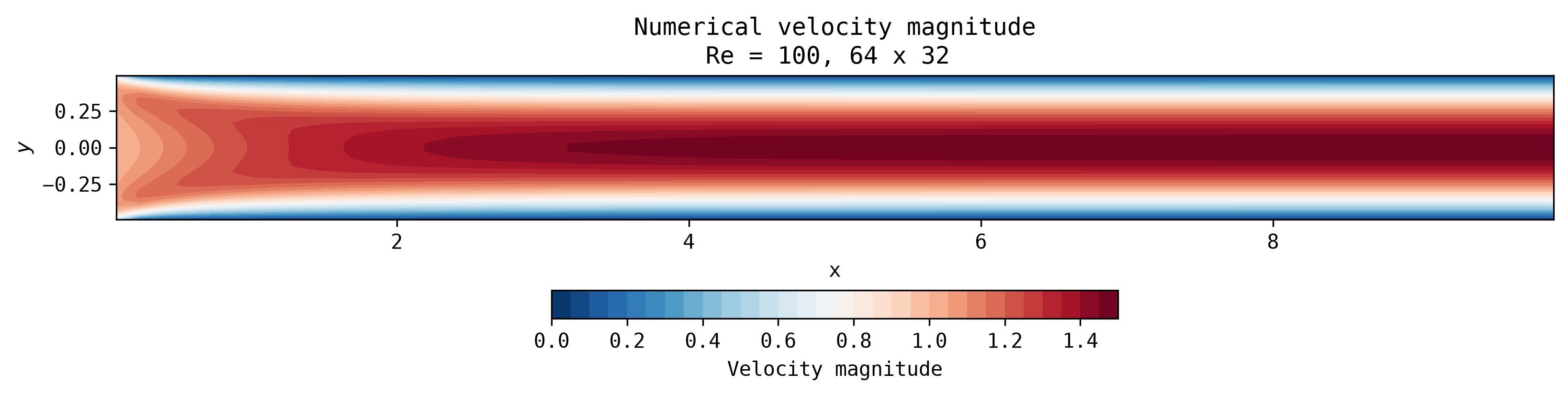}
	
	\vspace{0.5em}
	
	\includegraphics[width=0.95\textwidth]
	{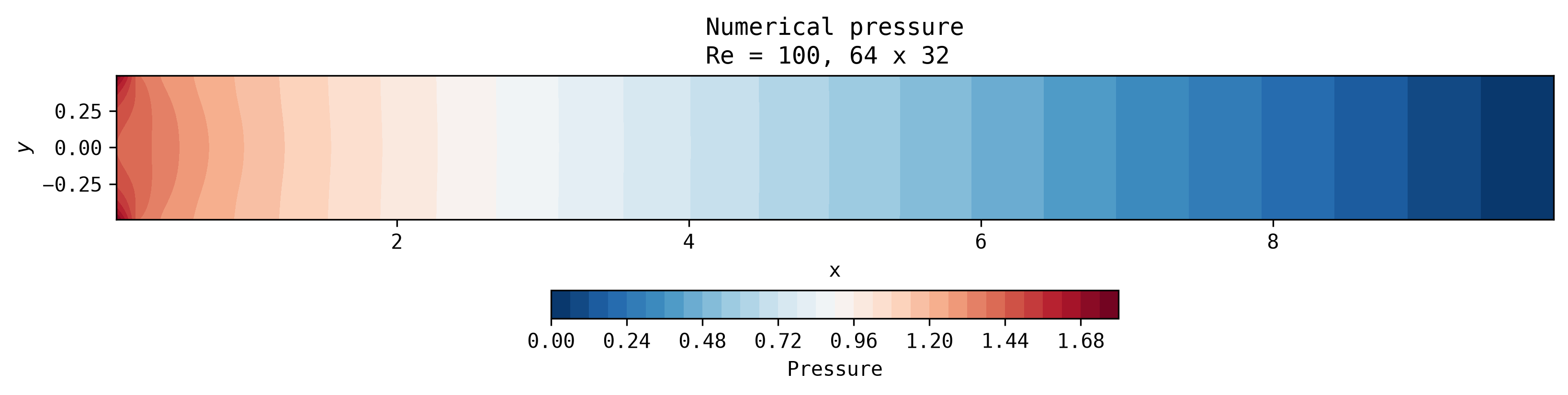}
	
	\caption{Plane channel flow at $Re=100$ and $t=8$ s: stretched computational
		grid with $s_y=0.6$ (top), velocity-magnitude contours (middle),
		and pressure contours (bottom).}
	\label{fig:channel_flow}
\end{figure}

\clearpage

\section{Further discussions}

The numerical experiments in the preceding sections expose several
properties of the projection formulation that deserve separate
consideration. These will be discussed here.

\subsection{Violation of discrete divergence on a nonuniform grid}
\label{ssec:violation}

The wall-bounded vortex considered in
Section~\ref{ssec:wall_bounded_vortex} is analytically divergence-free.
However, direct sampling of an analytical velocity field does not necessarily
preserve the corresponding discrete incompressibility constraint on a
nonuniform staggered grid.

To examine this effect, the analytical wall-bounded vortex solution was
sampled directly at the native staggered velocity locations. Uniform grids
were considered as a control, together with nonuniform grids generated using
$s_x=s_y=0.6$. Grid resolutions of $N=16$, 32, 64, 128, and 256 were used.
No time integration was performed.

The discrete-divergence defect was measured using the
pressure-control-volume-weighted $L_2$ norm,
\begin{equation}
	E_{\mathrm{div}}
	=
	\left[
	\frac{
		\displaystyle
		\sum_k V_k^{(p)}
		\left(\Dop\mathbf{V}\right)_k^2
	}{
		\displaystyle
		\sum_k V_k^{(p)}
	}
	\right]^{1/2},
\end{equation}
where $\left(\Dop\mathbf{V}\right)_k$ is the discrete divergence
evaluated in pressure control volume $k$ and $V_k^{(p)}$ is its volume, as usual.

A Standard projection was then applied directly to the sampled field,
\begin{equation}
	\Lop p
	=
	\frac{\rho}{\Delta t}
	\Dop\mathbf{V}^{*},
\end{equation}

\begin{equation}
	\mathbf{V}^{0}
	=
	\mathbf{V}^{*}
	-
	\frac{\Delta t}{\rho}
	\Gop p,
\end{equation}
where $\mathbf{V}^{*}$ denotes the analytically sampled initial velocity.
The operation therefore serves only to construct an initial field satisfying
the discrete incompressibility constraint.

\begin{table}[!htbp]
	\centering
	\caption{Discrete divergence of the analytically sampled vortex
		before and after the initial Standard projection. The nonuniform grids
		use $s_x=s_y=0.6$.}
	\label{tab:initial_divergence}
	\begin{tabular}{ccccc}
		\hline
		& & \multicolumn{2}{c}{$E_{\mathrm{div}}$} & \\
		\cline{3-4}
		Grid & $N$
		& Before projection
		& After projection
		& $E_{\mathrm{corr}}$ \\
		\hline
		Uniform
		& 16  & $3.799\times10^{-15}$ & $1.503\times10^{-15}$ & $1.351\times10^{-16}$ \\
		& 32  & $7.080\times10^{-15}$ & $3.046\times10^{-15}$ & $1.292\times10^{-16}$ \\
		& 64  & $1.535\times10^{-14}$ & $5.451\times10^{-15}$ & $1.318\times10^{-16}$ \\
		& 128 & $2.795\times10^{-14}$ & $1.124\times10^{-14}$ & $1.305\times10^{-16}$ \\
		& 256 & $4.660\times10^{-14}$ & $2.275\times10^{-14}$ & $1.300\times10^{-16}$ \\
		\hline
		Nonuniform
		& 16  & $1.578\times10^{-2}$ & $1.543\times10^{-15}$ & $9.446\times10^{-4}$ \\
		& 32  & $3.965\times10^{-3}$ & $3.735\times10^{-15}$ & $2.313\times10^{-4}$ \\
		& 64  & $9.924\times10^{-4}$ & $6.840\times10^{-15}$ & $5.753\times10^{-5}$ \\
		& 128 & $2.482\times10^{-4}$ & $1.383\times10^{-14}$ & $1.436\times10^{-5}$ \\
		& 256 & $6.205\times10^{-5}$ & $2.841\times10^{-14}$ & $3.590\times10^{-6}$ \\
		\hline
	\end{tabular}
\end{table}

As shown in Table~\ref{tab:initial_divergence}, direct sampling on the
uniform grids preserves discrete continuity to numerical precision. On the
nonuniform grids, the sampled field has a finite divergence defect that
converges at second order with grid refinement. The initial projection reduces
this defect to numerical-zero levels, while the associated relative velocity
correction, $E_{\mathrm{corr}}$, also decreases at second order.

This initial discrete-divergence defect can mask the formal temporal order
of a time-integration scheme if the analytically sampled field is used
without projection. This behavior was encountered in the Taylor--Green
vortex calculations on nonuniform grids.

\subsection{Positive-definiteness limit}
\label{sec:verification_perot_positive_definiteness}

The second-order projection introduces the operator $\Bop_2$ as an
approximation to the inverse of the implicit momentum operator. Unlike the
standard projection, this approximation introduces an additional numerical
condition on the time step. In particular, $\Bop_2$ is positive definite only
when the time step is sufficiently small \citep{perot1993analysis}.

Positive-definiteness requires that the matrix representation of $\Bop_2$ is symmetric and that its
eigenvalues are positive \citep{Ferziger2020}. Equivalently, for any nonzero vector $\mathbf{X}$,
\begin{equation}
	\mathbf{X}^{T}\Bop_2\mathbf{X} > 0.
\end{equation}

As the time step is
increased, the smallest eigenvalue decreases. The limiting time step is
reached when the smallest eigenvalue becomes zero. Beyond this point,
$\Bop_2$ is no longer positive definite.

To demonstrate this, consider the second-order projection used in Motif where  the approximate
inverse is
\begin{equation}
	\Bop_2
	=
	\Delta t
	\left(
	I+\frac{\nu\Delta t}{2}\Lop
	\right).
\end{equation}
Let $\lambda$ be an eigenvalue of the discrete velocity Laplacian
$\Lop$, with corresponding eigenvector $\mathbf{X}$, such that
\begin{equation}
	\Lop\mathbf{X}=\lambda\mathbf{X}.
\end{equation}
Application of $\Bop_2$ to the same vector gives
\begin{align}
	\Bop_2\mathbf{X}
	&=
	\Delta t
	\left(
	I+\frac{\nu\Delta t}{2}\Lop
	\right)\mathbf{X}
	\\
	&=
	\Delta t
	\left(
	I\mathbf{X}
	+
	\frac{\nu\Delta t}{2}\Lop\mathbf{X}
	\right)
	\\
	&=
	\Delta t
	\left(
	\mathbf{X}
	+
	\frac{\nu\Delta t}{2}\lambda\mathbf{X}
	\right)
	\\
	&=
	\Delta t
	\left(
	1+\frac{\nu\Delta t}{2}\lambda
	\right)\mathbf{X},
\end{align}
where $I\mathbf{X}=\mathbf{X}$ is used.

With the sign convention used in Motif, $\Lop$ represents
$+\nabla^2$ and its eigenvalues are non-positive. Therefore, the most
restrictive case is associated with its most negative eigenvalue,
$\lambda_{\min}$. Positive definiteness of $\Bop_2$ requires
\begin{equation}
	1+
	\frac{\nu\Delta t}{2}
	\lambda_{\min}
	>0.
\end{equation}
Since $\lambda_{\min}<0$, this gives
\begin{equation}
	\Delta t
	<
	\Delta t_{\mathrm{PD}},
\end{equation}
where
\begin{equation}
	\Delta t_{\mathrm{PD}}
	=
	\frac{2}
	{\nu\left|\lambda_{\min}\right|}.
	\label{eq:perot_dt_pd}
\end{equation}
The subscript ``PD'' denotes positive definiteness.
The largest magnitude eigenvalue of the discrete Laplacian scales
approximately as $1/\Delta^{2}$ for a uniform grid (see Section \ref{ssec:velocity_laplacian}). Then, Eq.~\eqref{eq:perot_dt_pd} gives
\begin{equation}
	\Delta t_{\mathrm{PD}} \sim \frac{\Delta^2}{\nu}.
\end{equation}
Thus, grid refinement makes this restriction progressively more severe.

This property was tested using the lid-driven cavity at
$\mathrm{Re}=100$. Uniform $64\times64$, $128\times128$, and
$256\times256$ grids were considered. For each grid,
$\Delta t_{\mathrm{PD}}$ was calculated from the discrete operators.
Calculations were then performed using time steps on both sides of this
limit. The standard projection was also tested at the same time steps.
Each calculation was requested to advance for 300 time steps.

Table~\ref{tab:perot_pd_limit} summarizes the behavior immediately below
and above the predicted limit. The smallest eigenvalue of $\Bop_2$ is
positive at $0.99\Delta t_{\mathrm{PD}}$ and negative at
$1.01\Delta t_{\mathrm{PD}}$.

\begin{table}[!htbp]
	\centering
	\caption{Positive-definiteness limit of the second-order projection
		for the lid-driven cavity at $\mathrm{Re}=100$.}
	\label{tab:perot_pd_limit}
	\begin{tabular}{cccccc}
		\hline
		Grid &
		$\Delta t_{\mathrm{PD}}$ &
		$\Delta t/\Delta t_{\mathrm{PD}}$ &
		$\lambda_{\min}(\Bop_2)$ &
		Standard &
		Perot \\
		\hline
		$64^2$
		& $6.105\times10^{-3}$
		& 0.99 & $ 6.04\times10^{-5}$ & completed & completed \\
		&
		& 1.01 & $-6.17\times10^{-5}$ & completed & completed \\
		
		$128^2$
		& $1.526\times10^{-3}$
		& 0.99 & $ 1.51\times10^{-5}$ & completed & completed \\
		&
		& 1.01 & $-1.54\times10^{-5}$ & completed & unstable \\
		
		$256^2$
		& $3.81\times10^{-4}$
		& 0.99 & $ 3.78\times10^{-6}$ & completed & completed \\
		&
		& 1.01 & $-3.85\times10^{-6}$ & completed & unstable \\
		\hline
	\end{tabular}
\end{table}

The limiting time step decreases by a factor of four when
the grid spacing is halved. The numerical experiment therefore confirms the expected
$\Delta t_{\mathrm{PD}}\propto\Delta^2$ scaling.

The change in behavior across the limit is particularly clear on the
$256\times256$ grid. At $0.99\Delta t_{\mathrm{PD}}$, both projection
methods completed all 300 steps. Their maximum Courant numbers were
approximately 0.093. At $1.01\Delta t_{\mathrm{PD}}$, the standard
projection again completed all 300 steps. The second-order projection however
became unstable after 47 steps. Its maximum velocity component grew to
$2.27\times10^4$ before the calculation was terminated.

These results also distinguish the positive-definiteness condition from
the explicit viscous stability restriction (Eq. \ref{eq:diffusion_stability}). Diffusion is treated
implicitly by Crank--Nicolson in the present method. The restriction arises
instead from the approximate inverse used by the second-order Perot
projection. Its $\Delta^2/\nu$ scaling resembles an explicit diffusion
restriction, but its origin is different.

The condition
$\Delta t<\Delta t_{\mathrm{PD}}$ is therefore adopted as a practical
time-step requirement for the second-order projection in Motif.

\subsection{The Standard projection under periodic boundary conditions}
\label{ssec:periodicproj}

The Standard projection contains a splitting approximation and is therefore
generally associated with a first-order splitting error. This does not,
however, imply that the computed velocity must always be first-order accurate
in time. \cite{perot1993analysis} showed that second-order temporal
accuracy of velocity can be recovered under periodic boundary conditions. This special behavior does not generally extend to problems with
physical boundaries, where the boundary treatment introduces additional
splitting errors.

The Taylor--Green vortex considered in Sections~\ref{sec:inviscid_atgv} and \ref{sec:taylor_green} 
provide a direct numerical demonstration of this property. The problem is
periodic in both coordinate directions and has an analytical solution.
Temporal accuracy tables show that the Standard projection exhibits approximately
second-order temporal convergence of the velocity on the uniform grid.
The second-order projection shows essentially the same convergence behavior.
Thus, for this periodic problem, use of the higher-order approximate inverse
in the second-order projection does not provide an even higher temporal order for velocity.

\chapter{DNS suitability}

\section{Purpose and assessment criteria}
\label{ssec:dns_suitability}

The verification cases considered in the preceding section establish the
accuracy of the spatial and temporal discretizations. They also compare the
numerical solution with analytical solutions and published benchmark data.
Motif is intended to produce DNS-grade flow fields. But DNS requires
more than accurate solutions of such benchmark problems. The unsteady
flow dynamics should also be resolved without numerical artifacts that
significantly create, remove, or incorrectly redistribute important physical
quantities.

There is no single set of universally accepted numerical criteria that
certifies a solver as suitable for DNS. Nevertheless, several useful
properties and diagnostics have been identified in the literature. They test
whether the discrete equations preserve important properties of the
continuous incompressible-flow equations. In addition, these tests inspect whether the resolved flow scales
are represented without excessive numerical contamination
\citep{Morinishi1998,Vasilyev2000,Ham2002}. The criteria used here are
introduced below.

\subsection{Discrete mass conservation}
\label{sssec:dns_mass}

For incompressible flow, discrete mass conservation requires
zero net volumetric flux through every pressure control volume. The local flux
imbalance is evaluated using the finite-volume divergence operator defined in
Section~\ref{ssec:divergence_op}. The diagnostic used here is
\begin{equation}
	\max
	\left|
	\left(\mathcal{D}_f\mathbf{V}\right)_{i,j}
	\right|.
\end{equation}
This is a local measure. It identifies the largest mass imbalance in
any pressure control volume. For Motif, this quantity should remain at
machine-zero level after projection.

\subsection{Momentum conservation}
\label{sssec:dns_momentum}
Because the momentum equations are solved directly, their discrete
conservation is a basic requirement. In a conservative finite-volume
discretization, momentum is transported as fluxes through control-volume
faces.  For example, the convective
flux of $x$-momentum through a surface satisfies
\begin{equation}
	\int_V
	\nabla\cdot\left(\rho u\mathbf{V}\right)\,dV
	=
	\int_A
	\rho u\left(\mathbf{V}\cdot\mathbf{n}\right)\,dA .
\end{equation}
A flux through a shared internal face enters the two neighboring control
volumes with opposite signs. These contributions therefore cancel
when the momentum is summed over the complete domain
\citep{Morinishi1998,Ham2002}. Therefore, changes in the momentum contained in the complete domain can 
result from interactions with the domain boundaries or from applied
source terms only.

For a periodic domain, the boundary contributions also cancel. Momentum can be redistributed within the domain, but the momentum contained
in the complete periodic domain cannot change. This makes the amount of momentum summed over the whole domain a good measure.

The momentum flux through a control-volume face and the momentum contained in a control volume are different quantities. The total $x$- and $y$-momentum contained in the complete domain $V$ are,
\begin{align}
	P_x &= \int_V \rho u\,dV,\\
	P_y &= \int_V \rho v\,dV.
\end{align}
On the staggered grid, $u$ and $v$ belong to different momentum control
volumes. The corresponding discrete forms are therefore
\begin{align}
	P_x
	&=
	\rho
	\sum_{i,j}
	u_{i+1/2,j} V^u_{i+1/2,j},
	\\
	P_y
	&=
	\rho
	\sum_{i,j}
	v_{i,j+1/2} V^v_{i,j+1/2},
\end{align}
where $V^u$ and $V^v$ are the areas of the staggered $u$- and $v$-momentum
control volumes, respectively. No interpolation to the pressure-cell centers
is required.

The changes in momentum are measured relative to their initial values. Using
the reference velocity and length scales of a given problem, they may be
reported in nondimensional form as
\begin{align}
	\Delta P_x^*
	&=
	\frac{P_x-P_{x,0}}
	{\rho U_{\mathrm{ref}}L_{\mathrm{ref}}^2},
	\\
	\Delta P_y^*
	&=
	\frac{P_y-P_{y,0}}
	{\rho U_{\mathrm{ref}}L_{\mathrm{ref}}^2}.
\end{align}
Both quantities should remain at machine-zero level for a periodic test problem that will be introduced below.

\subsection{Kinetic-energy conservation}
\label{sssec:dns_ke}

The incompressible Navier--Stokes equations contain equations for mass and
momentum, but no separate equation for kinetic energy. A kinetic-energy
equation is derived from the mass and momentum equations. But this does not mean that discrete
conservation of mass and momentum automatically guarantees discrete
kinetic-energy conservation \citep{Morinishi1998}.

The kinetic-energy equation is obtained by taking the dot product of the
momentum equation in Eq.~\eqref{eq:momentum} with $\mathbf{V}$ and using
incompressibility. Defining the kinetic energy per unit mass as
\begin{equation}
	e_k
	=
	\frac{1}{2}
	\mathbf{V}\cdot\mathbf{V},
\end{equation}
the resulting equation can be written as
\begin{equation}
	\frac{\partial e_k}{\partial t}
	+
	\mathbf{V}\cdot
	\nabla\cdot
	\left(
	\mathbf{V}\mathbf{V}
	\right)
	=
	-\frac{1}{\rho}
	\mathbf{V}\cdot\nabla p
	+
	\mathbf{V}\cdot
	\nabla\cdot
	\left(
	\nu\nabla\mathbf{V}
	\right).
\end{equation}
Viscosity physically dissipates kinetic energy through the last term. For an inviscid assumption, this term is zero and there is no dissipation. Next, consider the convective term,
\begin{equation}
	\mathbf{V}\cdot
	\nabla\cdot
	\left(
	\mathbf{V}\mathbf{V}
	\right).
\end{equation}
Using the incompressibility condition,
$\nabla\cdot\mathbf{V}=0$, the convective term can also be written as
\begin{equation}
	\mathbf{V}\cdot
	\nabla\cdot
	\left(
	\mathbf{V}\mathbf{V}
	\right)
	=
	\nabla\cdot
	\left(
	e_k\mathbf{V}
	\right).
\end{equation}
This representation is helpful. Convection appears as the divergence of a kinetic-energy flux. It can redistribute kinetic energy within the domain, but cannot change its total amount. Integrate over the complete domain and use Gauss theorem:
\begin{equation}
	\int_V
	\nabla\cdot
	\left(
	e_k\mathbf{V}
	\right)
	\,dV
	=
	\int_A
	e_k
	\left(
	\mathbf{V}\cdot\mathbf{n}
	\right)
	\,dA .
\end{equation}
The right-hand side accounts for the kinetic-energy flux through the domain
boundaries. For a periodic domain, the boundary contributions cancel. Thus,
convection cannot change the total kinetic energy. As shown in the next
section, the pressure term also makes no net contribution in a periodic
incompressible domain. Since viscous dissipation is also absent in an
inviscid flow, the kinetic energy in the domain must remain constant. A periodic
inviscid problem therefore provides a useful kinetic-energy conservation
test.

The total kinetic energy per unit density in the domain is
\begin{equation}
	K
	=
	\int_V e_k\,dV
	=
	\frac{1}{2}
	\int_V
	\mathbf{V}\cdot\mathbf{V}\,dV .
\end{equation}

On the staggered grid, Motif evaluates the discrete kinetic energy directly
on the native velocity control volumes as
\begin{equation}
	K
	\approx
	\frac{1}{2}
	\left[
	\sum_{i,j}
	u_{i+1/2,j}^2 V^u_{i+1/2,j}
	+
	\sum_{i,j}
	v_{i,j+1/2}^2 V^v_{i,j+1/2}
	\right].
\end{equation}
The relative change from the initial value $K_0$ is
\begin{equation}
	\frac{\Delta K}{K_0}
	=
	\frac{K-K_0}{K_0}.
\end{equation}
For the periodic inviscid test considered below, this quantity should remain
close to zero.

\subsection{Pressure-term kinetic-energy conservation}
\label{sssec:dns_pressure_energy}

The preceding kinetic-energy discussion considered the roles of convection
and viscosity. The pressure-gradient term also contributes to the
kinetic-energy equation:
\begin{equation}
	-\frac{1}{\rho}
	\mathbf{V}\cdot\nabla p .
\end{equation}
Using the product rule,
\begin{equation}
	-\frac{1}{\rho}
	\mathbf{V}\cdot\nabla p
	=
	-\frac{1}{\rho}
	\nabla\cdot
	\left(
	p\mathbf{V}
	\right)
	+
	\frac{p}{\rho}
	\nabla\cdot\mathbf{V}.
\end{equation}
For incompressible flow, $\nabla\cdot\mathbf{V}=0$. The pressure contribution
therefore reduces to
\begin{equation}
	-\frac{1}{\rho}
	\mathbf{V}\cdot\nabla p
	=
	-\frac{1}{\rho}
	\nabla\cdot
	\left(
	p\mathbf{V}
	\right).
\end{equation}
This term represents rate of pressure work per unit mass. Pressure may locally increase or decrease
kinetic energy by accelerating or decelerating the flow. However, it does not
produce a net change in the total kinetic energy of a periodic domain.

For a periodic domain, integration of the right hand side over the complete domain gives
\begin{equation}
	-\frac{1}{\rho}
	\int_V
	\nabla\cdot
	\left(
	p\mathbf{V}
	\right)\,dV
	=
	-\frac{1}{\rho}
	\int_A
	p
	\left(
	\mathbf{V}\cdot\mathbf{n}
	\right)
	\,dA
	=
	0,
\end{equation}
because the pressure and velocity are equal on corresponding periodic boundaries,
while the outward normals have opposite directions. Thus, this pressure should make no net
contribution to the total kinetic-energy rate in a periodic incompressible
flow
\citep{Morinishi1998,Vasilyev2000}.

Motif tests this property using the left hand side representation  which contains the pressure gradient. Its global contribution to the
kinetic-energy rate is evaluated directly on the staggered velocity control
volumes as
\begin{equation}
	W_p
	=
	-\frac{1}{\rho}
	\left[
	\sum_{i,j}
	u_{i+1/2,j}
	\left(
	\frac{\partial p}{\partial x}
	\right)_{i+1/2,j}
	V^u_{i+1/2,j}
	+
	\sum_{i,j}
	v_{i,j+1/2}
	\left(
	\frac{\partial p}{\partial y}
	\right)_{i,j+1/2}
	V^v_{i,j+1/2}
	\right].
\end{equation}
Evaluating this form tests the pressure-gradient operator used by the solver.
For a periodic, discretely divergence-free velocity field, $W_p$ should
remain at machine-zero level. 

For reporting the pressure-work contribution in nondimensional form, Motif
uses
\begin{equation}
	W_p^*
	=
	\frac{W_p}
	{U_{\mathrm{ref}}^3L_{\mathrm{ref}}},
\end{equation}
where $U_{\mathrm{ref}}$ and $L_{\mathrm{ref}}$ are the reference velocity
and length scales, respectively.

\subsection{Enstrophy behavior}
\label{sssec:dns_enstrophy}

Vorticity measures the local rotation of the fluid. For a two-dimensional
flow in the $x$--$y$ plane, only the out-of-plane component is nonzero,
\begin{equation}
	\omega
	=
	\left(
	\nabla\times\mathbf{V}
	\right)_z
	=
	\frac{\partial v}{\partial x}
	-
	\frac{\partial u}{\partial y}.
\end{equation}

In Motif, vorticity is evaluated at the physical pressure-cell vertices (corners).
The staggered arrangement makes this location convenient because on the neighboring faces of the vertices, the
$u$- and $v$-velocities are directly available. No interpolation
of either velocity component is required. 
The same formulation applies on stretched Cartesian grids because the local
pressure-center distances are used. At physical boundaries, the required
velocity values are supplied through the ghost layers.

Enstrophy is a measure based on the square of vorticity. It is not a measure
of kinetic energy. Instead, it gives greater importance to regions with
strong rotation even when those regions are small. The total enstrophy in the
domain is defined as
\begin{equation}
	\mathcal{E}
	=
	\frac{1}{2}
	\int_V
	\omega^2\,dV .
\end{equation}
Large values of $\mathcal{E}$ indicate that strong vorticity is present in
the flow.

For a two-dimensional incompressible flow, the vorticity equation is
\begin{equation}
	\frac{\partial\omega}{\partial t}
	+
	\mathbf{V}\cdot\nabla\omega
	=
	\nu\nabla^2\omega,
\end{equation}
which is found by taking the curl of the momentum equation. For an inviscid flow, $\nu=0$. The equation then becomes
\begin{equation}
	\frac{D\omega}{Dt}
	=
	\frac{\partial\omega}{\partial t}
	+
	\mathbf{V}\cdot\nabla\omega
	=
	0.
\end{equation}
Thus, the vorticity carried by a fluid particle remains constant. Vorticity
can be transported to another location, but its magnitude is not changed.

Multiplying the inviscid vorticity equation by $\omega$ and using
incompressibility gives
\begin{equation}
	\frac{\partial}{\partial t}
	\left(
	\frac{\omega^2}{2}
	\right)
	+
	\nabla\cdot
	\left(
	\frac{\omega^2}{2}\mathbf{V}
	\right)
	=
	0.
\end{equation}
Integration over the complete domain gives
\begin{equation}
	\frac{d\mathcal{E}}{dt}
	=
	-
	\int_A
	\frac{\omega^2}{2}
	\left(
	\mathbf{V}\cdot\mathbf{n}
	\right)
	\,dA .
\end{equation}
For a periodic domain, the boundary contributions on the right hand side cancel each other. The total
enstrophy therefore remains constant. Thus, a two-dimensional, inviscid,
incompressible flow in a periodic domain provides a simple test of enstrophy
conservation.

Viscosity changes this behavior. It smooths velocity gradients and causes
enstrophy to decrease. Three-dimensional flow also behaves differently
because vortex stretching can change the magnitude of vorticity.

Numerical vorticity contains estimations of spatial derivatives of velocity. Therefore it is more
sensitive to small-scale numerical errors than kinetic energy. Artificial
dissipation can reduce enstrophy, while numerical errors may also produce
spurious enstrophy.

The relative change from the initial value is measured as
\begin{equation}
	\frac{\Delta\mathcal{E}}{\mathcal{E}_0}
	=
	\frac{\mathcal{E}-\mathcal{E}_0}
	{\mathcal{E}_0}.
\end{equation}
For the periodic inviscid test considered below, this quantity should remain
close to zero.

\subsection{Kinetic energy spectrum}
\label{sssec:dns_spectral}

The kinetic-energy measure introduced above gives a single number for the complete
velocity field. But it does not show how this energy is distributed among flow
structures of different sizes. A kinetic-energy spectrum provides this
additional information.

Consider first a one-dimensional velocity component $u(x)$ in a periodic
domain of length $L$. A Fourier representation expresses this velocity as a
sum of spatial waves,
\begin{equation}
	u(x)
	=
	\sum_{n=-\infty}^{\infty}
	\widehat{u}_n
	e^{ik_nx}.
\end{equation}
Here, $n$ is an integer called the Fourier mode, $k_n$ is the
wavenumber of that mode and $\widehat{u}_n$ is its Fourier coefficient. 

The possible wavenumbers follow from the periodicity of the domain. This requires each Fourier contribution to have the same value at
$x$ and $x+L$. Thus,
\begin{equation}
	\widehat{u}_n e^{ik_n(x+L)}
	=
	\widehat{u}_n e^{ik_nx}.
\end{equation}
For a nonzero Fourier coefficient, this reduces to
\begin{equation}
	e^{ik_n(x+L)}
	=
	e^{ik_nx}.
\end{equation}
This condition is satisfied when
\begin{equation}
	k_n
	=
	\frac{2\pi n}{L},
	\qquad
	n=0,\pm1,\pm2,\ldots .
	\label{eq:kn}
\end{equation}
The periodic domain therefore permits a discrete set of spatial
wavenumbers. For $n\ne0$, the corresponding wavelength is
\begin{equation}
	\lambda_n
	=
	\frac{2\pi}{|k_n|}
	=
	\frac{L}{|n|}.
\end{equation}
Thus, the first mode fits one wavelength into the domain, the second fits
two, and so forth. Increasing $|n|$ therefore represents progressively
smaller spatial scales. The mode $n=0$ is the spatially uniform component
and represents the mean value of $u$.

Note that although $\widehat{u}_n$ and $e^{ik_nx}$ are generally complex, the summation
represents the real velocity $u(x)$. This follows from the relation between
the positive and negative Fourier modes. Since $u(x)$ is real, it can be shown that its Fourier
coefficients satisfy
\begin{equation}
	\widehat{u}_{-n}
	=
	\widehat{u}_n^*,
\end{equation}
where the superscript $*$ denotes the complex conjugate. Also,
$k_{-n}=-k_n$ due to Equation \ref{eq:kn}. The contributions of a pair of modes can therefore be written
as
\begin{align}
	\widehat{u}_n e^{ik_nx}
	+
	\widehat{u}_{-n}e^{ik_{-n}x}
	&=
	\widehat{u}_n e^{ik_nx}
	+
	\widehat{u}_n^*e^{-ik_nx}
	\\
	&=
	2\operatorname{Re}
	\left\{
	\widehat{u}_n e^{ik_nx}
	\right\},
\end{align}
which is real. Thus, the reconstructed velocity from the complex quantities remains real.

The Fourier coefficient of mode $n$ is generally a complex number and may be
written as
\begin{equation}
	\widehat{u}_n
	=
	a_n
	+
	b_n i,
\end{equation}
where $a_n$ and $b_n$ are its real and imaginary parts, respectively. Its
magnitude,
\begin{equation}
	\left|\widehat{u}_n\right|
	=
	\sqrt{a_n^2+b_n^2},
\end{equation}
describes how strongly that mode is present in the velocity field. Its phase
determines the spatial position of the corresponding wave. The squared
magnitude is
\begin{equation}
	\left|\widehat{u}_n\right|^2
	=
	a_n^2+b_n^2.
\end{equation}
The kinetic-energy contribution of a Fourier mode depends on this squared
magnitude.

The same idea applies to a two-dimensional velocity field. A Fourier mode
now has two wavenumber components, $k_x$ and $k_y$, and Fourier coefficients
$\widehat{u}(k_x,k_y)$ and $\widehat{v}(k_x,k_y)$. Its kinetic-energy
contribution is proportional to
\begin{equation}
	\frac{1}{2}
	\left[
	\left|\widehat{u}(k_x,k_y)\right|^2
	+
	\left|\widehat{v}(k_x,k_y)\right|^2
	\right].
\end{equation}
The wavenumber magnitude of this mode is
\begin{equation}
	k
	=
	\sqrt{k_x^2+k_y^2}.
\end{equation}
Modes with different pairs $(k_x,k_y)$ can therefore represent similar
spatial scales if their wavenumber magnitudes are similar. To describe the
energy distribution using a single wavenumber $k$, the contributions of such
modes must be collected together. Since the Fourier modes have discrete
wavenumbers, they are grouped into intervals of width $\Delta k$. For each
interval, the kinetic-energy contributions of the modes whose wavenumber
magnitudes fall within that interval are summed. Dividing this sum by $\Delta k$ gives the
one-dimensional kinetic-energy spectrum $S(k)$,
\begin{equation}
	S(k)
	=
	\frac{1}{\Delta k}
	\sum
	\frac{1}{2}
	\left[
	\left|\widehat{u}(k_x,k_y)\right|^2
	+
	\left|\widehat{v}(k_x,k_y)\right|^2
	\right].
\end{equation}
Thus, $S(k)$ shows how the kinetic energy is distributed with spatial scale.
Small $k$ represents large spatial scales, while large $k$ represents small
spatial scales.

With the FFT normalization used in Motif, summing the spectrum over all
resolved wavenumbers gives
\begin{equation}
	\sum_k S(k)\Delta k
	\approx
	\frac{1}{2V}
	\int_V
	\left(
	u^2+v^2
	\right)\,dV .
\end{equation}
The left-hand side obtains the domain-mean kinetic energy from the Fourier
coefficients, while the right-hand side obtains the same quantity directly
from the physical velocity field. Their equivalence follows from Parseval's
relation. This provides a consistency check for the computed spectrum.

The spectrum therefore shows how kinetic energy is distributed across the
resolved spatial scales. This provides information that cannot be obtained
from the total kinetic energy alone. Numerical dissipation appears as a loss
of energy at the affected wavenumbers. Numerical errors may instead produce
an artificial accumulation of energy, particularly near the highest resolved
wavenumbers.

In turbulence studies, the kinetic-energy spectrum is commonly formed from
velocity fluctuations. It then describes how turbulent kinetic energy is
distributed among spatial scales. The principle is the same here. The present
DNS-suitability test uses the complete instantaneous velocity field because
the objective is to examine how the numerical method represents the spatial
scales of the test flow.

\subsection{Plotting the spectrum}
\label{sssec:spectrum_interpretation}

The spectrum plotted as $S(k)$ provides a view of the spatial scales
represented by a calculation. As explained above, increasing wavenumber
corresponds to decreasing spatial scale since $\lambda=2\pi/k$. The left
portion of the spectrum therefore represents larger spatial structures,
while increasing $k$ represents progressively smaller structures. The
quantity $S(k)\Delta k$ gives the kinetic energy per unit mass associated
with a wavenumber interval of width $\Delta k$. Thus, the vertical axis
indicates how kinetic energy is distributed among spatial scales.

The wavenumber range has lower and upper limits. From
Eq.~\eqref{eq:kn}, the smallest nonzero wavenumber is
\begin{equation}
	k_{\min}
	=
	\frac{2\pi}{L}.
\end{equation}
It is determined by the domain size.

The upper limit is determined by the grid spacing. According to the Nyquist
sampling criterion, at least two grid points are required to distinguish a
spatial wave. A wavelength equal to $\Delta x$ would have the same value at
every grid point and would appear as a constant. The shortest wavelength
that can be represented is therefore
\begin{equation}
	\lambda_{\min}
	=
	2\Delta x.
\end{equation}
Since $k=2\pi/\lambda$, the corresponding Nyquist wavenumbers are
\begin{equation}
	k_{\mathrm{N},x}
	=
	\frac{\pi}{\Delta x},
	\qquad
	k_{\mathrm{N},y}
	=
	\frac{\pi}{\Delta y},
\end{equation}
where the subscript $\mathrm{N}$ denotes the Nyquist limit.

The two-dimensional FFT provides modes at discrete pairs $(k_x,k_y)$.
The spectrum $S(k)$ groups these modes according to their magnitude
\begin{equation}
	k
	=
	\sqrt{k_x^2+k_y^2}.
\end{equation}
For a given $k$, Fourier modes should be available in all directions in the
$(k_x,k_y)$ plane. This is possible only while $k$ remains below both
directional Nyquist limits. The largest wavenumber satisfying this condition
is therefore
\begin{equation}
	k_{\max}
	=
	\min
	\left(
	k_{\mathrm{N},x},
	k_{\mathrm{N},y}
	\right)
	=
	\min
	\left(
	\frac{\pi}{\Delta x},
	\frac{\pi}{\Delta y}
	\right).
\end{equation}

The kinetic energy is not distributed equally over the wavenumber range.
Large flow structures can contain much more energy than small structures.
Thus, $S(k)$ may span many orders of magnitude, and logarithmic axes are
useful. The spectrum also need not decrease monotonically. Particular
spatial scales may contain more energy than neighboring scales.

The high-wavenumber region is especially useful for assessing spatial
resolution. Spectra obtained on different grids should agree over adequately
resolved wavenumbers. As the grid is refined, grid-dependent behavior should
move toward larger $k$. Flattening or growth that changes with grid
refinement can therefore indicate that the corresponding small scales are
affected by insufficient spatial resolution.

\section{Periodic shear-layer roll-up test}

\subsection{Problem definition}

The periodic shear-layer roll-up provides a nonlinear test of the numerical
method in the absence of physical dissipation.  The problem is closely
related to the doubly periodic shear-layer calculations of
\cite{Bell1989}.  Variants of this configuration have subsequently been
used to assess the conservation and robustness properties of
incompressible-flow discretizations
\citep{Knikker2009,PalhaGerritsma2017}.  The present test follows the
inviscid formulation employed by \cite{PalhaGerritsma2017}.

The computational domain is the square
\begin{equation}
	\Omega
	=
	[0,2\pi]\times[0,2\pi],
\end{equation}
with periodic boundary conditions in both coordinate directions.  The flow is
inviscid,
\begin{equation}
	\nu=0,
\end{equation}
and no external forcing is applied.

The initial velocity components are prescribed as
\begin{align}
	u(x,y,0)
	&=
	\begin{cases}
		\tanh\left[
		\dfrac{y-\pi/2}{\delta}
		\right],
		& y\leq\pi,
		\\[8pt]
		\tanh\left[
		\dfrac{3\pi/2-y}{\delta}
		\right],
		& y>\pi,
	\end{cases}
	\\
	v(x,y,0)
	&=
	\epsilon\sin(x),
\end{align}
where
\begin{equation}
	\delta=\frac{\pi}{15},
	\qquad
	\epsilon=0.05.
\end{equation}
The streamwise velocity therefore contains two finite-thickness shear layers,
while the transverse velocity introduces a small periodic perturbation that
initiates their roll-up.  The initial field is divergence-free because
$u$ is independent of $x$ and $v$ is independent of $y$.

Since $\nu=0$, the calculation corresponds to the two-dimensional
incompressible Euler equations. 

\subsection{Qualitative vortex evolution}

The small transverse perturbation grows with time and initiates the roll-up
of the two shear layers.  Each layer develops into a periodic array of
large-scale vortices.  The shear layers subsequently wrap around the large vortex cores \citep{Bell1989,Knikker2009}.
Owing to the symmetry of the initial condition, the upper and lower shear
layers undergo corresponding evolutions, shifted relative to one another in
the periodic domain \cite{Bell1989}.

The qualitative evolution is
assessed using vorticity contours.  In particular, the contour levels
\begin{equation}
	\omega = \pm 1,\ \pm 2,\ \ldots,\ \pm 6
\end{equation}
are used. 

\begin{figure}[!htbp]
	\centering
	
	\begin{subfigure}[t]{0.32\textwidth}
		\centering
		\includegraphics[width=\linewidth]{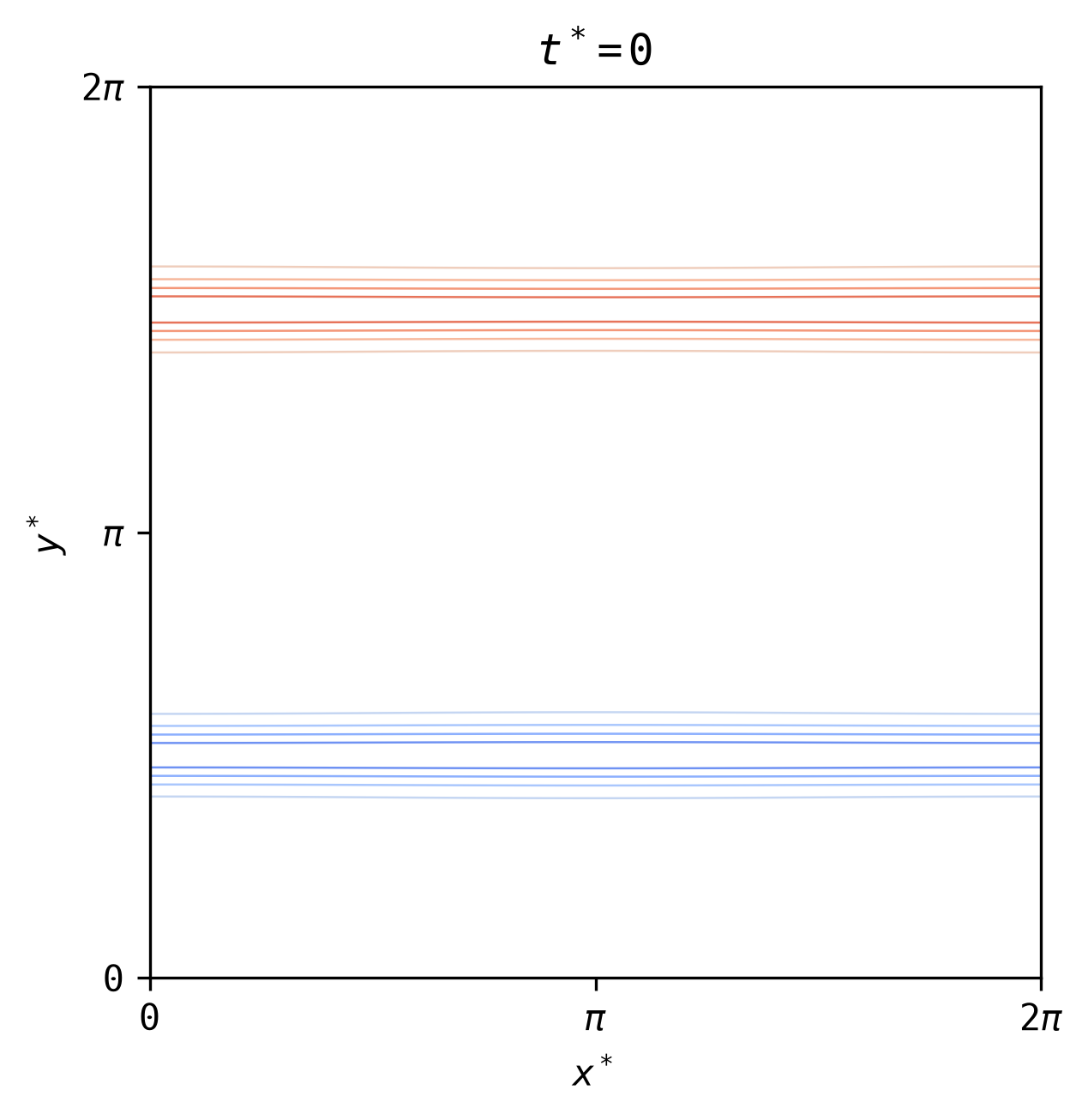}
	\end{subfigure}
	\begin{subfigure}[t]{0.32\textwidth}
		\centering
		\includegraphics[width=\linewidth]{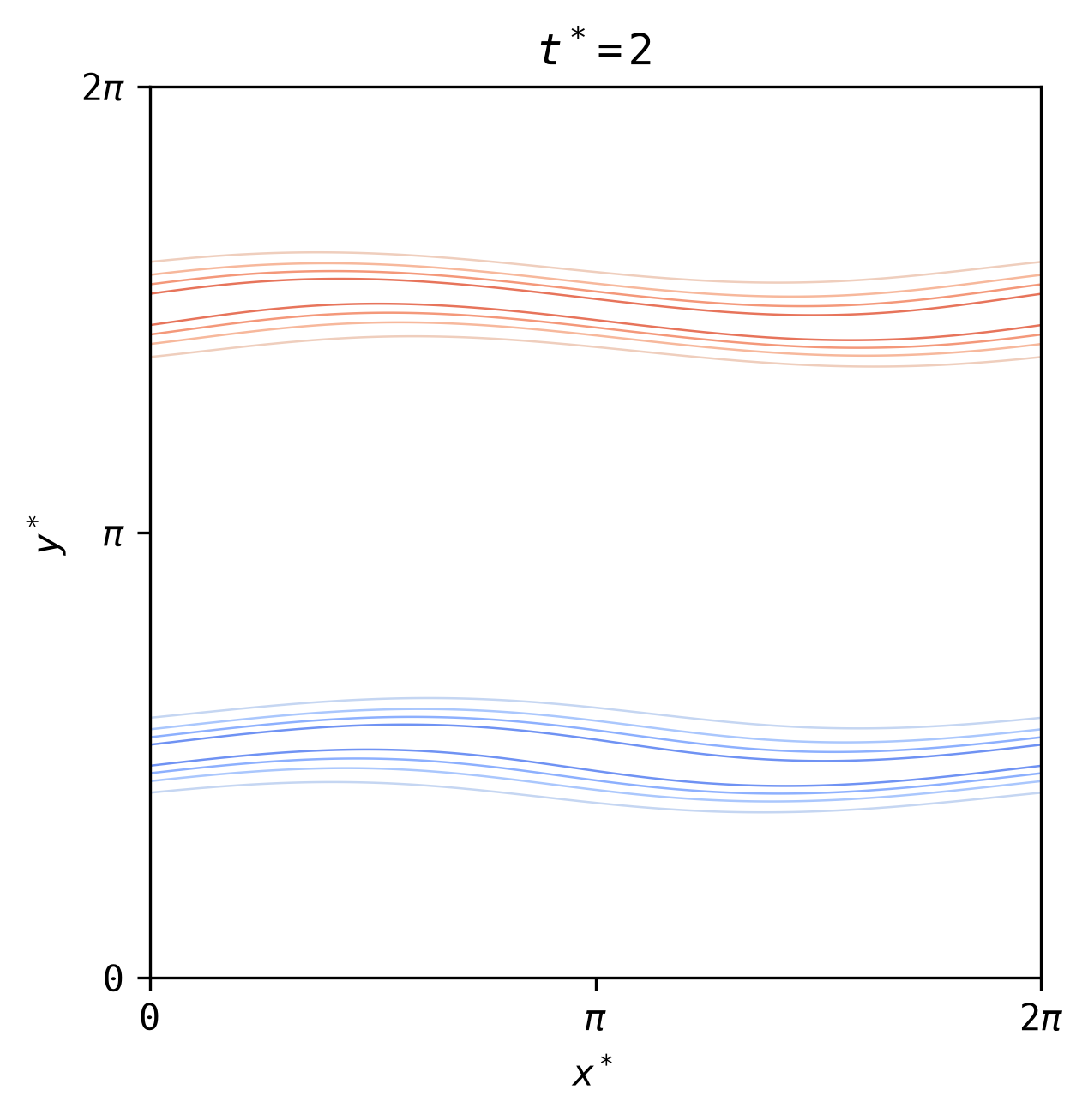}
	\end{subfigure}
	\begin{subfigure}[t]{0.32\textwidth}
		\centering
		\includegraphics[width=\linewidth]{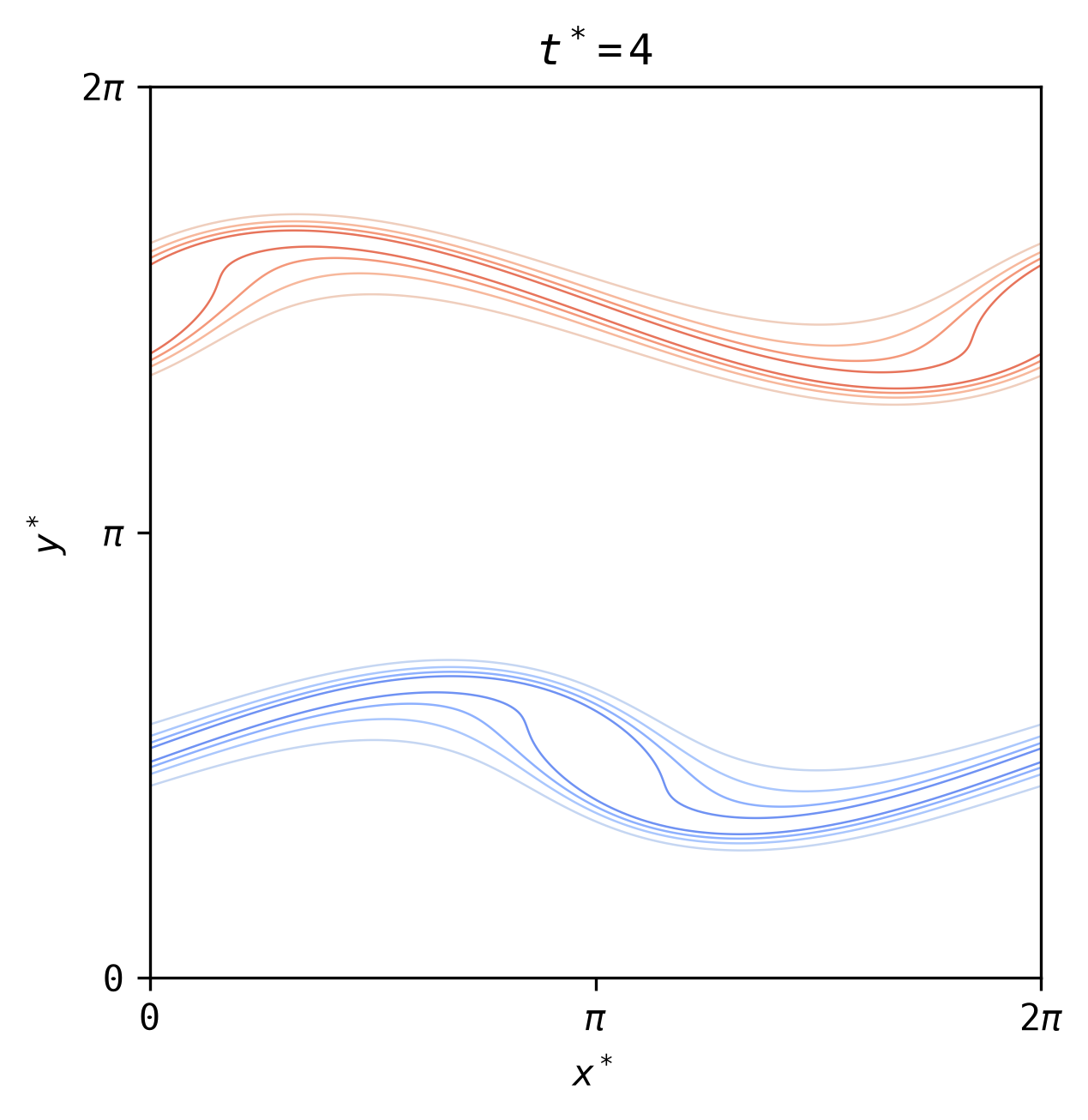}
	\end{subfigure}
	
	\vspace{0.4em}
	
	\begin{subfigure}[t]{0.32\textwidth}
		\centering
		\includegraphics[width=\linewidth]{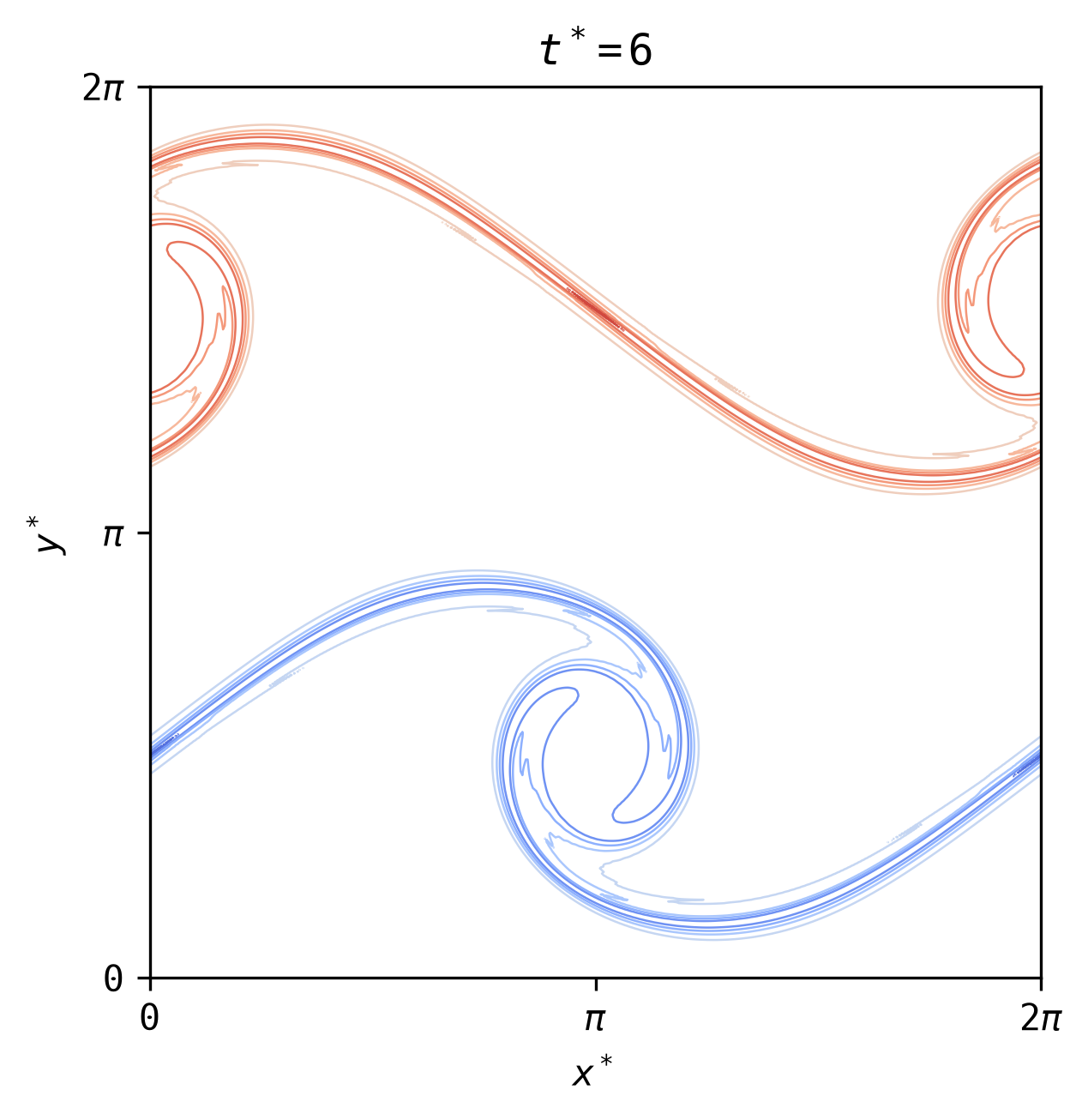}
	\end{subfigure}
	\begin{subfigure}[t]{0.32\textwidth}
		\centering
		\includegraphics[width=\linewidth]{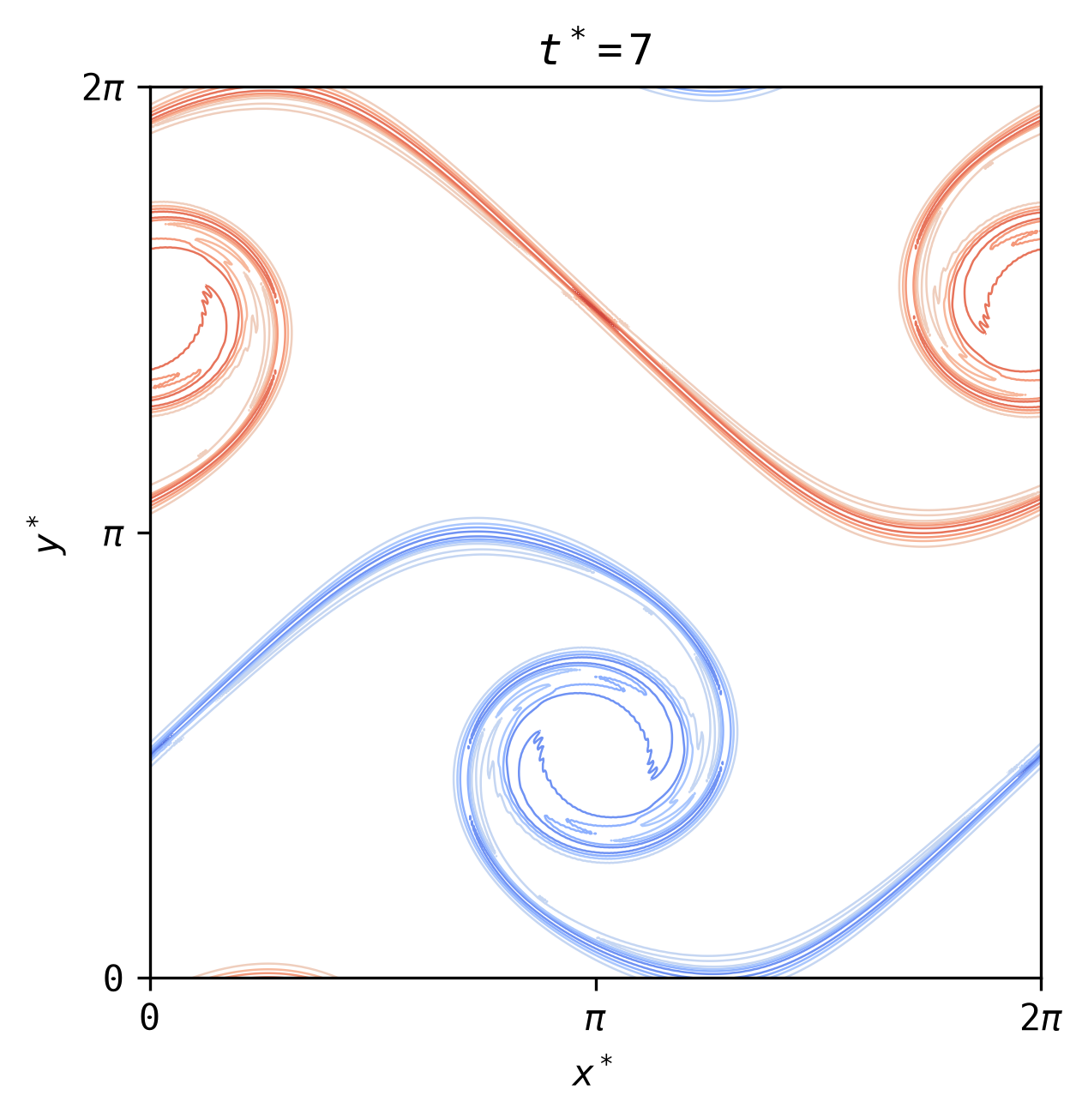}
	\end{subfigure}
	\begin{subfigure}[t]{0.32\textwidth}
		\centering
		\includegraphics[width=\linewidth]{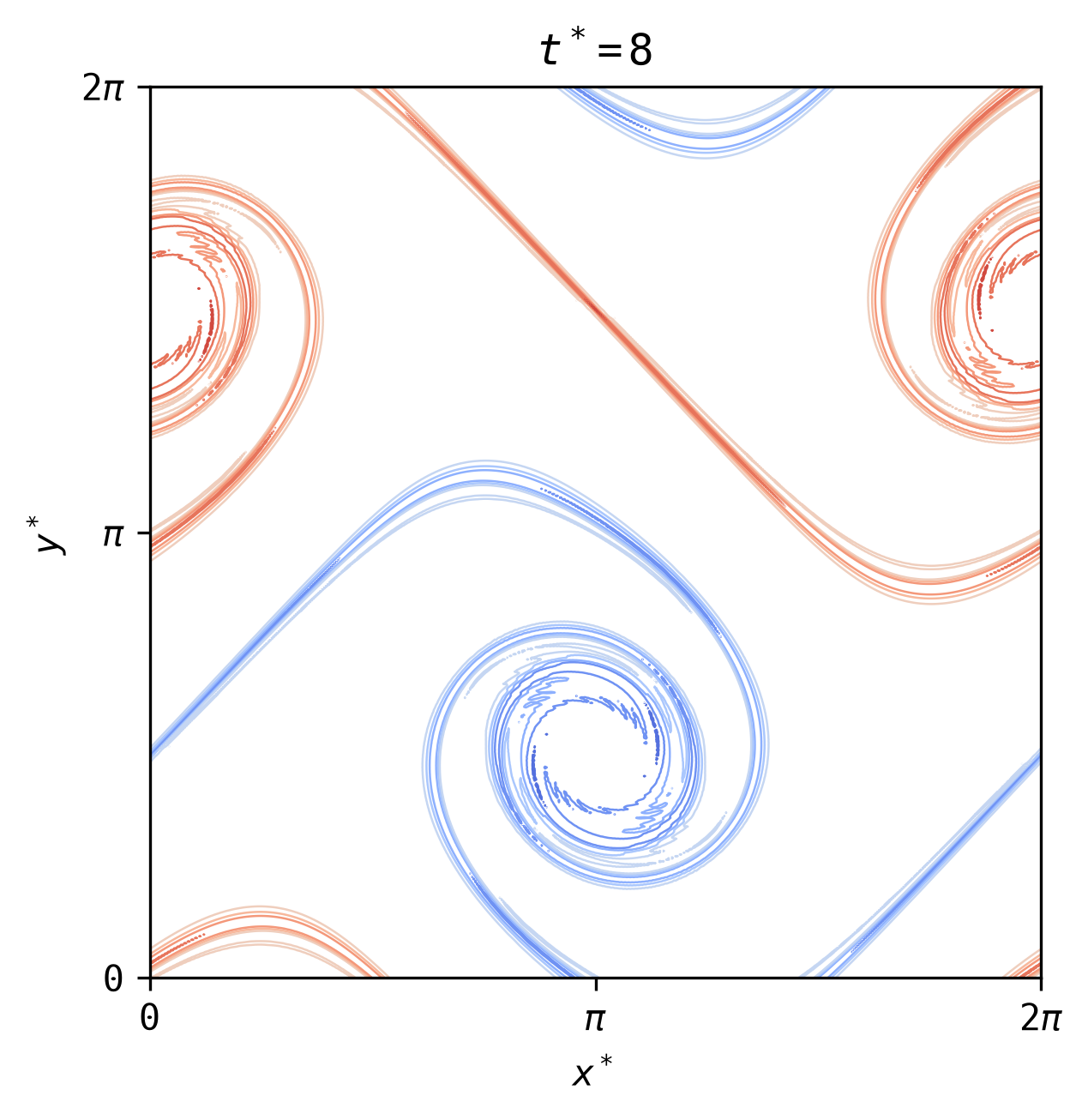}
	\end{subfigure}
	
	\caption{Evolution of the vorticity field in the periodic shear-layer
		roll-up test at $t^*=0$, $2$, $4$, $6$, $7$, and $8$. Blue and red contours denote negative (clockwise) and positive
		(counter-clockwise) vorticity, respectively.
		A $512\times512$ uniform grid is used.}
	\label{fig:shear_layer_evolution}
\end{figure}

\cite{PalhaGerritsma2017} present the vorticity field at the dimensionless
times $t^*=4$ and $t^*=8$, while \cite{Knikker2009} provides a corresponding
comparison at $t^*=8$. The Motif solution is shown in
Figure~\ref{fig:shear_layer_evolution} in the same fashion at
$t^*=0$, $2$, $4$, $6$, $7$, and $8$. These results are obtained using a
$512\times512$ uniform grid and a time-step size of
$\Delta t^*=0.00125$. Overall, the results are encouraging and show good
qualitative agreement with the reference solutions.

\section{Discrete conservation properties}

\subsection{Mass conservation}

Table~\ref{tab:shear_mass_momentum} reports the maximum local flux
imbalance during the shear-layer roll-up for the $512\times512$ uniform-grid
calculation with $\Delta t^*=0.00125$. The quantity remains of order
$10^{-15}$ throughout the calculation. Thus, the projection maintains the
discrete incompressibility constraint to approximately machine precision,
even as progressively smaller vortical structures develop.

\begin{table}[!htbp]
	\centering
	\caption{Discrete mass and momentum conservation during the periodic
		shear-layer roll-up test on the $512\times512$ uniform grid with
		$\Delta t^*=0.00125$.}
	\label{tab:shear_mass_momentum}
	\begin{tabular}{cccc}
		\hline
		$t^*$ &
		$\displaystyle
		\max\left|
		(\mathcal{D}_f\mathbf{V})_{i,j}
		\right|$ &
		$\Delta P_x^*$ &
		$\Delta P_y^*$ \\
		\hline
		0 & $0$                         & $0$                         & $0$ \\
		1 & $1.77849814\times10^{-16}$ & $ 1.59594560\times10^{-16}$ & $-2.22044605\times10^{-16}$ \\
		2 & $2.84955436\times10^{-16}$ & $ 1.04083409\times10^{-17}$ & $-2.22044605\times10^{-16}$ \\
		3 & $5.81782886\times10^{-16}$ & $-4.99600361\times10^{-16}$ & $-3.33066907\times10^{-16}$ \\
		4 & $1.02131845\times10^{-15}$ & $-5.30825384\times10^{-16}$ & $ 3.33066907\times10^{-16}$ \\
		5 & $1.52959242\times10^{-15}$ & $-5.27355937\times10^{-16}$ & $-1.11022302\times10^{-16}$ \\
		6 & $1.71694256\times10^{-15}$ & $-6.01949046\times10^{-16}$ & $-1.11022302\times10^{-16}$ \\
		7 & $2.16276649\times10^{-15}$ & $-6.03683770\times10^{-16}$ & $ 1.66533454\times10^{-15}$ \\
		8 & $1.65807038\times10^{-15}$ & $-6.45317133\times10^{-16}$ & $-1.11022302\times10^{-16}$ \\
		\hline
	\end{tabular}
\end{table}

\subsection{Momentum conservation}

Table~\ref{tab:shear_mass_momentum} reports the changes in the two
momentum components during the shear-layer roll-up for the
$512\times512$ uniform-grid calculation with $\Delta t^*=0.00125$.

Both momentum changes remain of order $10^{-15}$ or smaller throughout the
calculation. There is no systematic growth in either component. The
accumulated momentum in the periodic domain is therefore conserved to
approximately machine precision during the shear-layer roll-up.

\subsection{Kinetic-energy conservation}

For the $512\times512$ uniform-grid calculation with
$\Delta t^*=0.00125$, the initial kinetic energy was
$K(0)=1.713198991551\times10^{1}$. The relative changes were
$1.5649\times10^{-7}$ at $t^*=4$ and $1.1193\times10^{-7}$ at
$t^*=8$. The maximum absolute deviation over the saved states up to
$t^*=8$ was approximately $2.5118\times10^{-7}$.

Since AB2 does not conserve kinetic energy exactly in time, exact
kinetic-energy conservation is not expected for the fully discrete
calculation. This does not prevent its use for DNS. Rather, the temporal
kinetic-energy defect should decrease consistently with timestep refinement
and become sufficiently small at the timestep used for the calculation.
A timestep refinement study was therefore performed on the same
$512\times512$ grid. The
calculations were advanced to $t^*=2$ using
$\Delta t^*=0.0025$, $0.00125$, and $0.000625$.  The resulting relative
kinetic-energy changes are listed in Table~\ref{tab:shear_ke_dt}.
\begin{table}[!htbp]
	\centering
	\caption{Temporal sensitivity of the relative kinetic-energy change for
		the periodic shear-layer problem on the $512\times512$ grid.}
	\label{tab:shear_ke_dt}
	\begin{tabular}{ccc}
		\hline
		$\Delta t^*$ &
		$t^*=1$ &
		$t^*=2$
		\\
		\hline
		$0.0025$   & $1.6643\times10^{-8}$ & $5.6559\times10^{-8}$ \\
		$0.00125$  & $4.1643\times10^{-9}$ & $1.4155\times10^{-8}$ \\
		$0.000625$ & $1.0415\times10^{-9}$ & $3.5407\times10^{-9}$ \\
		\hline
	\end{tabular}
\end{table}

Halving the timestep reduces the kinetic-energy defect by approximately a
factor of four at both reporting times. The observed kinetic-energy error
therefore scales as $O(\Delta t^2)$, consistent with the second-order temporal
integration. The small residual kinetic-energy change is attributed to the time discretization rather than to a 
kinetic-energy defect of the spatial discretization. The temporal defect is
therefore convergent and can be controlled by timestep refinement. The
kinetic-energy behavior is consequently considered satisfactory for the
present DNS calculations.

\subsection{Pressure-term kinetic-energy conservation}

Table~\ref{tab:shear_pressure_work} reports the pressure contribution to the
kinetic-energy rate during the shear-layer roll-up for the
$512\times512$ uniform-grid calculation with $\Delta t^*=0.00125$.

\begin{table}[!htbp]
	\centering
	\caption{Pressure contribution to the kinetic-energy rate during the
		periodic shear-layer roll-up test on the $512\times512$ uniform
		grid with $\Delta t^*=0.00125$.}
	\label{tab:shear_pressure_work}
	\begin{tabular}{cc}
		\hline
		$t^*$ & $W_p^*$ \\
		\hline
		0 & $0$ \\
		1 & $-3.46944695\times10^{-18}$ \\
		2 & $-1.38777878\times10^{-17}$ \\
		3 & $-5.55111512\times10^{-17}$ \\
		4 & $0$ \\
		5 & $0$ \\
		6 & $ 4.44089210\times10^{-16}$ \\
		7 & $ 6.66133815\times10^{-16}$ \\
		8 & $ 4.44089210\times10^{-16}$ \\
		\hline
	\end{tabular}
\end{table}

The magnitude of $W_p^*$ remains below $6.7\times10^{-16}$ throughout the
calculation. Thus, the discrete pressure-gradient term makes no measurable
net contribution to the kinetic-energy rate and satisfies the expected
periodic-domain property to approximately machine precision. The small
kinetic-energy variations observed during the calculation therefore cannot
be attributed to net work by the discrete pressure term.

\subsection{Enstrophy behavior}

The sensitivity of this quantity to timestep refinement was first examined
on the $512\times512$ uniform grid.  The results are listed in
Table~\ref{tab:shear_enst_dt}.
\begin{table}[!htbp]
	\centering
	\caption{Temporal sensitivity of the relative enstrophy change $\Delta\mathcal{E}/\mathcal{E}_0$ for the
		periodic shear-layer problem on the $512\times512$ grid.}
	\label{tab:shear_enst_dt}
	\begin{tabular}{ccc}
		\hline
		$\Delta t^*$ &
		$t^*=1$ &
		$t^*=2$
		\\
		\hline
		$0.0025$   & $6.3741\times10^{-7}$ & $8.4090\times10^{-6}$ \\
		$0.00125$  & $4.8999\times10^{-7}$ & $8.0107\times10^{-6}$ \\
		$0.000625$ & $4.5311\times10^{-7}$ & $7.9111\times10^{-6}$ \\
		\hline
	\end{tabular}
\end{table}
In contrast to the kinetic-energy results, halving the timestep does not
produce a corresponding second-order reduction in the enstrophy defect.
The values instead approach a nonzero level as the timestep is reduced.
This indicates that, at this resolution, the enstrophy error is dominated
by the spatial discretization and by the discrete representation of
vorticity rather than by the temporal integration.

This interpretation is supported by spatial refinement at fixed
$\Delta t^*=0.0025$.  The corresponding results are given in
Table~\ref{tab:shear_enst_grid}.
\begin{table}[!htbp]
	\centering
	\caption{Grid sensitivity of the relative enstrophy change for the
		periodic shear-layer problem with $\Delta t^*=0.0025$.}
	\label{tab:shear_enst_grid}
	\begin{tabular}{ccc}
		\hline
		Grid &
		$t^*=1$ &
		$t^*=2$
		\\
		\hline
		$128\times128$ & $9.9870\times10^{-6}$ & $1.8204\times10^{-4}$ \\
		$256\times256$ & $2.1156\times10^{-6}$ & $3.5228\times10^{-5}$ \\
		$512\times512$ & $6.3741\times10^{-7}$ & $8.4090\times10^{-6}$ \\
		\hline
	\end{tabular}
\end{table}
The enstrophy defect decreases systematically with grid refinement.  At
$t^*=2$, successive grid doublings reduce the error by factors of
approximately $5.17$ and $4.19$, corresponding to observed orders of
approximately $2.37$ and $2.07$, respectively.  The asymptotic behavior is
therefore consistent with approximately second-order spatial convergence.

For the longer $512\times512$ calculation with
$\Delta t^*=0.0025$, the relative enstrophy changes were
$1.93\times10^{-4}$ at $t^*=4$ and $5.15\times10^{-4}$ at $t^*=8$.
The maximum absolute change over the saved states up to $t^*=8$ was
$1.36\times10^{-3}$.  

The increase of relative enstrophy change with time is consistent with the progressive
generation of finer vortical structures during roll-up. Enstrophy therefore
provides a substantially more demanding test of spatial resolution than
kinetic energy. However, the enstrophy defect decreases systematically
under grid refinement and approaches approximately second-order spatial
convergence. Therefore the numerical
enstrophy defect is bounded and converges toward zero with spatial
refinement. With this condition noted, Motif passes the present enstrophy test.

\section{Kinetic energy spectrum}

As explained in Section~\ref{sssec:spectrum_interpretation}, the present
$2\pi\times2\pi$ domain has $k_{\min}=1$. Since
$\Delta x=\Delta y=2\pi/N$, the largest wavenumber retained in the spectrum
is $k_{\max}=N/2$. The $128\times128$, $256\times256$, and
$512\times512$ grids therefore give $k_{\max}=64$, $128$, and $256$,
respectively.

\begin{figure}[!htbp]
	\centering
	\includegraphics[width=0.75\textwidth]
	{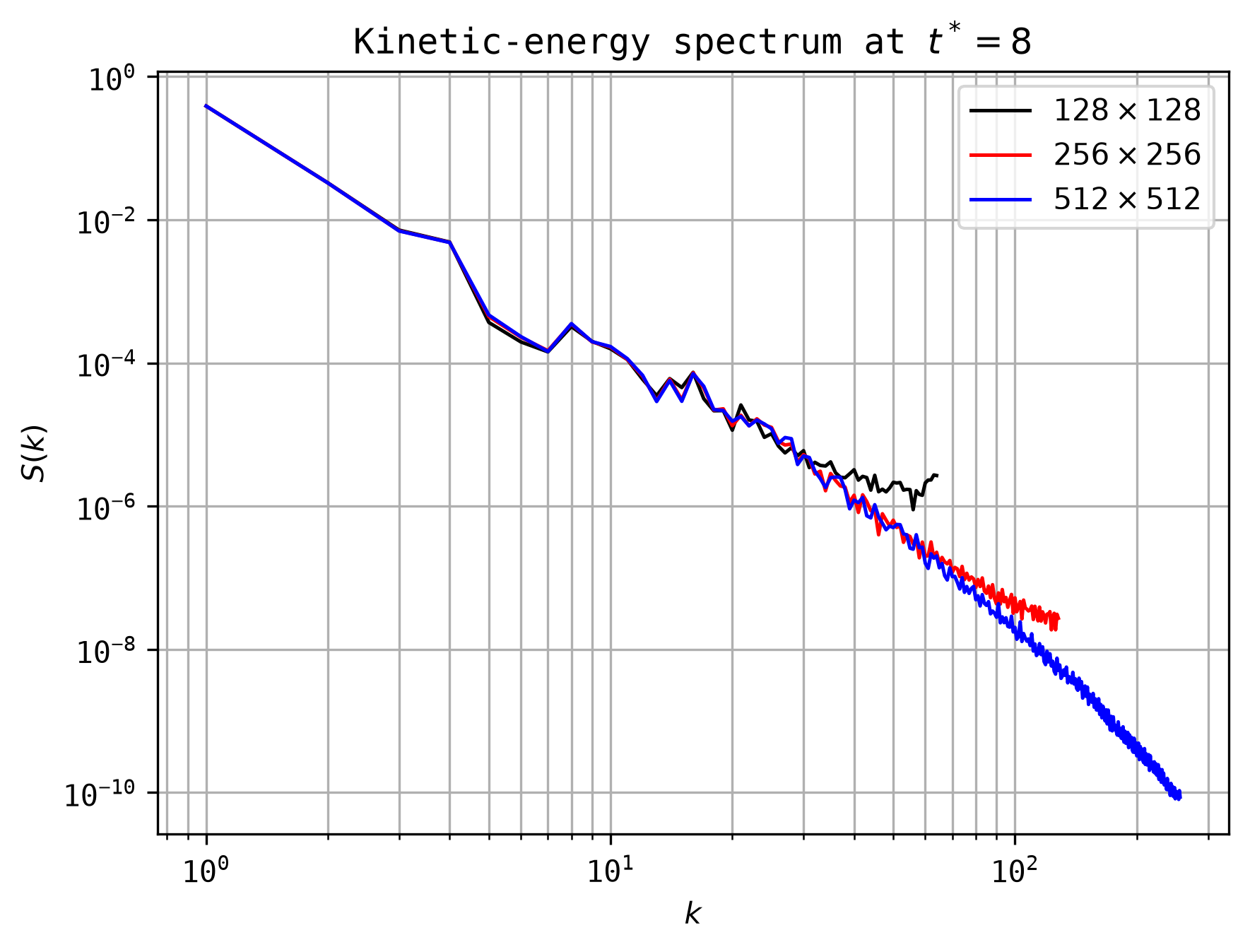}
	\caption{Kinetic-energy spectra at $t^*=8$ for the
		$128\times128$, $256\times256$, and $512\times512$
		uniform grids. The dimensionless time-step size is
		$\Delta t^*=0.00125$ for all three calculations.}
	\label{fig:shear_spectrum}
\end{figure}

Figure~\ref{fig:shear_spectrum} compares the kinetic-energy spectra at
$t^*=8$ obtained using the $128\times128$, $256\times256$, and
$512\times512$ uniform grids. The same dimensionless time-step size,
$\Delta t^*=0.00125$, is used in all three calculations.

For the present $2\pi\times2\pi$ periodic domain, the spacing between
Fourier wavenumbers is $\Delta k=1$. The spectrum is therefore formed using
wavenumber intervals of width $\Delta k=1$, centered at
$k=0,1,2,\ldots$. Each Fourier mode $(k_x,k_y)$ is assigned to the interval
nearest to its wavenumber magnitude $k=\sqrt{k_x^2+k_y^2}$. After excluding
the mean mode $k=0$, the $128\times128$, $256\times256$, and
$512\times512$ grids contain 64, 128, and 256 intervals, respectively,
within the retained wavenumber range.

We observe:

\begin{enumerate}
	
	\item
	The three spectra are nearly identical at low wavenumbers. They also remain
	close over a considerable intermediate-wavenumber range. Thus, the three
	grids predict essentially the same kinetic-energy distribution for the
	larger resolved flow structures.
	
	\item
	Differences become apparent as the wavenumber increases. The
	$128\times128$ spectrum departs from the two finer-grid results and
	becomes nearly horizontal. Such a plateau means that successive equal-width
	wavenumber intervals contain approximately the same amount of kinetic energy,
	although the represented spatial scales become progressively smaller. This
	behavior is not reproduced by the finer grids at the same wavenumbers. It is
	therefore a grid-dependent effect rather than a converged feature of the
	solution. The high-wavenumber plateau is consistent with the small-scale
	numerical structures observed in the corresponding $128\times128$ vorticity
	field.
	
	\item
	The $256\times256$ and $512\times512$ spectra remain close over a wider
	wavenumber range. The $256\times256$ spectrum eventually departs from the
	$512\times512$ result and begins to flatten. Thus, grid-dependent behavior
	moves toward larger wavenumbers as the grid is refined. The
	$512\times512$ spectrum shows no comparable plateau over the retained
	wavenumber range. Together with the systematic reduction of the enstrophy
	defect under grid refinement, these results support the use of the
	$512\times512$ grid for the present calculation through $t^*=8$.
	
	\item
	The spectral test therefore behaves as expected under grid refinement:
	grid-dependent behavior moves toward larger wavenumbers and smaller spatial
	scales. The $512\times512$ calculation provides the best-resolved result
	examined here and shows no comparable high-wavenumber plateau over the
	retained range. The spectral behavior is therefore considered satisfactory
	for the present DNS-suitability assessment through $t^*=8$. Since the problem
	is inviscid, continued evolution may generate still smaller scales and
	eventually require further grid refinement.
	
\end{enumerate}

\chapter{Conclusions}

This work presented Motif, a finite-volume solver developed for the
two-dimensional incompressible Navier--Stokes equations on orthogonal
Cartesian grids. The solver combines a staggered pressure--velocity
arrangement, conservative finite-volume discretization, ghost-layer
boundary conditions, and projection-based pressure--velocity coupling.
Uniform and nonuniform grids, explicit Euler and second-order
Adams--Bashforth/Crank--Nicolson time integration, and both Standard and
Perot-type approximate projections were considered. Through systematic
verification and numerical assessment, the present work examined not only
the formal accuracy of these components, but also their conservation
properties, numerical errors, and suitability for direct numerical
simulation.

The principal conclusions of the present work are as follows.

\begin{enumerate}
	
	\item
	The formulation was verified using Stokes' first problem, advection of a
	sine-wave velocity profile, inviscid and viscous advecting Taylor--Green
	vortices, a wall-bounded vortex, lid-driven cavity flow, and plane channel
	flow. These problems exercise convection, diffusion, pressure--velocity
	coupling, boundary conditions, and calculations on uniform and stretched
	grids. The verification studies demonstrate second-order spatial accuracy
	of both velocity and pressure. Second-order temporal accuracy of
	velocity is obtained in all tests with the Perot projection.
	
	\item
	The sine-wave advection test revealed negligible numerical dissipation but
	increasing dispersion as the wavelength became less resolved. This behavior
	is a well-known characteristic of centered convection discretizations
	\citep{LeVeque2002,Hirsch2007}. Upwind-biased, high-resolution, and
	dispersion-optimized formulations provide alternative means of controlling
	such errors, particularly in compressible-flow CFD. Motif deliberately
	retains the centered formulation to avoid introducing artificial numerical
	dissipation. Consequently, adequate
	spatial resolution remains essential when a broad range of spatial scales
	must be resolved.
	
	\item
	The temporal accuracy of pressure requires particular care in projection
	methods and has been discussed extensively in the literature
	\citep{kim1985application,VanKan1986,Bell1989,
		perot1993analysis,BrownCortezMinion2001}.
	The present results show that the computed pressure must first be
	interpreted at its proper midpoint time level. With this interpretation,
	the Standard projection exhibits first-order temporal convergence of
	pressure, whereas the Perot projection exhibits second-order convergence.
	The latter behavior is observed for both periodic and wall-bounded
	problems and on both uniform and smoothly stretched grids.
	
	\item
	Particular attention was given to the suitability of the method for DNS.
	The periodic shear-layer roll-up provides a demanding nonlinear test in
	which progressively smaller spatial scales are generated without physical
	viscous dissipation. Discrete mass and momentum are conserved to
	approximately machine precision. The pressure term produces no measurable
	net kinetic-energy contribution in the periodic domain.
	
	\item
	The numerical choices present in Motif do not imply exact conservation of
	every property examined. In particular, Adams--Bashforth time integration
	does not conserve kinetic energy exactly. The remaining kinetic-energy
	defect decreases approximately as $O(\Delta t^2)$ and is therefore
	associated with the temporal discretization. If exact discrete
	kinetic-energy conservation is required, an energy-conserving
	time-integration scheme may be used in place of Adams--Bashforth. The enstrophy defect decreases approximately as $O(\Delta x^2)$.
	
	\item
	The kinetic-energy spectra provide an additional measure of spatial
	resolution. High-wavenumber behavior which is grid dependent moves toward
	larger wavenumbers as the grid is refined. The $512\times512$ calculation
	shows no high-wavenumber plateau over the retained spectral range through
	$t^*=8$. Together with the conservation and refinement tests, these results
	support the suitability of the present numerical formulation for the DNS
	calculations considered here. This conclusion does not imply that a given
	grid or timestep is sufficient for every DNS problem. Spatial and temporal
	resolution must still be established for each calculation.
	
	\item
	The tests demonstrate that numerical accuracy should not be inferred only
	from the formal order of the individual discretizations. Conservation,
	temporal error, dispersion, small-scale behavior, and grid dependence must
	be examined for the complete numerical method. Motif was designed so that
	these properties can be tested directly and individual numerical components
	can be modified without changing the underlying grid and finite-volume
	structure.
	
\end{enumerate}

Extension to three dimensions is a natural next step for turbulent DNS. The same framework can
also support additional pressure--velocity coupling algorithms and governing
equations. As these capabilities are added, the emphasis will remain on
transparent discretization, conservation, verification, and direct
assessment of numerical error.

Motif is an ongoing development, and
this document will be updated as the solver evolves.

\bibliographystyle{plainnat}
\bibliography{references}

\end{document}